\documentclass[a4paper,11pt]{article}
\pdfoutput=1 

\usepackage{jcappub} 
\usepackage[T1]{fontenc}
\usepackage{tikz}
\usetikzlibrary{decorations.pathreplacing,decorations.markings, positioning}
\usepackage{fix-cm}
\DeclareMathOperator{\im}{Im}
\DeclareMathOperator{\re}{Re}
\usepackage{mathtools}
\usepackage{comment}
\usepackage{scalerel,tikz}
\usetikzlibrary{svg.path}
\definecolor{orcidlogocol}{HTML}{A6CE39}
\tikzset{orcidlogo/.pic={
 \fill[orcidlogocol] svg{M256,128c0,70.7-57.3,128-128,128C57.3,256,0,198.7,0,128C0,57.3,57.3,0,128,0C198.7,0,256,57.3,256,128z};
 \fill[white] svg{M86.3,186.2H70.9V79.1h15.4v48.4V186.2z}
 svg{M108.9,79.1h41.6c39.6,0,57,28.3,57,53.6c0,27.5-21.5,53.6-56.8,53.6h-41.8V79.1z M124.3,172.4h24.5c34.9,0,42.9-26.5,42.9-39.7c0-21.5-13.7-39.7-43.7-39.7h-23.7V172.4z}
 svg{M88.7,56.8c0,5.5-4.5,10.1-10.1,10.1c-5.6,0-10.1-4.6-10.1-10.1c0-5.6,4.5-10.1,10.1-10.1C84.2,46.7,88.7,51.3,88.7,56.8z};
}}
\newcommand\orcidicon[1]{\href{https://orcid.org/#1}{\mbox{\scalerel*{
\begin{tikzpicture}[yscale=-1,transform shape]
\pic{orcidlogo};
\end{tikzpicture}
}{|}}}}

\title{Dynamical Friction in a\\ Fuzzy Dark Matter Universe}

\author[a,1]{Lachlan Lancaster~\orcidicon{0000-0002-0041-4356},\note{Corresponding author.}}
\author[b]{Cara Giovanetti~\orcidicon{0000-0003-1611-3379},}
\author[a,2]{Philip Mocz~\orcidicon{0000-0001-6631-2566},\note{Einstein Fellow}}
\author[c,d]{Yonatan Kahn~\orcidicon{0000-0002-9379-1838},}
\author[b]{Mariangela Lisanti~\orcidicon{0000-0002-8495-8659},}
\author[a,b,e]{David N. Spergel~\orcidicon{0000-0002-5151-0006}}

\affiliation[a]{Department of Astrophysical Sciences, Princeton University, Princeton, NJ, 08544, USA}
\affiliation[b]{Department of Physics, Princeton University, Princeton, New Jersey, 08544, USA}
\affiliation[c]{Kavli Institute for Cosmological Physics, University of Chicago, Chicago, IL, 60637, USA}
\affiliation[d]{University of Illinois at Urbana-Champaign, Urbana, IL, 61801, USA}
\affiliation[e]{Center for Computational Astrophysics, Flatiron Institute, NY, NY 10010, USA}

\emailAdd{lachlanl@princeton.edu}
\emailAdd{pmocz@astro.princeton.edu}
\emailAdd{caralg@princeton.edu}
\emailAdd{yfkahn@illinois.edu}
\emailAdd{mlisanti@princeton.edu}
\emailAdd{dspergel@flatironinstitute.org}

\newcommand{\rhobar}{\overline{\rho}}
\newcommand{\lambdabar}{{\mkern0.75mu\mathchar '26\mkern -9.75mu\lambda}}
\newcommand{\vrel}{v_{\rm rel}}
\newcommand{\kt}{\tilde{k}}
\newcommand{\Rt}{\widetilde{R}}
\newcommand{\zt}{\tilde{z}}
\newcommand{\bt}{\tilde{b}}
\newcommand{\diff}{\mathrm{d}}
\definecolor{DarkBlue}{rgb}{0.0, 0.0, 0.55}

\abstract{We present an in-depth exploration of the phenomenon of dynamical 
friction in a universe where the dark matter is composed entirely of so-called 
Fuzzy Dark Matter (FDM), ultralight bosons of mass $m\sim\mathcal{O}(10^{-22})\,$eV. 
We review the classical treatment of dynamical friction before presenting 
analytic results in the case of FDM for point masses, extended mass distributions, 
and FDM backgrounds with finite velocity dispersion. We then test these results 
against a large suite of fully non-linear simulations that allow us to assess the 
regime of applicability of the analytic results. We apply these results  to a variety of astrophysical problems of interest, including infalling satellites in a galactic dark matter background, and determine that 
\emph{(1)}~for FDM masses $m\gtrsim 10^{-21}\, {\rm eV}\, c^{-2}$, the timing problem of the Fornax dwarf 
spheroidal's globular clusters is no longer solved and 
\emph{(2)}~the effects of FDM on the process of dynamical friction for satellites of total mass $M$ and relative velocity $v_{\rm rel}$ should require detailed numerical simulations for 
$\left(M/10^9~M_{\odot}\right) \left(m/10^{-22}~{\rm eV}\right)\left(100~{\rm km}~{\rm s}^{-1}/v_{\rm rel}\right) \sim 1$, parameters which would lie outside the validated range of applicability of any currently developed analytic theory, due to transient wave structures in the time-dependent regime.}

\begin{document}
\maketitle
\flushbottom

\section{Introduction}
\label{sec:intro}

The standard cosmological model, developed over the past several decades, 
has been extraordinarily successful in explaining the Universe that we 
observe around us on the largest scales. The Dark Energy ($\Lambda$) and 
Cold Dark Matter (CDM) model simultaneously explains the power spectrum of the 
Cosmic Microwave Background (CMB) radiation \cite{WMAP03,Planck18} and the distribution 
of large-scale structure \cite{BAODR12},
while detailed numerical cosmological simulations of dark matter and baryons are able to self-consistently create a diverse population of realistic galaxies
\cite{2018MNRAS.475..676S}. 
Yet, for all these successes, we lack 
a fundamental understanding of the nature of both dark energy and dark matter \citep{2018RPPh...81f6201R}.
In the case of dark matter, there is an additional mismatch between 
the predictions of the theory of CDM and observations on cosmologically small 
scales \cite[and references therein]{MBKJBRev17}. Though some of 
these discrepancies have standard astrophysical explanations 
\cite{PontzenGovernato12,Read16,Zolotov12,Wetzel16,Hopkins18,Dutton16},
they have also motivated the development of alternative theories of dark 
matter that can solve the small-scale discrepancies while remaining consistent 
with CDM on large scales \cite{HOTW,Goodman00,Hu00fdm,SIDM01,Berezhiani15}.

A model that has become popular recently as an alternative to CDM treats the  
dark matter as an `ultralight' boson field~\cite{Hu00fdm,HOTW}. This model 
posits that the dark matter is so light, with mass 
$m \sim \mathcal{O}\left(10^{-22} \, {\rm eV}\right)$, that it has a de 
Broglie wavelength on kiloparsec scales and thus exhibits wave phenomenon 
on galactic length scales~\citep{Schive14,Schive14b,Mocz17,BarOr18,Church19,2019PhRvL.123n1301M}. 
Following the work of \cite{Hu00fdm} and \cite{HOTW}, we shall refer to this 
model as Fuzzy Dark Matter (FDM). Though FDM seems promising in explaining a 
number of small-scale observational inconsistencies with CDM theory, it can 
drastically change the dynamics of galaxies as compared to CDM 
\cite{HOTW,Mocz17,Schive14,Schive14b,MarshSilk14,2019PhRvL.123n1301M}.  Some ways in which 
galactic dynamics in FDM varies most strongly from CDM are dynamical heating 
and friction due to FDM's wavelike substructure~\citep{HOTW,BarOr18,Church19}. 

In this paper, we investigate how the phenomenon of dynamical 
friction, which controls the merging of galaxies \cite{Tremaine75,Tremaine76}, 
the slowing down of spinning galactic bars \cite{TW84,Weinberg85}, and the 
coalescence of supermassive black holes (SMBH) in galaxy mergers \cite{BBR80,Yu02},
changes in an FDM paradigm.\footnote{See also \cite{Berezhiani:2019pzd} for a detailed investigation of dynamical friction in superfluid backgrounds with nonzero sound speed $c_s$; our FDM analysis here is equivalent to the case $c_s = 0$.}  The presence of quantum mechanical 
pressure and `quasi-particles' arising in velocity-dispersed media in the FDM 
paradigm alters the formation of the dark matter wake and therefore 
warrants a detailed investigation of all relevant cases, including finite-size effects for the infalling objects.  Dynamical interactions may also reshape the structure of
the dark matter subhalos and reduce the tension between the predicted profiles of isolated halos~\cite{Schive14} and the 
observed dwarf galaxy profiles~\cite{Safarzadeh2019}. A key achievement of this paper is the comparison of analytic perturbation theory results to fully non-linear numerical simulations in order to clearly identify where the 
perturbation theory results are applicable. We additionally identify how the application of perturbation theory in the non-linear regime can bias inferences.

This paper is organized as follows. In Section 
\ref{sec:Scales}, we begin by summarizing our conventions and notation, including specifying the key dimensionless numbers which control dynamical friction effects for our systems of interest.  In Section \ref{sec:ClassicalDF}, we give an overview of 
the classical treatment of dynamical friction both in the context of dark 
matter and in collisional media, to provide a more familiar context for 
the discussion that follows. Next, in Section \ref{sec:Analytics}, we review the analytic theory of dynamical friction in an  FDM background for a point mass, and derive new results for finite-size satellites and FDM backgrounds with velocity dispersion. We then perform a series of simulations of the formation of the dark 
matter wake in several different contexts to test our analytic 
predictions. These simulations are described in Section \ref{sec:Numerical}, and
we compare the results of our analytic calculations to these numerical 
simulations in Section \ref{sec:Results}. Finally, in Section
\ref{sec:Application}, we discuss the consequences of our results for 
several concrete astrophysical observables and the associated constraints on the FDM model, in particular deriving an upper bound on the FDM mass needed to explain the infall times of the Fornax globular clusters. We conclude in Section \ref{sec:Conclusions}. The three Appendices contain technical details of our analytic calculations and numerical simulations.

\section{Notation and Scales}
\label{sec:Scales}
We begin by defining our notation and specifying the physical systems of interest. In this paper, we investigate how a mass $M$ moving 
through an infinite background of average mass density $\rho$ with velocity 
$\vrel$ relative to the background feels an effective drag force from the 
gravitational wake accumulated during its motion.  The physical situation we 
have in mind is a large object like a globular cluster, a satellite galaxy, 
or an SMBH moving in a galactic dark matter halo, so we will refer to 
$M$ as a satellite. As we describe in Sec.~\ref{sec:madelung}, we can treat the FDM background as a condensate, and we will often refer 
to it as such.  We will define the `overdensity' $\alpha(\mathbf{x})$ of the background medium as the fractional change of the background density $\rho(\mathbf{x})$ from the mean density $\rhobar$:
\begin{equation}
    \label{eq:overdense}
    \alpha(\mathbf{x}) \equiv \frac{\rho(\mathbf{x}) - \rhobar}{\rhobar} \, .
\end{equation}

There are several important reference scales that we will use throughout the 
paper. The first of these is the de Broglie wavelength associated with the relative 
velocity of the satellite and the condensate, which we will refer to as the 
background de Broglie wavelength 
\footnote{Note, the more typical de Broglie wavelength is $2\pi$ times this quantity.}:
\begin{equation}
    \label{eq:lambda_rel}
    \lambdabar = \frac{\hbar}{m v_{\rm rel}} \, ,
\end{equation}
where $\hbar$ is the reduced Planck's constant and $m$ is the mass of the 
FDM particle.
In what follows, we will often put length scales in units of $\lambdabar$; we will 
indicate this by placing tildes above variables which are represented in 
units of $\lambdabar$ (\emph{i.e.}, $\tilde{x}\equiv x / \lambdabar$). We can also 
put wave vector quantities in these units as $\tilde{k} = k \lambdabar$.

There is a characteristic quantum size associated with the mass of the 
satellite $M$, which can be interpreted as a gravitational Bohr radius, defined as
\begin{equation}
L_{\rm Q} = \frac{(\hbar/m)^2}{G M}\, ,
\end{equation}
where $G$ is Newton's constant. Its related velocity scale is 
\begin{equation}
v_{\rm Q} = \frac{\hbar}{mL_{\rm Q}}\, .
\end{equation}
For satellites of finite size, there is an additional length scale 
$\ell$ corresponding to the classical size of the satellite (for 
example, its core radius). 

Finally, we will consider cases where there is some finite background 
velocity dispersion in the condensate, denoted by $\sigma$. The de 
Broglie wavelength associated with this dispersion is given by
\begin{equation}
    \label{eq:lambda_sigma}
    \lambdabar_{\sigma}\equiv \frac{\hbar}{m \sigma} \, .
\end{equation}

Using ratios of the above quantities we can fully describe the most 
general system we will consider in terms of three dimensionless 
quantities:
\begin{itemize}
    \item[1.] The quantum Mach number,
    \begin{equation}
        \label{eq:MQ}
        \mathcal{M}_{\rm Q} \equiv \frac{v_{\rm rel}}{v_{\rm Q}} \, ,
    \end{equation}
    which is equivalent to the inverse of the parameter $\beta$ 
    discussed in Appendix D of \cite{HOTW}. Since $\mathcal{M}_Q$ is 
    inversely proportional to $M$, we expect perturbation theory to 
    work best in the limit of $\mathcal{M}_Q \gg 1$. To give an idea 
    of the order-of-magnitude scale, we can write the quantum Mach 
    number as:
    \begin{equation}
        \label{eq:MQ_calc_aid}
        \mathcal{M}_{\rm Q} = 44.56  
        \left(\frac{v_{\rm rel}}{1~{\rm km}~{\rm s}^{-1}}\right)
        \left(\frac{m}{10^{-22}~{\rm eV}}\right)^{-1}
        \left(\frac{M}{10^5 M_\odot}\right)^{-1}\, .
    \end{equation}
    
    \item[2.] The classical Mach number,
    \begin{equation}
        \label{eq:M_sig}
        \mathcal{M}_{\sigma} \equiv \frac{\lambdabar_{\sigma}}{\lambdabar}
        = \frac{\vrel}{\sigma} \, .
    \end{equation}
    While the first expression defines $\mathcal{M}_\sigma$ as the ratio 
    of two de Broglie wavelengths, the second expression makes clear that 
    $\mathcal{M}_\sigma$ is purely classical (and independent of the satellite 
    mass), facilitating comparisons to dynamical friction in systems with 
    classical backgrounds.
    
    \item[3.] The dimensionless satellite size,
    \begin{equation}
        \label{eq:ltil}
        \tilde{\ell} \equiv \frac{\ell}{\lambdabar} \, ,
    \end{equation}
    which we define as the ratio of the satellite size to the background 
    de Broglie wavelength. In this work, we consider for the first time 
    effects that depend on nonzero $\tilde{\ell}$, which allows us to apply 
    our results to realistic systems. Again, to give an idea of the scale 
    of this dimensionless parameter, we may write:
    \begin{equation}
        \label{eq:tl_calc_aid}
        \tilde{\ell} = 5.22 \times 10^{-5}
        \left(\frac{\ell}{1~{\rm pc}} \right)
        \left(\frac{\vrel}{1~{\rm km}~{\rm s}^{-1}} \right)
        \left(\frac{m}{10^{-22}~{\rm eV}} \right).
    \end{equation}
\end{itemize}

When we derive the dynamical friction forces below,
it will be helpful to define a reference force value in terms 
of the dimensionful constants that we have listed above. We therefore define
\begin{equation}
    \label{eq:F_defs}
    F_{\rm rel} \equiv 4\pi \rhobar  \left( \frac{G M}{\vrel}\right)^2 \, .
\end{equation}
From this, we may define the dimensionless dynamical friction coefficient,
\begin{equation}
    \label{eq:Cdef}
    C_{\rm rel} \equiv \frac{F_{\rm DF}}{F_{\rm rel}} \, ,
\end{equation}
where $F_{\rm DF}$ is the total dynamical friction force experienced by the 
satellite in any given scenario.

\section{Classical Treatment of Dynamical Friction}
\label{sec:ClassicalDF}

In this section, we review the classical theory of dynamical friction.
There are two fundamental ways of tackling this problem. The first 
consists of treating the background as an infinite medium of `field' 
particles of mass $m_{\rm f}$ and number density $n_{\rm f}$ such that the 
background mass density is $\rho = m_{\rm f} n_{\rm f}$. We then consider 
the aggregate effect of many two-body interactions between the field masses 
and the satellite mass under the assumption that the satellite mass $M$ 
satisfies $M \gg m_{\rm f}$.  As first discussed by \cite{1943ApJ....97..255C}, 
these assumptions lead to an estimate of the diffusion of the satellite or 
subject particle through phase space. We will discuss this approach in 
Section \ref{subsec:Chandrasekhar_df}.

The second approach consists of calculating the form of the 
gravitational wake from the equations of motion of the background, treating the background 
as a continuous fluid \citep{Marochnik68,Kalnajs72}. The dynamical friction force is 
then calculated by integrating the gravitational force of the 
over-dense wake on the satellite. This approach has been used to calculate 
the dynamical friction in various other contexts \citep{Ostriker99}; 
we will discuss it in Section \ref{subsec:overdensity_calculation}.

\subsection{Phase Space Diffusion}
\label{subsec:Chandrasekhar_df}

In this approach, we use the Fokker-Planck approximation to model how the 
satellite, often referred to as the `subject' particle in this approach,
interacts via many two-body interactions with `field' 
particles~\cite[Sec. 7.4]{2008gady.book.....B}.  This approximation works 
in the context of the collisional Boltzmann Equation:
\begin{equation}
    \label{eq:fokker_planck}
    \frac{\diff f}{\diff t} = \Gamma \left[f \right] \, ,
\end{equation}
where $f$ is the phase-space distribution function of the satellite
(treated as a point mass) and $\Gamma \left[ f \right]$ is the encounter 
operator, which describes how collisions with field particles change the 
satellite's `normal' or `collisionless' path through phase space. The 
encounter operator can be written in terms of the transition probability function 
$\Psi\left(\mathbf{w},\Delta\mathbf{w} \right)\diff^6\left(\Delta\mathbf{w}\right)$, 
which describes the probability per unit time that the satellite at phase-space 
coordinate $\mathbf{w}$ is scattered into the volume of phase space 
$\diff^6\left(\Delta\mathbf{w}\right)$ centered on $\mathbf{w} + \Delta\mathbf{w}$. 

Under the Fokker-Planck approximation, we approximate the encounter 
operator in terms of the first two moments of the transition 
probability, $D\left[\Delta w_i\right]$ and $D\left[\Delta w_i \Delta w_j\right]$:
\begin{equation}
    \label{eq:encounter_op_fp}
    \Gamma \left[ f\right] \approx - \sum_{i=1}^6 \, \frac{\partial}{\partial w_i}
    \left\lbrace D \left[ \Delta w_i\right]f(\mathbf{w})\right\rbrace 
    + \frac{1}{2} \sum_{i,j=1}^6 \frac{\partial^2}{\partial w_i \partial w_j}
    \left\lbrace D\left[ \Delta w_i \Delta w_j\right] f(\mathbf{w})\right\rbrace \, ,
\end{equation}
where 
\begin{equation}
    D\left[ \Delta w_i \right] \equiv \int \diff^6 \left( \Delta \mathbf{w}\right)
    \Psi\left(\mathbf{w}, \Delta \mathbf{w} \right) \Delta w_i 
\end{equation}
quantifies the steady drift through phase space, and
\begin{equation}
    D\left[ \Delta w_i \Delta w_j \right] \equiv \int \diff^6 \left( \Delta \mathbf{w}\right)
    \Psi\left(\mathbf{w}, \Delta \mathbf{w} \right) \, \Delta w_i \,\Delta w_j
\end{equation}
quantifies the amount by which the star undergoes a random walk through phase space.

The second-order diffusion coefficient is scaled by a factor of $m_{\rm f}/M$ 
relative to the first-order diffusion coefficient.   In the case where the satellite 
is much more massive than the field particles ($M\gg m_{\rm f}$), as we have here, 
the second-order diffusion coefficient is much smaller than the first-order 
diffusion coefficient, so we can ignore it.  
To evaluate $D\left[ \Delta w_i \right]$, we adopt simple Cartesian 
positions and velocities $(x_i, v_i)$ as our coordinates on phase space.  By the 
symmetries of the problem, the only diffusion coefficients that are non-zero are 
those in the direction of motion of the satellite. If we assume that the 
satellite is moving in the $\mathbf{\widehat{x}_{||}}$ direction, the only 
diffusion coefficients that we need to worry about are 
$D\left[ \Delta v_{||}\right]$ and $D\left[ \Delta x_{||}\right]$. In the case of 
dynamical friction, we can assume that the interactions between the satellite and 
field particles take place over a short enough period of time so that they only 
affect the velocity of the satellite and not its position, to first 
approximation \citep{Weinberg86,2008gady.book.....B}. Thus, we only need to 
worry about $D\left[ \Delta v_{||} \right]$ moving forward.

The evaluation of diffusion coefficients is dependent upon the distribution 
function of field particles $f_{\rm f}(\mathbf{w})$ and is provided 
in full in Appendix L of \cite{2008gady.book.....B}. 
The result needed here is
\begin{equation}
    \label{eq:Dv_parallel}
    D \left[\Delta v_{||} \right] = - \frac{16 \pi^2 m_{\rm f} \left(m_{\rm f} + M\right) \ln \Lambda}{\vrel^2} \int_0^{\vrel}\diff v_{\rm f} \, v_{\rm f}^2 \,f_{\rm f}(v_{\rm f}) \, ,
\end{equation}
where we have assumed that the distribution of the field particles is 
homogeneous in position space and isotropic in velocity space, but have 
not yet specified $f_{\rm f}(v_{\rm f})$, the velocity distribution of the field particles. 
Note that field particles moving faster than the 
satellite do not factor into the diffusion coefficient, as indicated 
by the limits of the integral above.  We have also introduced the Coulomb 
logarithm, $\Lambda$, which is defined as 
\begin{equation}
    \label{eq:Coulomb_log_def}
    \Lambda \equiv \frac{b}{\ell_{90}} \, ,
\end{equation}
where $b$ is the maximum distance at which the field particles are 
still interacting with the satellite and $\ell_{90}$ is the distance 
at which a field particle has to approach the satellite to be deflected 
by 90$^{\circ}$. Note that this definition assumes  
that \emph{(1)}~the medium through which the satellite travels is infinite
and \emph{(2)}~the satellite is a point mass. The first assumption means 
that if we include arbitrarily large scales, the dynamical friction 
would be infinite, as there would be an infinite number of field stars 
acting on the satellite.  As a result, we must stop counting stars after 
a certain distance. 
The second assumption implies that a field star interacting with the 
satellite can have a large effect on the satellite's motion, changing 
its relative motion by a factor of $2 \vrel$ for an impact parameter 
of 0. In this limit, our assumption of small velocity changes (a diffusion 
through phase space) breaks down, so we must regulate this assumption by 
only including contributions with impact parameters great than $\ell_{90}$.

For the case of a Maxwellian velocity distribution of the field stars, 
we can evaluate the diffusion coefficient above as:
\begin{equation}
    \label{eq:diffusion_coeff_maxwell}
     D \left[\Delta v_{||} \right] = - \frac{4 \pi^2 M G^2 \,\rhobar \,\ln \Lambda}{\sigma^2}
     \mathbb{G}\left( X\right) \, ,
\end{equation}
where $\sigma$ is the one-dimensional velocity dispersion of the Maxwellian 
distribution of the field stars, $\rhobar$ is their mean density, $X \equiv \vrel /\sqrt{2}\sigma$, and 
\begin{equation}
    \label{eq:GX_def}
    \mathbb{G}(X) = \frac{1}{2X^2} \left[{\rm erf}(X) - \frac{2X}{\sqrt{\pi}}e^{-X^2} \right] \, ,
\end{equation}
where ${\rm erf}(X)$ is the error function.  We can then use 
Eqs.~\ref{eq:fokker_planck},  \ref{eq:encounter_op_fp}, 
and~\ref{eq:diffusion_coeff_maxwell} to relate the dynamical friction 
force to the diffusion coefficient as
\begin{equation}
    \label{eq:chandrasekhar_form}
    F_{\rm DF} = - M D\left[\Delta v_{||} \right]
    = \frac{4 \pi^2 M^2 G^2 \, \rhobar \, \ln \Lambda}{\sigma^2} 
    \mathbb{G}\left( X\right)
    = 2\pi F_{\rm rel} X^2 \ln \Lambda \mathbb{G}\left( X\right)\, .
\end{equation}
This result was first derived by \cite{1943ApJ....97..255C}.

\subsection{Overdensity Calculation}
\label{subsec:overdensity_calculation}

An independent derivation of the dynamical friction force involves directly 
calculating the overdensity in the medium induced by the satellite's gravity. We can then 
determine the dynamical friction force on the satellite, or `perturber' as it is often referred to in this approach, by integrating the force from each 
mass element of the overdense medium. This approach has been employed in modeling classical collisionless particles with some velocity dispersion 
\cite{Marochnik68,Kalnajs72}, as well as collisional gases~\cite{Ostriker99}.  
We will employ these methods in the majority of the rest of the paper to find 
the mathematical form of the overdensity, defined in Eq.~\ref{eq:overdense}.

For consistency, we briefly outline here the case of a medium of collisionless 
`field' particles with a Maxwellian velocity distribution, as done 
in Section \ref{subsec:Chandrasekhar_df}. For this case, the relevant 
evolution equation is the collisionless version of Eq.~\ref{eq:fokker_planck}:
\begin{equation}
    \label{eq:CBE}
    \frac{\diff f_{\rm f}}{\diff t} = \frac{\partial f_{\rm f}}{\partial t} + 
    \mathbf{v} \cdot \frac{\partial f_{\rm f}}{\partial \mathbf{x}} 
    - \frac{\partial U}{\partial \mathbf{x}} \cdot
    \frac{\partial f_{\rm f}}{\partial \mathbf{v}} = 0 \, ,
\end{equation}
where $f_{\rm f}$ is again the distribution function of the field particles  
and $U(\mathbf{x})$ is the gravitational potential of the satellite.  To 
solve for the overdensity that the field stars make in response to the 
potential of the perturber, we first linearize Eq.~\ref{eq:CBE} around the zeroth-order solution of a uniform medium with a Maxwellian velocity 
distribution at every point in space. We then solve the equation for the 
first-order deviation from this zeroth-order solution. We then simply 
integrate the first-order distribution function over all of velocity 
space to obtain the overdensity. This process is shown in detail in 
Appendix A of \cite{Weinberg86}.

After obtaining the overdensity, we can compute the dynamical friction 
force using 
\begin{equation}
    \label{eq:df_def}
    F_{\rm DF} = \rhobar \int \diff^3 \mathbf{x} \, 
    \alpha(\mathbf{x}) \,\left(\mathbf{\hat{x}}_{||}\cdot \nabla \right)  U\left(\mathbf{x} \right) \, ,
\end{equation}
where $U(\mathbf{x})$ is the potential of the satellite and $\mathbf{\hat{x}}_{||}$ is the unit vector pointing in the direction of the satellite's motion.  We will go 
through the details of this calculation in much more depth below for 
the case of an FDM wake.

\section{Dynamical Friction in a Condensate: Analytic Theory}
\label{sec:Analytics}

In this section, we will modify the discussion of Sec.~\ref{sec:ClassicalDF} 
for a condensate background, rather than a classical background. 
Here, when we say ``condensate'', we simply mean a complex field $\psi$ whose equation of motion is the 
Schr\"odinger equation with a gravitational source term given by
\begin{equation}
\label{eqn:S1}
i\frac{\partial \psi}{\partial t}
= \left[-\frac{\hbar}{m}\frac{\nabla^2}{2} + \frac{m}{\hbar}U \right]  \psi  \, ,
\end{equation}
where $m$ is the mass 
of the constituent particle of the condensate. ``Condensate'' is 
intended to be synonymous with the FDM; we use this more general term 
rather than ``superfluid'' (which has also appeared in the literature) 
because we are agnostic as to the nature of the sign of a possible 
self-interaction term between FDM particles. 
It should be noted that this is a non-relativistic approximation to 
the underlying field theory that describes the field $\psi$ and as such 
is not applicable on short enough time/length scales, though it is valid 
for all scales probed in this paper \cite{GHPW15}.
The wave function is typically normalized so that the 
mass density of the condensate is given by $\rho = \left| \psi \right|^2$.
We will assume that $U$ is 
dominated by the gravitational potential of the satellite. 
In principle, at high enough background densities, the 
condensate's own self-gravity becomes important, and 
the $U$ in Eq.~\ref{eqn:S1} becomes a combination of the 
gravity of the satellite and the self-gravity of the background.

As in the classical problem reviewed above, dynamical friction occurs 
due to the gravitational force between the satellite and the `wake' in 
the condensate that forms behind it as it moves through the condensate background. 
Below, we will review how this wake forms in various treatments of the 
problem, varying the method of approach (exact solution versus linear perturbation 
theory), the mass distribution of the satellite (point source versus extended), 
and the nature of the velocity distribution of the condensate (plane wave 
versus velocity-dispersed).

\subsection{Madelung Formalism}
\label{sec:madelung}

We begin by reviewing the key formalism for the treatment of this 
problem in linear perturbation theory (LPT): the Madelung formalism, 
developed nearly 100 years ago \citep{Madelung27}. In this formalism 
the wave function $\psi$ is decomposed in terms of its magnitude
and phase, 
\begin{equation}
    \label{eq:wave_decomp}
    \psi = \sqrt{\rho} \, e^{i\theta} \, ,
\end{equation}
where as noted above, $\rho$ is interpreted as the mass density of the 
condensate and $\theta$ is the phase of the wave function. Defining
\begin{equation}
    \label{eq:velocity_def}
    \mathbf{u} = \frac{\hbar}{m} \nabla \theta \, ,
\end{equation}
we may rewrite the Schr\"odinger equation \ref{eqn:S1} as two real 
partial differential equations in $\rho$ and $\mathbf{u}$ 
(equivalently $\theta$),
\begin{equation}
    \label{eq:madelung1}
    \frac{\partial \rho}{\partial t} + 
    \nabla \cdot \left(\rho \mathbf{u} \right) = 0
\end{equation}
and 
\begin{equation}
    \label{eq:madelung2}
    \frac{\partial \mathbf{u}}{\partial t} + 
    \left(\mathbf{u} \cdot \nabla \right) \mathbf{u} = 
    - \nabla U - \nabla U_Q \, ,
\end{equation}
which are simply the equations for an incompressible fluid under the 
potential $U$ with an additional term given by the gradient of a 
``quantum pressure''
\begin{equation}
    \label{eq:vq_def}
    U_Q \equiv - \frac{\hbar^2}{2m^2} \frac{\nabla^2 \sqrt{\rho}}{\sqrt{\rho}} \, .
\end{equation}
The correspondence between these equations and the classical 
fluid equations has been studied in depth in the literature 
\cite{WidrowKaiser93,2018PhRvD..97h3519M}. A particularly in-depth and recent numerical study of the correspondence between 
the Schr\"odinger-Poisson equation and the Madelung formalism 
is given in \cite{Li19}.

This formalism provides a convenient starting point from which to carry out 
the perturbation theory analysis, which follows \cite{Lora12}. We consider the problem in the rest frame of the satellite, and assume that 
$\rho$ and $\mathbf{u}$ have mean solutions $\rhobar$ and 
$\mathbf{v}_{\rm rel}$, respectively, which are independent of time and space, and have small perturbations $\delta \rho$ and $\delta \mathbf{v}$, which are sourced by the satellite. This mean background solution would be expressed in 
the wave function as
\begin{equation}
    \label{eq:ic}
    \psi_0 (\mathbf{x}) = \sqrt{\rhobar} \,
    e^{i\frac{m}{h}\mathbf{x} \cdot \mathbf{v}_{\rm rel}} \, .
\end{equation}
If we make the replacements $\rho \to \rhobar + \delta \rho$ and 
$\mathbf{u} \to \mathbf{v}_{\rm rel} + \delta \mathbf{v}$ 
in Eqs.~\ref{eq:madelung1} and \ref{eq:madelung2}, keeping only 
terms linear in the perturbations, Eq.~\ref{eq:madelung1} becomes (using the overdensity defined in Eq.~\ref{eq:overdense})
\begin{equation}
    \label{eq:madelung1_lpt}
    \frac{\partial \alpha}{\partial t} + 
    \left(\mathbf{v}_{\rm rel} \cdot \nabla \right) \alpha + 
    \nabla \cdot \delta \mathbf{v} = 0 \, ,
\end{equation}
and Eq.~\ref{eq:madelung2} becomes
\begin{equation}
    \label{eq:madelung2_lpt}
    \frac{\partial \delta \mathbf{v}}{\partial t} + 
    \left(\mathbf{v}_{\rm rel} \cdot \nabla \right) \delta \mathbf{v} = 
    - \nabla U + \frac{\hbar^2}{4m^2} \nabla \left(\nabla^2 \alpha \right) \, .
\end{equation}
We now choose coordinates so that 
$\mathbf{v}_{\rm rel} = -\vrel \mathbf{\hat{z}}$; that is, the satellite is moving in the $+\mathbf{\hat{z}}$ direction, so in the rest frame of the satellite, the mean velocity of the condensate is in the $-\mathbf{\hat{z}}$ direction. Taking the time derivative of Eq.~\ref{eq:madelung1_lpt} and the 
divergence of Eq.~\ref{eq:madelung2_lpt}, combining them, and 
simplifying, we arrive at:
\begin{equation}
    \label{eq:re_evol_lpt}
    \frac{\partial^2 \alpha}{\partial t^2} - 
    2 \vrel \frac{\partial^2 \alpha}{\partial z \partial t} + 
    \vrel^2 \frac{\partial^2 \alpha}{\partial z^2} - \nabla^2 U
    +\frac{\hbar^2}{4m^2}\nabla^4 \alpha = 0 \, .
\end{equation}
Note that the only dependence on the mass of the satellite appears in Eq.~\ref{eq:re_evol_lpt} as $\nabla^2 U$, which by Poisson's equation is equal to $4\pi G \rho_S$, where $\rho_S$ is the mass density of the satellite. Thus this LPT formalism is applicable for an arbitrary $\rho_S$, as long as the total mass associated with the satellite is small enough for a linear regime treatment to be valid.

We will study analytically the case where $\alpha$ is 
independent of time, which can be understood as the infinite-time limit. That is, we imagine that at $t = -\infty$, the satellite
is at rest at the origin in an (initially uniform) sea of quantum 
condensate moving with velocity $\mathbf{v}_{\rm rel}$ and with dynamics determined by Eq.~\ref{eqn:S1}, and we are interested in the configuration of the condensate as $t\to \infty$. Strictly speaking, this is not a self-consistent approximation, as can be seen by examining Eq.~\ref{eqn:S1} and noting that spatial gradients will typically drive 
evolution in time. That said, we may anticipate that as $t \to \infty$, this 
temporal evolution will become oscillatory, and our time-independent 
solution is something like an average over these oscillations. Indeed, as we will show in Sec.~\ref{sec:Results}, the assumption of time independence, carefully interpreted, gives excellent agreement with finite-time simulations.\footnote{We also point out that Eq.~\ref{eq:re_evol_lpt}, transformed into the lab-frame rather than the moving perturber frame, is
the biharmonic wave equation with a moving load \citep{roetman1967biharmonic}
\begin{equation}
    \frac{\partial^2 \alpha}{\partial t^2} 
    +\frac{\hbar^2}{4m^2}\nabla^4 \alpha = 4\pi G \rho_S(t) \, ,
\end{equation}
which describes the elastic behavior and deflections of a beam in Euler-Bernoulli beam theory \citep{timoshenko1953history}.
This correspondence has also been mentioned in \cite{Lora12}.}


\subsection{Linear Perturbation Theory: Point Source}
\label{sec:lpt_point_source}

We will begin by treating the case in which the satellite is a point 
mass $M$ with a Keplerian potential
\begin{equation}
    \label{eq:kepler_pot}
    U_K = - \frac{G M}{r} \, ,
\end{equation}
where $r$ is the Euclidean distance from the satellite. The corresponding mass density is $\rho_K = M \delta^3(\mathbf{r})$. Assuming time-independence as discussed above, Eq.~\ref{eq:re_evol_lpt} becomes
\begin{equation}
    \label{eq:tind_point_lpt}
    \vrel^2 \frac{\partial^2 \alpha}{\partial z^2} +
    \frac{\hbar^2}{4m^2} \nabla^4 \alpha = 4 \pi G M \delta^3(\mathbf{r}) \, .
\end{equation}
To solve this, we proceed as in \cite{Lora12} by Fourier-transforming Eq.~\ref{eq:tind_point_lpt}. In Fourier space,  Eq.~\ref{eq:tind_point_lpt} becomes
\begin{equation}
    \label{eq:four_tind_psoure_lpt}
    \left(\frac{\hbar^2 k^4}{4m^2} - \vrel^2 k_z^2 \right) \hat{\alpha}
    = 4\pi G M \, .
\end{equation}
Note that $\hat{\alpha}$, the Fourier transform of the overdensity, has dimensions 
of $[\rm length]^3$. Using the background de Broglie wavelength, we define a dimensionless wave vector
$\widetilde{\mathbf{k}} = \lambdabar \mathbf{k}$, so
Eq.~\ref{eq:four_tind_psoure_lpt} becomes
\begin{equation}
    \label{eq:four_tind_psource_lpt_dimensionless}
    \left(\kt^4 - 4{\kt}_z^2 \right) \hat{\alpha} = \frac{16 \pi \lambdabar^3}{\mathcal{M}_Q} \, .
\end{equation}
Defining a dimensionless length $\widetilde{\mathbf{r}} = \mathbf{r}/\lambdabar$, we may easily solve Eq.~\ref{eq:four_tind_psoure_lpt} algebraically for $\hat{\alpha}$ and perform the inverse Fourier transform to solve for $\alpha(\widetilde{\mathbf{r}})$:
\begin{equation}
    \label{eq:alpha_sol1}
    \alpha (\widetilde{\mathbf{r}}) = \frac{16 \pi}{\mathcal{M}_Q} \int 
    \frac{\diff^3 \widetilde{\mathbf{k}}}{(2\pi)^3} 
    \frac{e^{i\widetilde{\mathbf{k}}\cdot \widetilde{\mathbf{r}}}}{(\kt^4 - 4\kt_z^2)} \, .
\end{equation}

The only preferred direction in the problem is set by the velocity 
$\vrel \hat{\mathbf{z}}$, so we work in dimensionless cylindrical coordinates 
$(\widetilde{R}, \phi, \tilde{z})$. The integral over $\tilde{k}_\phi$ can be 
performed analytically, giving
\begin{equation}
    \label{eq:alpha_sol2}
    \alpha (\widetilde{\mathbf{r}}) = \frac{16 \pi}{(2\pi)^2\mathcal{M}_Q} 
    \int_0^{\infty} \diff \kt_R \int_{-\infty}^{\infty} \diff \kt_z 
    \frac{\kt_R \, J_0 (\kt_R \widetilde{R}) \, e^{i\kt_z \tilde{z}}}{(\kt^4 - 4\kt_z^2)}\, ,
\end{equation}
where $\kt_R$ is the dimensionless radial wavenumber and $J_0$ is the 
zeroeth-order Bessel function of the first kind. We can further simplify this 
result by integrating over $k_z$ using contour integration, which we perform in Appendix~\ref{app:ints_lpt}.\footnote{In Appendix~\ref{app:ints_lpt}, we correct an error 
in the choice of contour used in \cite{Lora12}.} 

In our setup, computing $\alpha(\widetilde{\mathbf{r}})$ is an intermediate step towards
the dynamical friction force that results from this overdensity. By cylindrical 
symmetry, we know that the drag force will be in the $\mathbf{\hat{z}}$ direction. 
The formula for the net force in this direction is then (from Eq.~\ref{eq:df_def})
\begin{equation}
    \label{eq:df_force}
    F_{\rm DF} = 2\pi G M \rhobar \lambdabar
    \int_0^{\infty} \diff \Rt \int_{-\infty}^{\infty} \diff \zt
    \frac{\Rt \, \zt}{(\Rt^2 + \zt^2)^{3/2}} \, \alpha(\Rt,\zt) \, .
\end{equation}

The drag force as defined above will be infinite, as the assumption of  time-independence necessitates that there has been an infinite 
amount of time for the wake to accumulate. This is one manifestation of the ubiquitous Coulomb logarithm in gravitational scattering problems. We must therefore impose a cutoff, which we define physically as a maximum distance $b$ from the satellite beyond which we do not consider the mass that has accumulated. This is equivalent to the situation where the satellite has only been 
traveling for some \emph{finite} time and therefore mass has only been 
able to accumulate out to a certain distance. Using the results from Appendix \ref{app:ints_lpt} and imposing a cutoff of 
$\bt = b/\lambdabar$, we have
\begin{equation}
    \label{eq:df_point_source_lpt_int}
    C_{\rm rel} = -\int\displaylimits_0^{\bt}\!\! \diff \Rt
    \int\displaylimits^{\sqrt{\bt^2 - \Rt^2}}_0 \!\!\!\!\!\!\!\diff \zt \int\displaylimits_0^2 \!\! \diff x \, 
    \frac{\Rt \, \zt \, J_0\left(\sqrt{2x-x^2} \Rt \right)}{\left(\Rt^2 + \zt^2 \right)^{3/2}} \frac{\sin(x \zt)}{x} \, .
\end{equation}
While a closed-form solution to this integral does not exist, $C_{\rm rel}$ can easily be evaluated numerically. For a point source, there is an alternate closed-form solution for $\alpha$ which permits a closed-form solution for $C_{\rm rel}$, which we will exploit below as a cross-check of this result. However, the formalism used in this section may be easily generalized to sources with finite extent $\ell$, unlike the case of the exact solution for the point mass.

\subsection{Exact Solution of a Point Source}
\label{sec:exat_sol_point_source}

\begin{figure}
    \centering
    \includegraphics{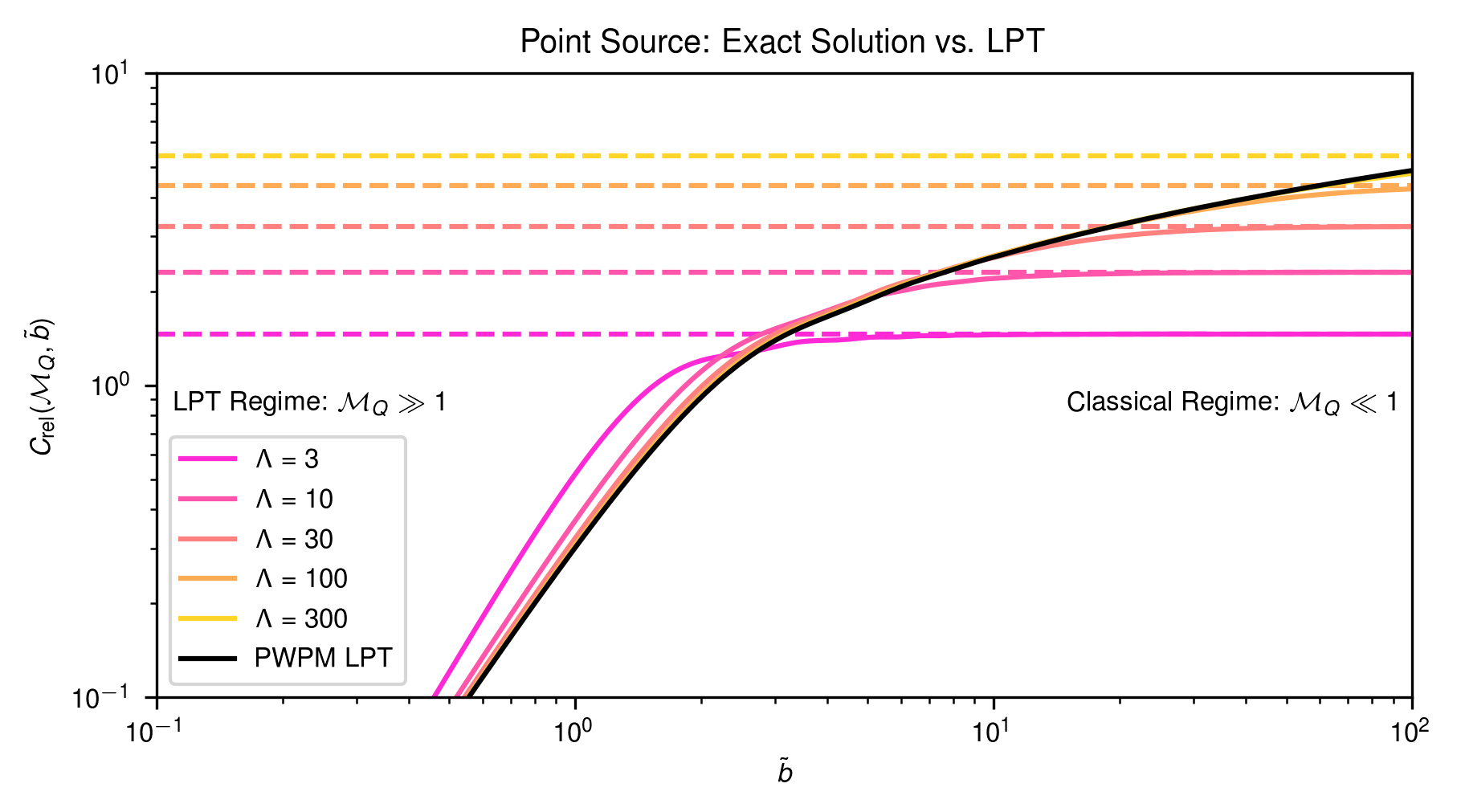}
    \caption{Dynamical friction coefficients for 
    a point source moving in a uniform density background 
    at fixed $\Lambda = \bt \mathcal{M}_Q$ as a function of 
    the cutoff scale $\bt$. 
    Solid colored lines plot the exact solution (Eq.~\ref{eq:df_exact}), 
    compared to the solution in LPT 
    (plane wave, point mass, linear perturbation theory: PWPM LPT), 
    as given in Eq.~\ref{eq:Crel_lpt}, in black. We use the plane wave, point mass
    linear perturbation theory curve as a reference curve in many of the other figures 
    in the remainder of the paper.
    As $\bt$ becomes large at fixed $\Lambda$ 
    ($\mathcal{M}_Q \ll 1$) the dynamical friction approaches the 
    classical limit (dashed lines), which depends only on $\Lambda$ and is given above 
    by the horizontal, dashed lines \cite{HOTW}. As $\bt$ becomes 
    small for a given $\Lambda$ 
    ($\mathcal{M}_Q \gg 1$) we approach the perturbation theory result (left side of the plot).
    This Figure is a recreation of Figure 2 from \cite{HOTW}.}
    \label{fig:point_source_exact_LPT_comparison}
\end{figure}

\begin{figure}
    \centering
    \includegraphics{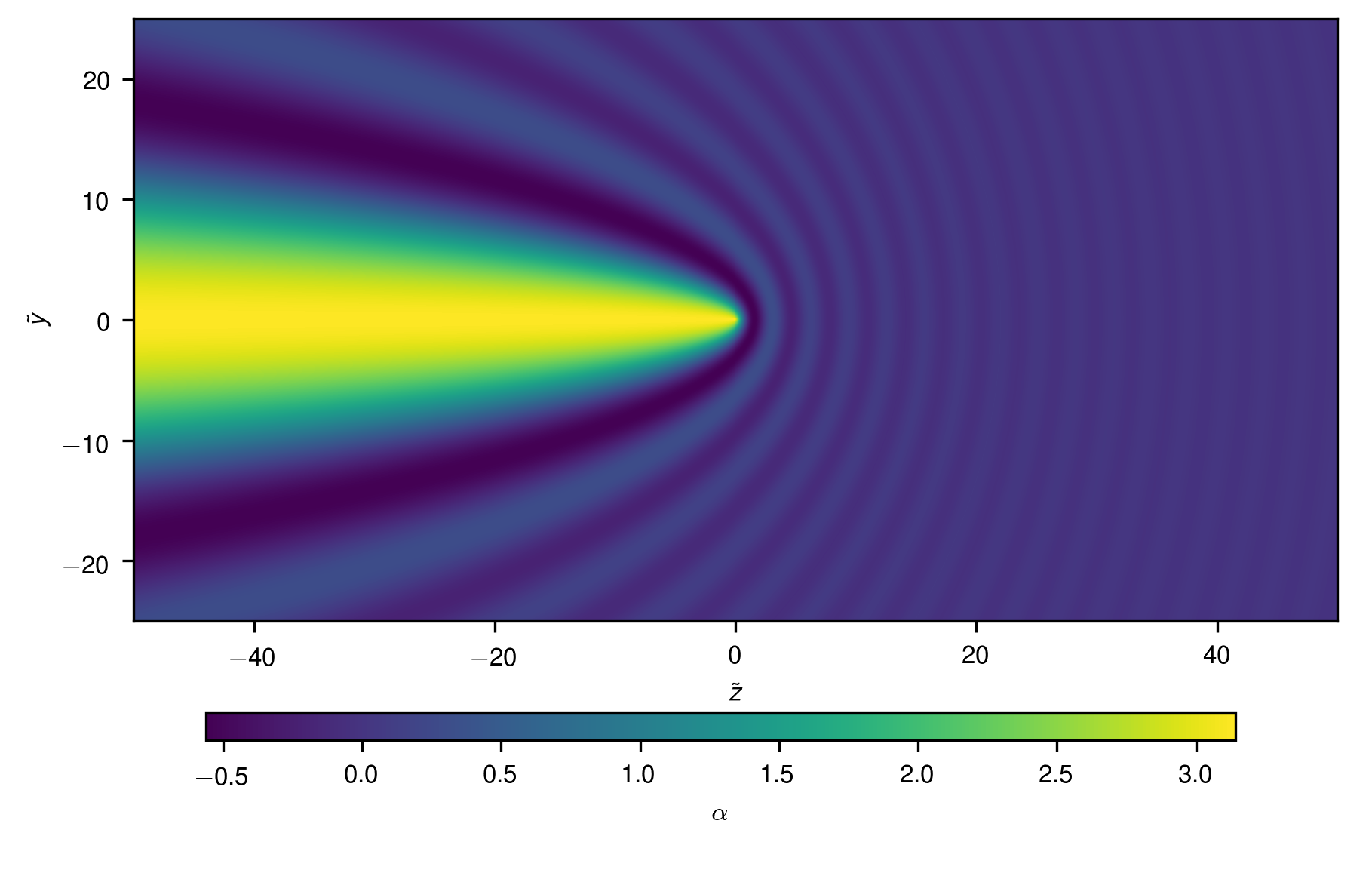}
    \caption{The density contrast, $\alpha\equiv \delta \rho /\rhobar$, 
    generated by a point-mass satellite at the 
    origin in an asymptotically uniform FDM medium moving in the 
    $z$-direction with relative velocity $\vrel$, in the limit that the equations 
    of motion can be approximated as time independent and the satellite mass is 
    small ($\mathcal{M}_Q \to \infty$ or the LPT). The 
    solution given in Eq.~\ref{eq:exact_lpt_alpha} is plotted here, where we 
    set $\mathcal{M}_Q = 1$ for simplicity; the gravitational wake behind 
    the satellite is clearly visible.}
    \label{fig:LPT_solution}
\end{figure}

For the special case of a point-mass satellite, the time-independent problem can 
actually be solved exactly, without resorting to perturbation theory. It should not be particularly 
surprising that this problem has been studied in some depth, as 
it is identical to the Coulomb scattering problem, which describes, for example, an 
incident beam of electrons scattered off a point-like nucleus through the Coulomb 
interaction \cite{HOTW}. In our case, the dark matter condensate plays the role of the electron beam and the satellite plays the role of the nucleus, but the mathematics (and quantum mechanics) of the attractive Coulomb potential between oppositely charged particles are identical to the universally-attractive Newtonian potential between two point masses.

The exact solution to this problem is given by the scattering states of the Coulomb potential, which can be found in a 
number of older quantum mechanics texts  
\cite{bethe2013quantum, AtomicCollisions}. The exact wave function which solves Eq.~\ref{eqn:S1} in the 
time-independent regime with a gravitational potential given by 
Eq.~\ref{eq:kepler_pot}, normalized so that it approaches $\sqrt{\rhobar}$ at 
large distances, is 
\begin{equation}
    \label{eq:point_source_exact_sol}
    \psi(\Rt, \zt) = \sqrt{\rhobar} \, e^{i \zt}e^{\frac{\pi}{2\mathcal{M}_Q}}
    \left|\Gamma\left(1-\frac{i}{\mathcal{M}_Q}\right)\right|
    {}_1 F_1 \left(\frac{i}{\mathcal{M}_Q},1,i\left(\sqrt{\Rt^2 + \zt^2} +\zt\right) \right) \, ,
\end{equation}
where $\Gamma(x)$ is the gamma function, 
${}_1F_1(a,b,c)$ is the confluent hypergeometric function \cite{HOTW}, and the tildes represent variables in units of $\lambdabar$. Note that $\psi$ is a function only of $\Rt$ and $\zt$, as expected from cylindrical symmetry.

We compute the full density distribution by taking the squared norm of 
Eq.~\ref{eq:point_source_exact_sol} and obtain the 
overdensity as defined in Eq.~\ref{eq:overdense} through this density 
distribution.
We derive the dynamical friction force experienced by the perturbing 
mass exactly as in Eq.~\ref{eq:df_force}. Defining $\tilde{q} \equiv \sqrt{\Rt^2 + \zt^2} + \zt$, the dynamical friction coefficient is
\begin{equation}
    \label{eq:df_exact}
    C_{\rm rel} = \frac{\mathcal{M}_Q}{2} \, e^{\pi/\mathcal{M}_Q} \,
    \left|\Gamma\left(1-\frac{i}{\mathcal{M}_Q}\right)\right|^2
    \int_0^{2\bt} \diff \tilde{q} \,
    \left| {}_1 F_1 \left(\frac{i}{\mathcal{M}_Q},1,i\tilde{q} \right) \right|^2
    \left( \frac{\tilde{q}}{\bt} - 2 - \log \frac{\tilde{q}}{2\bt}\right) \, .
\end{equation}
This is the same as Equation D8 given in \cite{HOTW}.
In Fig.~\ref{fig:point_source_exact_LPT_comparison}, we compare 
$C_{\rm rel}$ as calculated from Eq.~\ref{eq:df_exact} with both 
the linear theory (equivalently derivable from 
Eq.~\ref{eq:df_point_source_lpt_int} or the derivation below) and 
the classical limit.

As noted in Sec.~\ref{sec:Scales}, the LPT analysis performed in Section \ref{sec:lpt_point_source} applies in the regime where the quantum Mach number is much greater than 1, $\mathcal{M}_Q \gg 1$. In slightly more detail, we can write $\mathcal{M}_Q$ as 
\begin{equation}
    \label{eq:MQ_massdef}
    \mathcal{M}_Q \equiv \frac{\vrel}{v_Q} 
    = \frac{\lambdabar\vrel^2}{G M} \, .
\end{equation}
If we assume that the size of the system is some (probably large) multiple 
of $\lambdabar$ and use $\vrel$ as a proxy for the circular 
velocity at some radius, then we can say 
$M_{\rm system} \approx A \lambdabar\vrel^2/G$ for 
some dimensionless $A$. We see that $\mathcal{M}_{Q}$ is proportional to the ratio of the background system mass to the satellite mass. For a linear 
theory argument, we are clearly interested in the regime where this ratio is 
much greater than 1. One can also see from Eq.~\ref{eq:MQ_massdef} that 
$\bt \mathcal{M}_Q = b \vrel^2/GM$, where $GM/\vrel^2 \approx \ell_{90}$, the 
distance at which field particles passing by the satellite are deflected by 
90$^{\circ}$, as mentioned in Eq.~\ref{eq:Coulomb_log_def}. This is where the 
definition of the analogous $\Lambda$ in 
Fig.~\ref{fig:point_source_exact_LPT_comparison} comes from.

Expanding the hypergeometric and gamma functions in inverse powers of $\mathcal{M}_Q$, we have
\begin{align}
    \label{eq:chf_series_F}
    {}_1 F_1 \left(\frac{i}{\mathcal{M}_Q},1,iq\right) &= 1 
    - \frac{{\rm Si}(q)}{\mathcal{M}_Q} - i\frac{{\rm Cin}(q)}{\mathcal{M}_Q} 
    + \mathcal{O}\left(\mathcal{M}_Q^{-2} \right) \\
    \label{eq:chf_series_G}
    \Gamma\left(1-\frac{i}{\mathcal{M}_Q}\right) & = 1 + \frac{i \gamma}{\mathcal{M}_Q} + \mathcal{O}\left(\mathcal{M}_Q^{-2}\right) \, ,
\end{align}
where ${\rm Si}(q) \equiv \int_0^q \sin(t)\diff t/t $ and ${\rm Cin}(q) \equiv \int_0^q (1-\cos(t))\diff t /t$ are the sine and cosine integrals, respectively, and $\gamma$ is the Euler-Mascheroni constant. The corresponding density contrast is
\begin{equation}
    \label{eq:exact_lpt_alpha}
    \alpha(\tilde{q}) = \frac{\pi - 2\, {\rm Si}(\tilde{q})}{\mathcal{M}_Q} + \mathcal{O}\left(\mathcal{M}_Q^{-2}\right) \, ,
\end{equation}
which exactly matches Eq.~\ref{eq:alpha_sol2} to machine precision to leading order in $1/\mathcal{M}_Q$. 
We plot this linear-regime overdensity in Fig.~\ref{fig:LPT_solution}
in the $y-z$ plane; by cylindrical symmetry, the full three-dimensional result is obtained by rotating around the $z$-axis.

Inserting the expansion from Eqs.~\ref{eq:chf_series_F}-\ref{eq:chf_series_G}
into the expression for the dynamical friction in Eq.~\ref{eq:df_exact}, 
we obtain an analytic expression for the dynamical friction 
coefficient in LPT:
\begin{equation}
    \label{eq:Crel_lpt}
    C_{\rm rel} = {\rm Cin}(2\bt) + \frac{\sin(2\bt)}{2\bt} - 1 
    + \mathcal{O}\left(\mathcal{M}_Q^{-1}\right) \, .
\end{equation}
Again, this formula has been previously shown in Equation D14 of \cite{HOTW}. Note the dependence of the result on $\bt$, the cutoff we impose to make the integral finite. In Fig.~\ref{fig:point_source_exact_LPT_comparison}, we compare this exact result with the LPT results in Sec.~\ref{sec:lpt_point_source} and the classical limit from Sec.~\ref{sec:ClassicalDF}. From Fig.~\ref{fig:point_source_exact_LPT_comparison} it 
is clear that the exact solution approaches the perturbation theory result in the limit that $\mathcal{M}_Q \gg 1$, as expected.

We also refer the reader to calculations of a fixed point mass in a static FDM background with full self-gravity carried out in \cite{2019arXiv190804790Y}, which give rise to soliton-like solutions that resemble the ground state of the hydrogen atom in the limit that the point mass is large.

\subsection{Linear Perturbation Theory: Extended Source}
\label{sec:lpt_extended}

We would now like to generalize to the case where the satellite has an extended mass distribution, rather than simply being a point mass. 
We will work with the same LPT approach as in Section 
\ref{sec:lpt_point_source}, starting from 
Eq.~\ref{eq:re_evol_lpt}.  We will assume a time-independent solution and take the satellite mass distribution to be a Plummer sphere \citep{Plummer}:
\begin{equation}
    \label{eq:plummer_pot}
    U_P = - \frac{G M}{\ell} 
    \left(1 + \left(\frac{r}{\ell} \right)^2 \right)^{-1/2} \, .
\end{equation}
This choice will facilitate comparison with the numerical simulations described in Sec.~\ref{sec:Numerical}; we show in Appendix~\ref{app:dist} that the qualitative features of a finite size $\ell$ do not depend sensitively on the precise mass distribution chosen.  Specifically, in that Appendix, we compare the results for the Plummer sphere with that of the truncated isothermal sphere profile.

Using this potential and Fourier transforming the partial differential equation, we arrive at a slightly modified version of Eq.~\ref{eq:four_tind_psoure_lpt} for the Plummer profile,
\begin{equation}
    \label{eq:four_tind_extended_psoure_lpt}
    \left(\frac{\hbar^2 k^4}{4m^2} - \vrel^2 k_z^2 \right) \tilde{\alpha}
    = 4\pi G M \ell k K_1\left(\ell k\right) \, ,
\end{equation}
where $K_1(x)$ is the first modified Bessel function of the second kind.
Changing to dimensionless variables in units of $\lambdabar$ gives
the extended-source version of 
Eq.~\ref{eq:four_tind_psource_lpt_dimensionless} as
\begin{equation}
    \label{eq:four_tind_extended_psoure_lpt_dimensionless}
    \left( \kt^4 - 4 \kt_z^2\right) \tilde{\alpha} = 
    \frac{16\pi \lambdabar^3}{\mathcal{M}_Q} \tilde{\ell}\kt 
    K_1(\tilde{\ell}\kt) \, ,
\end{equation}
where we have also written $\ell$ relative to $\lambdabar$, denoted $\tilde{\ell}$  and defined in 
Eq.~\ref{eq:ltil}. We can then recover $\alpha$ by performing the 
inverse Fourier transform; our version of Eq.~\ref{eq:alpha_sol2} 
is then 
\begin{equation}
    \label{eq:alpha_sol2_extended}
    \alpha (\widetilde{\mathbf{r}}) = \frac{16 \pi \tilde{\ell}}{\mathcal{M}_Q(2\pi)^2} 
    \int_0^{\infty} \diff \kt_R \int_{-\infty}^{\infty} \diff \kt_z 
    \frac{\kt_R \, J_0 (\kt_R \Rt)\,e^{i\kt_z \zt}}{(\kt^4 - 4\kt_z^2)} 
    \kt K_1\left(\tilde{\ell} \kt\right) \, ,
\end{equation}
where $\kt = \sqrt{\kt_R^2 + \kt_z^2}.$

We then carry out the $\kt_z$ integral in Eq.~\ref{eq:alpha_sol2_extended} 
using contour integration. The dynamical friction
is given by 
\begin{equation}
    \label{eq:df_extended_lpt_int}
    C_{\rm rel} = -2\tilde{\ell} \int\displaylimits_0^{\bt}\!\! \diff \Rt
    \int\displaylimits^{\sqrt{\bt^2 - \Rt^2}}_0 \!\!\!\!\!\!\!\diff \zt \int\displaylimits_0^2 \!\! \diff x 
    \frac{\Rt\, \zt \,J_0\left(\sqrt{2x-x^2}\, \Rt\right)}{\left(\tilde{\ell}^2 + \Rt^2 + \zt^2 \right)^{3/2}}\, \frac{\sin(x\zt)}{\sqrt{2x}} \,
    K_1\left(\tilde{\ell} \sqrt{2x}\right) \, ,
\end{equation}
which is similar to Eq.~\ref{eq:df_point_source_lpt_int} 
except for the regulation by $K_1(x)$.  As a consistency check, we note that Eq.~\ref{eq:df_extended_lpt_int} becomes 
Eq.~\ref{eq:df_point_source_lpt_int} in the limit where $\tilde{\ell} \to 0$.

\begin{figure}
    \centering
    \includegraphics{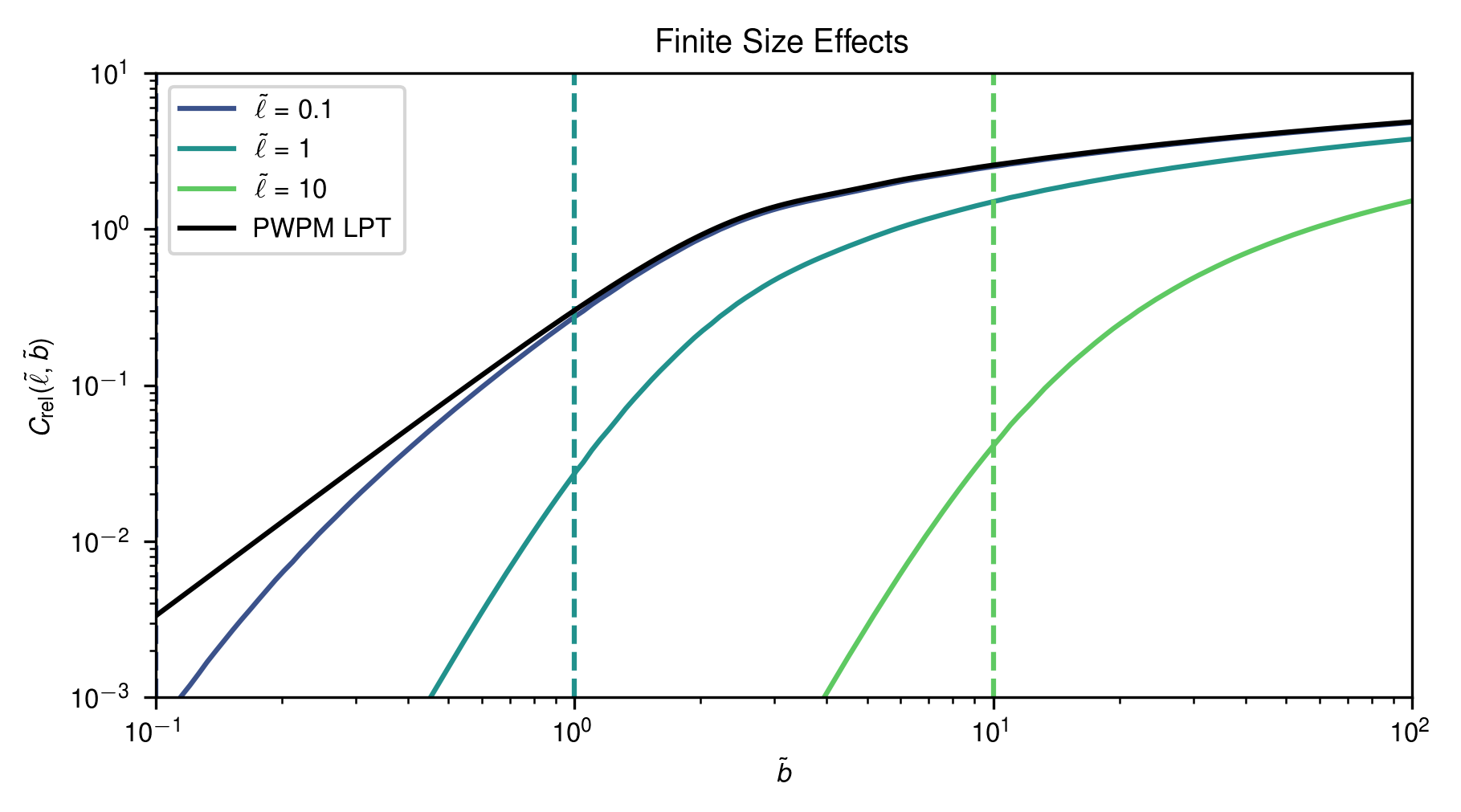}
    \caption{Effects of a finite-size satellite in the regime of linear 
    perturbation theory, where $C_{\rm rel}$ is completely independent of 
    the quantum mach number $\mathcal{M}_Q$. The perturbation theory 
    calculation for a point mass is shown in black (PWPM LPT) along with the perturbation 
    theory calculations for a satellite with a Plummer profile 
    (Eq.~\ref{eq:plummer_pot}) with varying scale length $\tilde{\ell}$ in 
    units of the background de Broglie wavelength --- these scales are also 
    indicated by the vertical dashed lines for comparison. We can see how the 
    finite size of the satellite suppresses dynamical friction on scales 
    comparable to the finite size, though on scales on the order of $\sim10^2$ 
    larger, the dynamical friction only differs by a factor of 
    order unity.}
    \label{fig:extended_comparison}
\end{figure}

In Fig.~\ref{fig:extended_comparison}, we compare the results given by Eq.~\ref{eq:df_extended_lpt_int} 
for several values of $\tilde{\ell}$ to the LPT 
result given equivalently in Eqs.~\ref{eq:df_point_source_lpt_int} and 
\ref{eq:Crel_lpt}. Again, we can see graphically that the extended-source result approaches the point-source result for all values of 
$\bt$ as $\tilde{\ell}\to 0$. As one would expect, the extended-source 
result approaches the point-source result for $\bt \gg \tilde{\ell}$,
where the source comes to `look like' a point source. For 
$\tilde{b}\ll \tilde{\ell}$, we find that $C_{\rm rel} \propto \bt^5$, which 
one can verify from Eq.~\ref{eq:df_point_source_lpt_int}. One should \textit{not} 
interpret the $\bt^5$ scaling as an extremely strong dynamical friction force on `fluffy' 
objects; the dynamical friction only increases as $\bt^5$ for scales smaller than 
the size of the object. This growth is simply a cause of the response of 
the condensate to the presence of the satellite.

\subsection{Velocity-Dispersed Condensate}
\label{subsec:vel_disp_analytic}

In a realistic scenario, the FDM medium should have a 
distribution of velocities and cannot be modeled as a single plane wave, as was 
done above. In particular, while the core of an FDM halo may be dispersionless, the outer part of the halo is expected to have a  Navarro-Frenk-White (NFW)-like profile with nonzero dispersion \cite{HOTW,NFW97}. This motivates us to investigate the behavior of the dynamical friction coefficient when the FDM medium has some velocity dispersion.

We can model this velocity dispersion by constructing the background 
wave function $\psi_0$, in analogy to Eq.~\ref{eq:ic}, as a linear 
combination of plane waves \cite{BarOr18}. In the absence of the 
perturbing satellite or any relative velocity, this 
background distribution would have a dependence on time and space 
given by
\begin{equation}
    \label{eq:baror_distribution}
    \psi_0(\mathbf{x},t) = \int \diff^3 \mathbf{k} \, \varphi(\mathbf{k}) \,
    e^{i\mathbf{k}\cdot \mathbf{x} - i \omega(\mathbf{k})t} \, ,
\end{equation}
where the dispersion relation is given by the free Schr\"odinger 
equation:
\begin{equation}
    \label{eq:quantum_disp}
    \omega(\mathbf{k}) = \frac{\hbar k^2}{2m} \, .
\end{equation}
The function $\varphi(\mathbf{k})$ determines the 
distribution of velocities by 
weighting the individual plane waves. For an isotropic 
distribution, $\varphi$ is only a function of the magnitude $k$ and 
each plane wave is imbued with some random, uncorrelated phase shift. This behavior is expected for any realistic halo, formed through the collapse of many uncorrelated proto-halos and further randomized by violent relaxation and phase mixing \cite{WhiteRees78}.  This randomness would ensure that the function $\varphi(\mathbf{k})$ is a random field.
We can also make the connection between $\varphi(\mathbf{k})$ and the 
actual velocity distribution function of the medium, $f(\mathbf{v})$,
by writing
\begin{equation}
    \label{eq:DistFuncc_def}
    \left\langle \varphi(\mathbf{k}) \varphi(\mathbf{k'}) \right\rangle_{\mathbf{k}} 
    = f\left(\frac{\hbar \mathbf{k}}{m}\right) \delta(\mathbf{k} - \mathbf{k'}) \, .
\end{equation}

For example, suppose that we would like to make our wave function 
mimic a classical Maxwell-Boltzmann distribution
(in which case $\varphi(k)$ would be a Gaussian random 
field):
\begin{equation}
    \label{eq:MB_dist}
    f(\mathbf{v}) = \frac{\rhobar}{(2\pi \sigma^2)^{3/2}} \exp\left(\frac{-v^2}{2\sigma^2}\right) \, .
\end{equation}
In the context of a numerical implementation where 
we only have a finite number of Fourier modes $\mathbf{k}_i$ to 
sum over, the coarse-grained version of Eq.~\ref{eq:baror_distribution} would look 
like
\begin{equation}
    \label{eq:disp_numerical_ic}
    \psi_0(\mathbf{x}) \propto \sum_{j} 
    \sqrt{f(\mathbf{x},\mathbf{v}_j)} {\rm
    e}^{i m \mathbf{x}\cdot\mathbf{v}_j/\hbar + i \phi_{{\rm rand},\mathbf{v}_j}}
    \, \left( \Delta v\right)^{3/2} \, ,
\end{equation}
where the sum is over the discrete 
Fourier modes (which can be written equivalently in terms of $\mathbf{k}$ 
or the velocities $\mathbf{v}$), the phase angles 
$\phi_{{\rm rand},\mathbf{v}}\in[0,2\pi)$ are the manifestation of 
ensuring that the modes have random, uncorrelated phases, and 
$f(\mathbf{v})$ is the desired distribution function.
The normalization is such that the average density is still $\rhobar$ --- see \cite{2018PhRvD..97h3519M} for details and \cite{Foster:2017hbq} 
for an example where this construction is used to model the axion dark 
matter field for direct-detection experiments. 
Eq.~\ref{eq:disp_numerical_ic} shows that the classical and quantum phase space are closely related:
the quantum wave function is a superposition of constant $v$ slices of the classical phase space, with amplitude $\mathbf{v}$ and random phases that give rise to interference patterns.
If the initial condition had just a single velocity $\mathbf{v}_0$ at each location $\mathbf{x}$, then there is no interference and the classical and quantum densities agree exactly: \[\rho(\mathbf{x})=|\psi(\mathbf{x})|^2 = \int
\left(\sqrt{f(\mathbf{x},\mathbf{v})}\right)^2\,\diff^3\mathbf{v}\, . \]
Through the remainder of the text, when working with a medium 
that has a distribution of velocities, we will use an isotropic distribution 
of the form of Eq.~\ref{eq:MB_dist}, determined solely by the 
velocity dispersion, $\sigma$.

Such a system as described above would have a classical Mach number
$\mathcal{M}_{\sigma}$ defined in Eq.~\ref{eq:M_sig}.
We can then describe the dynamical friction in such a velocity-dispersed 
medium by summing up the contributions 
from each plane wave, weighted by the distribution function of these 
plane waves \cite{BarOr18}:
\begin{equation}
    \label{eq:df_vdisp_baror}
    F_{\rm DF} = - 4 \pi G^2M^2 \int \diff^3{\mathbf{v}} \,
    \frac{\vrel - \mathbf{v}\cdot\mathbf{\hat{z}}}{|\mathbf{v}_{\rm rel} - \mathbf{v}|^3} \,
    f(\mathbf{v})\, C_{\rm rel, pw}\left(\mathcal{M}_Q, \tilde{b}\right) \, ,
\end{equation}
where the $C_{\rm rel,pw}(\mathcal{M}_Q,\bt)$ is the dimensionless 
dynamical friction coefficient for a plane wave, which in general depends on both the quantum Mach number and the cutoff scale of interactions. In practice, one can plug in either a fully non-linear function for $C_{\rm rel, pw}(\mathcal{M}_Q,\bt)$ such as Eq.~\ref{eq:df_exact}, or the LPT theory Eq.~\ref{eq:Crel_lpt}, or even include finite-size effects by using Eq.~\ref{eq:df_extended_lpt_int}, we leave this treatment to future work.
Taking the distribution function to be Maxwellian, and changing variables to the dimensionless velocity $\tilde{\mathbf{v}} = \mathbf{v}/\vrel$, we 
may state the above in terms of the dynamical friction coefficient as
\begin{equation}
    \label{eq:df_vdisp_coeff}
    C_{\rm rel} = -\frac{\mathcal{M}_{\sigma}^3}{(2\pi)^{3/2}} 
    \int \diff^3\tilde{\mathbf{v}} \, \frac{1 - \tilde{\mathbf{v}}\cdot \mathbf{\hat{z}}}{|\mathbf{\hat{z}} - \tilde{\mathbf{v}}|^3} \,
    \exp\left(-\frac{u^2}{2} \mathcal{M}_{\sigma}^2\right)
    C_{\rm rel, pw}\left(\mathcal{M}_Q, b\right) \, ,
\end{equation}
where, again, $C_{\rm rel, pw}\left(\mathcal{M}_Q, b\right)$ is the dynamical friction coefficient in the case of a plane wave, and $C_{\rm rel}$ is the overall dynamical friction coefficient in the velocity-dispersed case.

\begin{figure}
    \centering
    \includegraphics{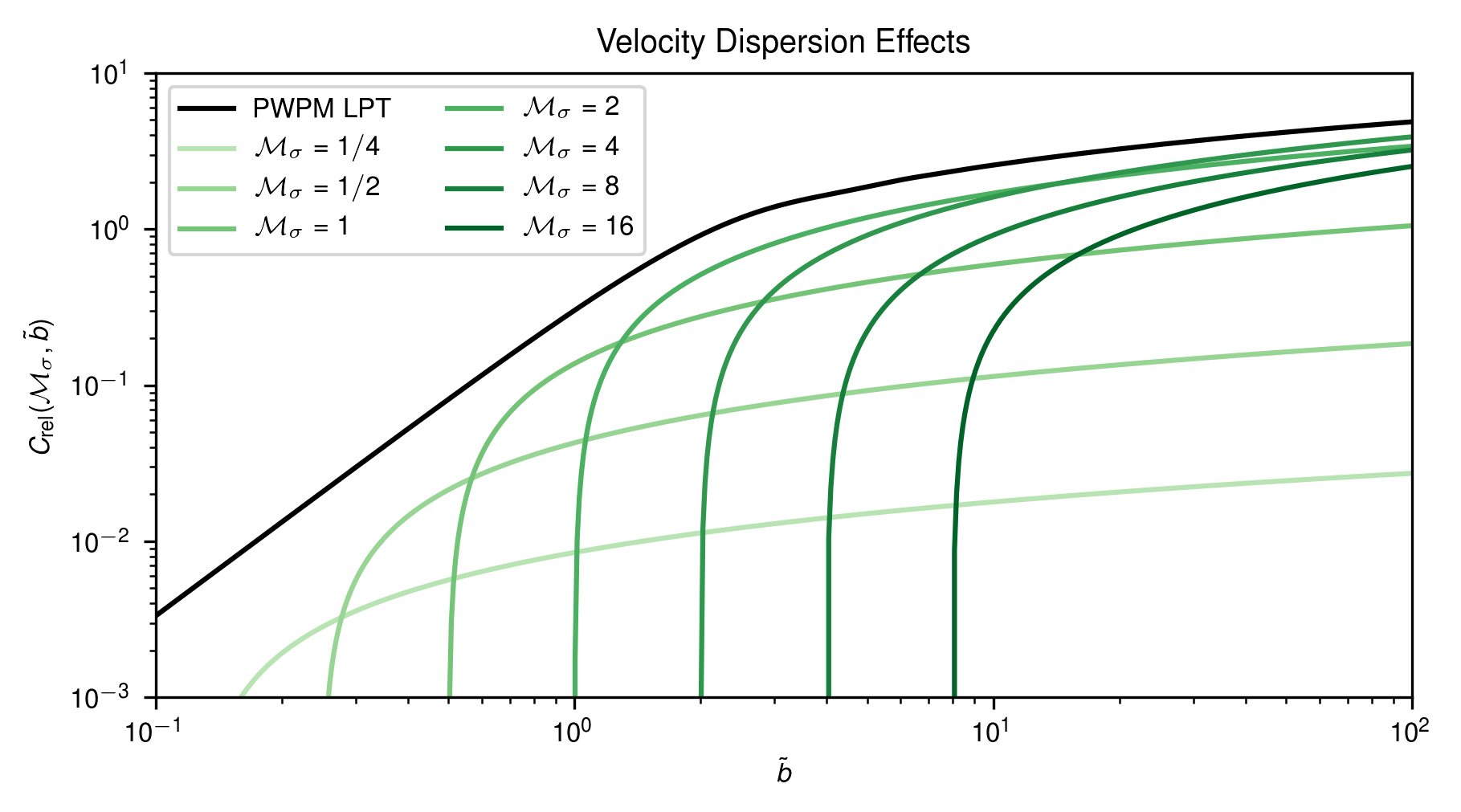}
    \caption{The dynamical friction coefficient $C_{\rm rel}$ for a point source, as derived 
    using linear perturbation theory (LPT) and defined in Eq.~\ref{eq:Crel_lpt}, 
    is shown in black. The green curves represent the coefficient as calculated 
    by Eq.~\ref{eq:Csig_baror} for several different values of the 
    classical Mach number $\mathcal{M}_{\sigma}$. See text for a detailed discussion of the effects shown here.}
    \label{fig:sigma_ps_comp}
\end{figure}

As we touched on in Sections \ref{sec:lpt_point_source} and 
\ref{sec:exat_sol_point_source}, when $\mathcal{M}_Q \ll 1$ the 
time-independent dynamical friction calculation for a point source 
approaches the classical answer.  However, in the limit 
of LPT, when $\mathcal{M}_Q \gg 1$, we can 
apply our perturbation theory argument to replace $C_{\rm rel, pw}$ 
above. Using these assumptions, along with the assumption that 
the cutoff scale $\tilde{b}$ is much larger than the dispersion de Broglie wavelength $\lambdabar_{\sigma}$, one can 
show~\citep{BarOr18} that the dynamical friction is given by
\begin{equation}
    \label{eq:Csig_baror}
    C_{\rm rel} = \mathcal{M}_{\sigma}^2
    \log \left( \frac{2\tilde{b}}{\mathcal{M}_{\sigma}}\right)
    \mathbb{G}\left(\frac{\mathcal{M}_{\sigma}}{\sqrt{2}} \right) \, ,
\end{equation}
where $\mathbb{G}(X)$ is defined in Eq.~\ref{eq:GX_def}. 
Indeed, this result is the same as the classical Chandrasekhar result 
\citep{1943ApJ....97..255C,2008gady.book.....B} except with the 
Coulomb logarithm defined in terms of  
$\mathcal{M}_{\sigma}$. In Fig.~\ref{fig:sigma_ps_comp}, 
we compare the dynamical friction found in this case to that 
found in the case of a point mass in a uniform medium as given in 
Eq.~\ref{eq:Crel_lpt}. We can see that the formula given here 
breaks down when $b<\lambdabar_{\sigma}$, or,
when $b$ is expressed in units of $\lambdabar$, $\tilde{b}<\mathcal{M}_{\sigma}$. 
We expect the velocity-dispersed solution to approach the point-source LPT solution in the limit that $\sigma \to 0$, or equivalently $\mathcal{M}_{\sigma}\to \infty$.  However, this also implies that the de Broglie wavelength corresponding to the velocity dispersion $\lambdabar_{\sigma} = \hbar/m\sigma$ becomes very large, and the assumption that $b \gg \lambdabar_{\sigma}$ is no longer valid. In this regime, we can assume that $\lambdabar_{\sigma}$ is so large that we essentially have a uniform density background.

In practice, then, care must be taken to consider the comparative size of $\lambdabar_{\sigma}$ and the size of the system $b$. The results derived in this section only apply in the case that $b\gg \lambdabar_{\sigma}$. On the other hand, when $b \ll \lambdabar_{\sigma}$, we may apply the uniform density results. In the intermediate regime of $b \approx \lambdabar_{\sigma}$, individual over-densities and under-densities can strongly influence the satellite's motion, and the true dynamical friction force becomes uncertain. The calculation behind Eq.~\ref{eq:Csig_baror} assumes that these over-/under-densities can be treated in a statistically averaged way, so it does not take this effect in to account. This is why we do not see any strange behavior at $\bt \approx \mathcal{M}_{\sigma}$ in Fig.~\ref{fig:sigma_ps_comp}.

\section{Numerical Simulations}
\label{sec:Numerical}

\begin{figure}
    \centering
    \includegraphics{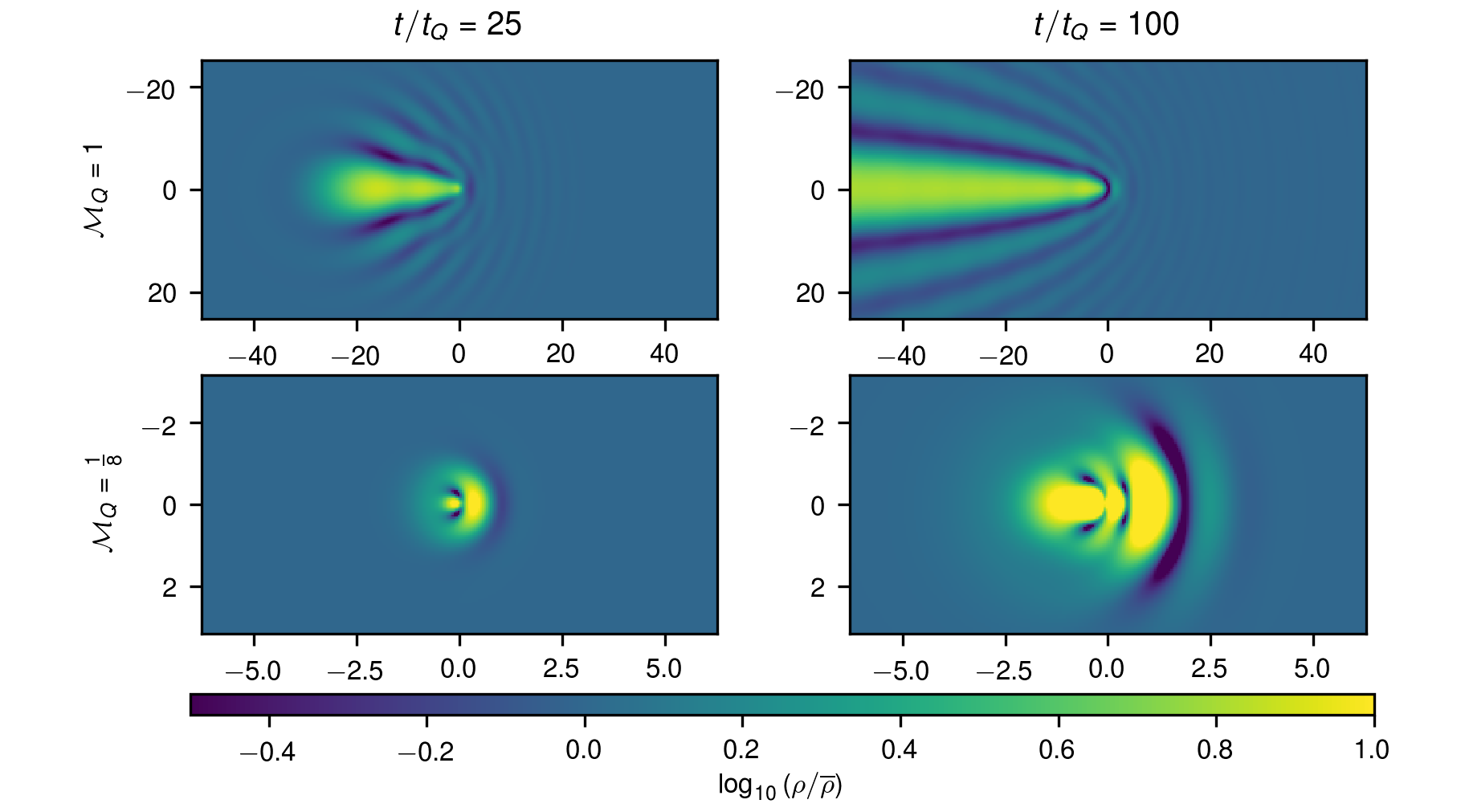}
    \caption{Density contrast in the $x=0$ plane at various time steps of our 
    simulations, for a satellite moving in the $\mathbf{\hat{z}}$ direction, 
    which points towards the right. The density scale is logarithmic and all plots 
    are on the same scale. The simulations shown here have zero background 
    velocity dispersion and a satellite with size $\ell/L_Q=0.5$, the 
    smallest size we explore numerically and thus the 
    closest to a point mass. The panels on the left show a snapshot of the 
    simulations at earlier times, with later times in the right panel. 
    The top panels show a simulation at a quantum Mach number of 
    $\mathcal{M}_Q=1$ while the bottom panels show $\mathcal{M}_Q= 0.125$.  
    We see that at higher Mach number, the  resulting wake is much less dense, 
    indicative of the decrease in the dynamical friction. Spatial scales 
    are indicated in units of $\lambdabar$, where the horizontal axis is the $z$-axis and the vertical axis is the $y$ axis.}
    \label{fig:sim_slices}
\end{figure}

\begin{figure}
    \centering
    \includegraphics{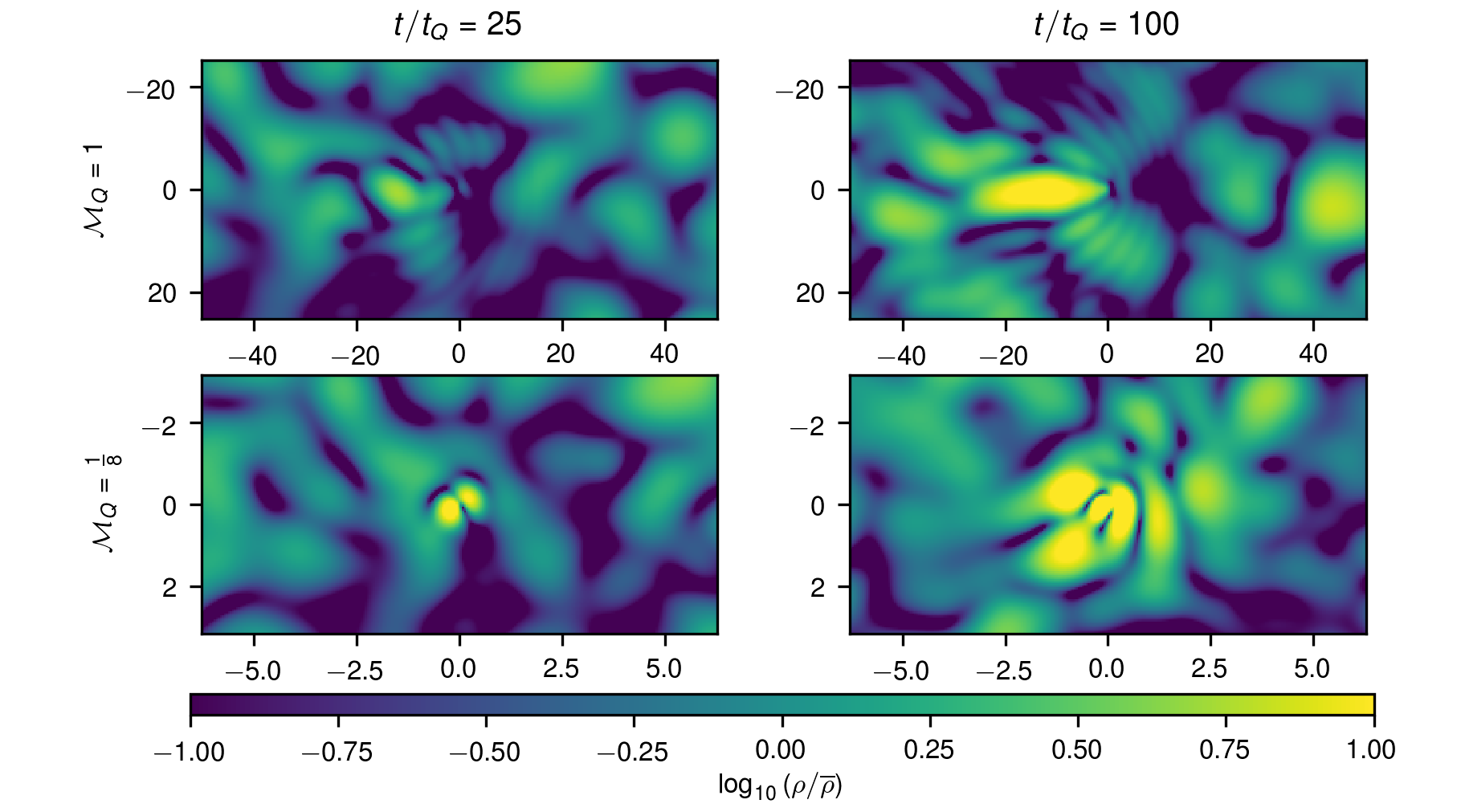}
    \caption{Density contrast as in Fig.~\ref{fig:sim_slices}, but for 
    the case of a background medium with some finite velocity 
    dispersion. $\mathcal{M}_{\sigma}=4$ for all four panels and 
    all other parameters are the same as Fig.~\ref{fig:sim_slices}. Spatial scales 
    are in units of $\lambdabar$. Note that the logarithmic density scale for these plots is different than Fig.~\ref{fig:sim_slices}. }
    \label{fig:sims_slice_veldisp}
\end{figure}

Now that we have explored the analytic results in 
different regimes, we move on to numerically calculate the time-dependent solutions 
in each of these regimes and compare them to the analytic results.
We carry out time-dependent numerical simulations of the response of the wave 
function to a massive satellite with a Plummer mass density profile in order 
to measure the dynamical friction 
coefficient as a function of dimensionless parameters 
$\mathcal{M}_{\rm Q}$,  $\tilde{\ell}$, and $\mathcal{M}_{\sigma}$ 
using the unitary spectral method of
\cite{Mocz17}. See Appendix~\ref{app:num} for details of the numerical implementation, which solves Eq.~\ref{eqn:S1}.

The perturbing satellite moves through half the distance of a periodic 
box of size $L_{\rm box} = 64\pi L_{\rm Q}$. In this amount of time, the 
simulation is unaffected by the boundary conditions. Our numerical 
resolution is $512^3$ grid points. We simulate $65$ different relative 
velocities, corresponding to choices of $\mathcal{M}_{\rm Q}$ in $[0,2]$.
We also simulate four different satellite sizes given by the Plummer profile, defined with respect to the quantum length 
scale as $\ell/L_Q=\frac{1}{2},1,2,4$. Finally, we simulate three different 
cases of background velocity dispersion, again defined with respect to the 
quantum length scale as $\lambdabar_{\sigma}/L_Q=\infty,8,4$, where 
$\infty$ corresponds to no velocity dispersion. This results in running a 
total of $780$ simulations. We verified that our solutions are numerically 
converged by comparing with simulations at $256^3$ resolution. For the 
velocity-dispersed simulations, we run ten additional simulations for each 
Mach number and satellite size, with different initial random phases, in order 
to obtain an \textit{ensemble average} for the calculation of the dynamical friction.

In Fig.~\ref{fig:sim_slices}, we show representative slices through 
the plane $x=0$ for a few of our simulations that do not contain any 
velocity dispersion. The morphological similarities between these 
slices and the solution for the time-independent LPT result 
in Fig.~\ref{fig:LPT_solution} further increases our 
confidence in both results. We can also see from the 
$\mathcal{M}_Q = 1/8$ simulations that the deviation from the LPT 
solution is significant at small $\mathcal{M}_Q$, as we expect.

In Fig.~\ref{fig:sims_slice_veldisp}, we show a version of 
Fig.~\ref{fig:sim_slices} except now with some finite velocity 
dispersion, specifically $\mathcal{M}_{\sigma} = 4$. Here, we can 
see how the overdensities induced in the fluid from the finite 
velocity dispersion act to disrupt the gravitational wake and 
therefore decrease the effect of dynamical friction.

\begin{figure}[tbp]
\centering
\includegraphics{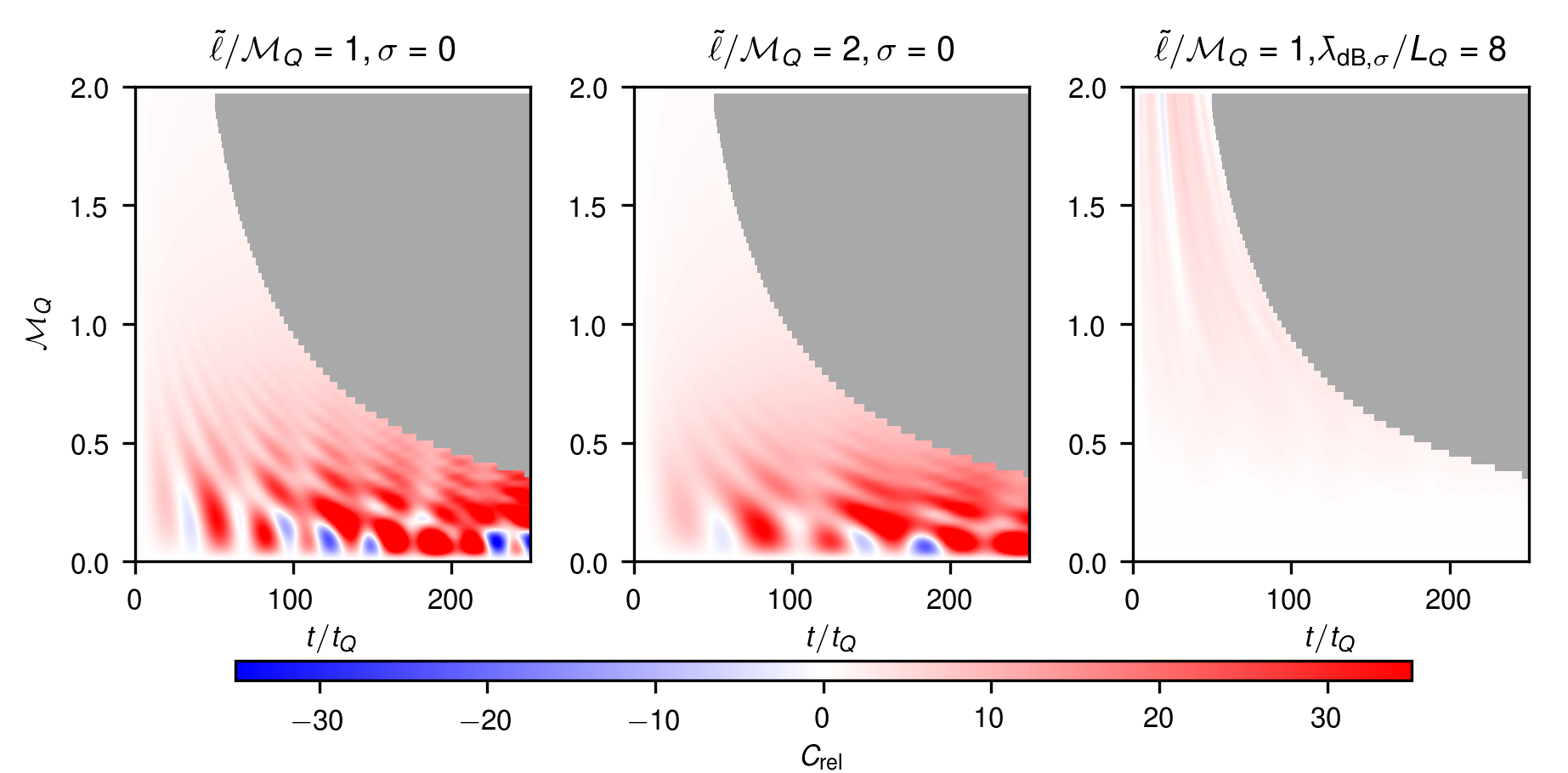}
\caption{\label{fig:DF} The instantaneous dynamical 
friction $C_{\rm rel}$ (defined in Eq.~\ref{eq:F_defs})
as both a function of time and Mach number $\mathcal{M}_Q$ from our 
simulations. Across 
the three panels, we vary the size of the satellite relative to $L_Q$ 
and the presence of velocity dispersion in the background medium. 
The gray areas at the top right of each panel indicate where the satellites have 
traveled half the length of the simulation domain, and we therefore stop 
tracking the evolution.
\textit{Left Panel}: Fiducial case of no background velocity dispersion 
and a satellite size $\ell/L_Q = 1$. \textit{Middle Panel}:
Same as the left panel, but with a satellite of twice the size. 
\textit{Right Panel}: Same as the left panel, 
but with a velocity dispersion $\lambdabar_{{\rm dB},\sigma}/L_Q=8$ (see Sec.~\ref{sec:Scales} for definitions).
}\label{fig:sim_summary}
\end{figure}

\section{Interpretation of Numerical Simulations}
\label{sec:Results}

We extract dynamical friction coefficients as a function of time in our simulations of a finite-size satellite traveling with Mach number $\mathcal{M}_Q$ in constant and velocity-dispersed backgrounds.
These were obtained by integrating the gravitational force of the perturbed dark matter density field acting on the satellite, taking into account the satellite's extended mass distribution as well; implementation details are found in Appendix~\ref{app:num}.
Fig.~\ref{fig:DF} shows the instantaneous dynamical friction coefficients for three different setups and Mach numbers $0\leq \mathcal{M}_Q\leq 2$. The three setups compare a fiducial case of an object of size $\ell/L_Q=1$ and no velocity dispersion $\sigma=0$ against a larger extended object of size $\ell/L_Q=2$ as well as with a velocity-dispersed background with dispersion $\lambdabar_{\sigma}/L_Q = 8$. 

The overall growth of the drag force is logarithmic in time, as expected from time-independent analytic theory.
But a key feature of the time-dependent dynamical friction coefficients are oscillations.
The duration of these oscillations occurs on the scale of the satellite size: it is seen in Fig.~\ref{fig:DF} that the period of oscillations approximately doubles in time when the satellite size is doubled. 
The size of the oscillations are strongest at low quantum Mach numbers $2\pi\mathcal{M}_Q \sim 1$. The overall dynamical friction force is also strongest at $2\pi \mathcal{M}_Q \sim 1$, and sharply drops to $0$ at $\mathcal{M}_Q=0$ which is when the object is at rest with respect to the medium.  
We note that the oscillations may be large enough at low quantum Mach numbers that the dynamical friction drag force can instantaneously be negative at times, e.g., when an interference crest ahead of the satellite pulls the satellite forward.
However, on average, the addition of velocity dispersion in the background reduces the drag force, as the background interferes with the wake.

Below, we will compare a number of time-\emph{independent} expressions 
for the dynamical friction in a given situation to the time-\emph{dependent} 
simulations that we have run.  This comparison, \textit{a priori}, does not 
seem physical. However, we know that the time-independent calculations 
depend upon the dynamical friction cutoff scale $\tilde{b}$.
We also know 
that, absent the quadratic dispersion relation given in 
Eq.~\ref{eq:quantum_disp}, we can think of the wake forming outwardly from 
the satellite propagating roughly at a speed $\vrel$. Thus we can treat the 
scale $\vrel t$ as a stand-in for the cutoff scale $b$ that we use in 
our time-independent calculations, similar to the formalism of \cite{Ostriker99}.

With this in mind, all of the comparisons discussed below will give the total
dynamical friction in the simulation at time $t$ and relate it to the 
time-independent calculation integrated out to the cutoff scale $\vrel t$. 
This scale will simply be referred to by the variable $b$ for the cutoff 
scale (typically in units of $\lambdabar$, indicated by $\tilde{b}$).

\begin{figure}
    \centering
    \includegraphics{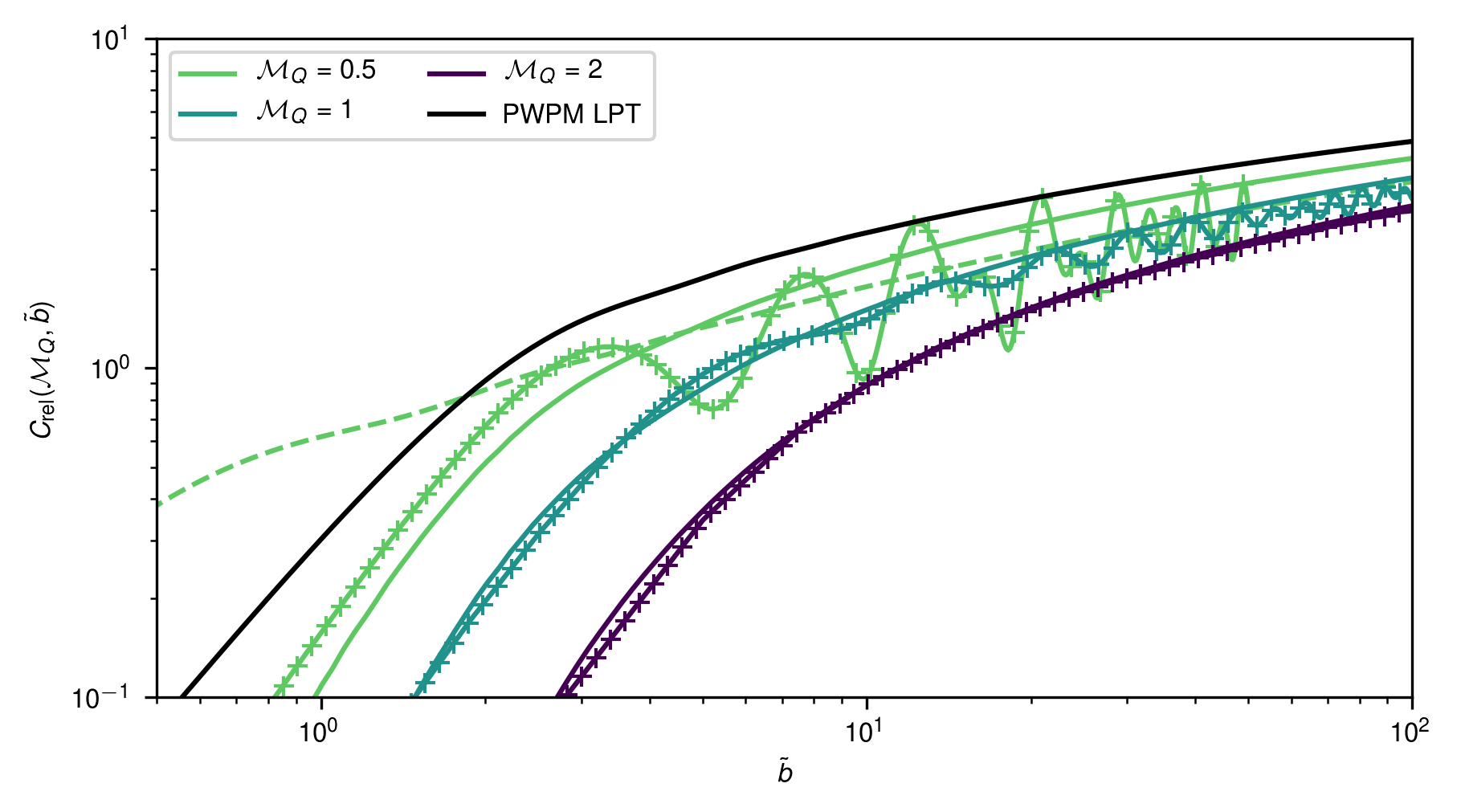}
    \caption{Dynamical friction coefficient 
    $C_{\rm rel}$ as computed from LPT in Eq.~\ref{eq:df_extended_lpt_int} (solid lines),
    along with the values calculated in our simulations (`+' marks), for 
    several different values of the quantum Mach number $\mathcal{M}_Q$. 
    For the most strongly non-linear case ($\mathcal{M}_Q=0.5$) we also 
    include the $C_{\rm rel}$ calculated from the exact solution for 
    a point source (Eq.~\ref{eq:df_exact}) with $\mathcal{M}_Q = 0.5$ (dashed line).
    In our simulations, $\tilde{b}$ is associated with $\vrel t$, as 
    described in the text.
    All of the above pertain to a finite-size satellite with $\ell / L_Q = 1$. We also show the LPT result for  
    a point source as computed from Eq.~\ref{eq:Crel_lpt} (black solid line).
    The LPT result for extended sources matches the $\mathcal{M}_Q =2$ 
    simulation to better than 3\% for $\tilde{b}\gtrsim 7$. For the non-linear 
    case of $\mathcal{M}_Q=0.5$, it is clear that the exact point mass formula 
    (Eq.~\ref{eq:df_exact}) better matches the mean behaviour of the simulations.
    However, it clearly does not capture the significant oscillations away from 
    this mean behavior, which are most likely due to the simulations taking place 
    over a finite time, whereas Eq.~\ref{eq:df_exact} is computed in an infinite-time 
    limit.
    }
    \label{fig:extended_LPT_comp}
\end{figure}

\subsection{Comparison of Finite-Size Calculations to Simulations}
\label{subsec:comp_fs}

Beginning first with the case of zero velocity dispersion, we investigate the accuracy of the LPT for 
finite-size satellites as discussed in Section \ref{sec:lpt_extended}. We expect our results to be most accurate in the LPT regime where $\mathcal{M}_Q \gg 1$, which is probed only to a limited extent by the simulations that we have run in this work, due to numerical resolution limitations. The results of this comparison can be seen in Fig.~\ref{fig:extended_LPT_comp}. 

As expected, the LPT 
predictions perform exceedingly well against the 
simulations at higher Mach number. In particular, 
the $\mathcal{M}_Q = 2$ case agrees to better than 3\% for 
$\tilde{b}\gtrsim 7$. This case corresponds to $\tilde{\ell} = 2$, 
since the simulations in Fig.~\ref{fig:extended_LPT_comp} all have 
a satellite size of $\ell = L_Q$. On the other hand, in the non-linear regime of 
$\mathcal{M}_Q=0.5$, we see that while the LPT predicts the general 
trend of the dynamical friction force quite well, the true dynamical 
friction fluctuates much more strongly than the LPT case. The LPT result also tends 
to systematically overestimate the dynamical friction force in this 
non-linear regime, by a factor that becomes larger as one moves deeper into the non-linear regime. In the weakly non-linear regime of $\mathcal{M}_Q=0.5$ that is shown in Fig.~\ref{fig:extended_LPT_comp}, the over-estimate of $C_{\rm rel}$ is only of order unity.

\begin{figure}
    \centering
    \includegraphics{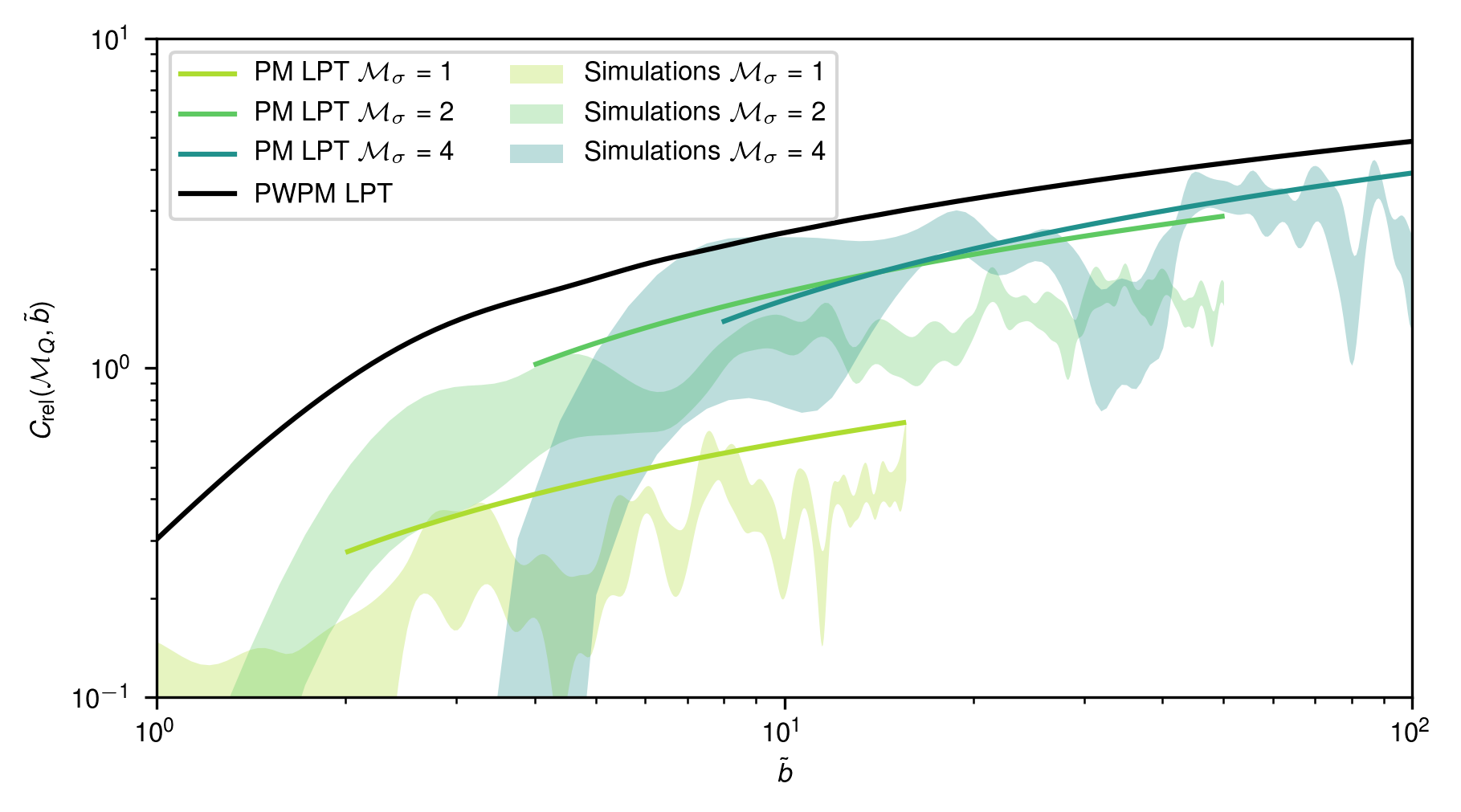}
    \caption{Dynamical friction coefficient 
    $C_{\rm rel}$ as computed from Eq.~\ref{eq:Csig_baror} using LPT for a velocity-dispersed medium (solid lines),
    along with the values calculated in our simulations (shaded regions), for 
    several different values of the quantum Mach number $\mathcal{M}_Q$. The
    shaded regions indicate the one-standard-deviation range of dynamical 
    friction forces over the ensemble of simulations for a given $\mathcal{M}_Q$, with the other parameters set to $\ell / L_Q = 1$ and $v_Q/\sigma=4$.
    Note that while the simulations are fully non-linear and use 
    finite-size satellites, we are comparing to the theory for a point mass in the linear regime, and only show these analytic results in the regime $\tilde{b}>2\mathcal{M}_{\sigma}$ where they are applicable.
    }
    \label{fig:veldisp_LPT_comp}
\end{figure}

\subsection{Comparison of Velocity-Dispersed Calculations to Simulations}

We can also compare our simulations of dynamical friction in a velocity-dispersed FDM medium to the theory that we have developed for that 
scenario in Section \ref{subsec:vel_disp_analytic}. When making this 
comparison, we must keep in mind that any individual realization of a velocity-dispersed FDM medium will have a particular set of over- and under-densities (see 
Fig.~\ref{fig:sims_slice_veldisp}) that will affect the temporal evolution 
of the dynamical friction. As mentioned in Section \ref{sec:Numerical}, we 
mitigate these effects by running an ensemble of simulations for a given 
set of parameters ($\mathcal{M}_Q$, $\mathcal{M}_{\sigma}$, $\mathcal{\ell}$) 
and then comparing our analytic theory with the range of values within one standard deviation over the ensemble of simulations. This comparison is shown in Fig.~\ref{fig:veldisp_LPT_comp} for $\ell/L_Q = 1$, $v_Q/\sigma=4$, and $\mathcal{M}_Q= 0.25,0.5,1$. 

While the simulations are 
of course fully non-linear in their treatment of the dynamical friction force 
and use finite-size satellites, we compare these simulations to the theory developed for point masses in a linear regime as given in Eq.~\ref{eq:Csig_baror}. Nonetheless, the analytic results give decent agreement with the simulations. In particular, for $\mathcal{M}_Q = 1$ which is the 
closest to the linear regime for the simulations shown, Eq.~\ref{eq:Csig_baror} 
does quite well at capturing the trends in the simulations.
However, as with the results of Section \ref{subsec:comp_fs}, we see that 
the LPT results overestimate the dynamical friction force in the non-linear 
regime ($\mathcal{M}_Q=0.5$ here). Some of this difference could also be 
due to finite-size effects. These simulations have $\tilde{\ell} = 0.5$, respectively, which
corresponds to the $\mathcal{M}_Q = 0.5$ cases shown in 
Fig.~\ref{fig:extended_LPT_comp}, from which we can see that the inclusion of finite size only changes the resulting dynamical friction force by a factor of order unity for $\bt \gtrsim 10$.


\section{Applications to Astrophysical Systems}
\label{sec:Application}

Now that we have thoroughly investigated the theory of dynamical friction 
in a universe composed of FDM, we would like to apply the 
theory to a few systems of interest. To do this, we must first identify 
the systems we are interested in and determine in what regime of dynamical 
friction they reside. Towards that end, we may refer to the formulae
for inferring the scale of the dimensionless parameters of interest, namely 
$\mathcal{M}_Q$, $\mathcal{M}_{\sigma}$, and $\tilde{\ell}$, given in 
Section \ref{sec:Scales}. It is important to note here that the regime that 
especially lies outside the validated regime of any analytic theory we have developed
here is $\mathcal{M}_Q \approx \mathcal{O}(1/2\pi)$, based on our simulations. 
This restriction is equivalent to:
\begin{equation}
\label{eq:IntermediateMassEstimate}
    \left(\frac{M}{10^9 \, M_{\odot}} \right)
    \left(\frac{m}{10^{-22} \, {\rm eV}} \right)
    \left(\frac{100\, {\rm km}\, {\rm s}^{-1}}{\vrel} \right)
    \sim 1
\end{equation}{}

Below we will treat the cases of the globular clusters around Fornax in depth, and illustrate why many massive satellites such as the Sagittarius dwarf and the Magellanic Clouds are likely well-described by the classical Chandrasekhar formula. The infall of supermassive black holes (SMBH) in an FDM halo has 
already been thoroughly discussed in \cite{HOTW} and \cite{BarOr18}, though our 
estimate above of the regime in which detailed simulations may be warranted suggests 
that it may be worth revisiting this case as well.

\subsection{Fornax Globular Clusters}

The Fornax dwarf spheroidal (dSph) galaxy is the most massive 
galaxy of its type orbiting the Milky Way (that shows no strong 
signs of tidal disruption) and has thus been extensively studied 
in the literature 
\citep{Coleman08,Poretti08,Larsen12,Cole12,DeBoer16,DelPino17,
Reid19Fornax,Kowalczyk19,Boldrini18,WangMeiDES18}. 
The tendency for dSph galaxies to be heavily dynamically dominated by 
dark matter \citep{Aaronson83,Mateo98,Gilmore07} has made them ideal 
test beds for the nature of dark matter on cosmologically small scales
\citep{WP11,MBKJBRev17}.

One of Fornax's unique features is the set of 
six globular clusters associated with it \citep{Cole12,WangMeiClusterSix}. 
These globular clusters have long puzzled astronomers as they all appear 
to be old ($\sim$10 Gyr), yet dynamical friction calculations show that 
globular clusters with similar orbits should have long ago sunk to the 
center of the Fornax dSph, assuming that they had been in these orbits 
for a significant fraction of their lifetimes~\citep{Tremaine76,Hernandez98,Oh2000,Goerdt06,Cole12}. Specifically, 
the classical treatment of the problem as first raised in \cite{Tremaine76} 
and subsequently discussed in~\cite{Hernandez98,Oh2000} indicated that the 
timescale for the infall of the globular clusters around Fornax to the 
center of the dSph due to dynamical friction should be on the order of 
$\tau_{\rm DF} \sim 1\, $Gyr \cite{Oh2000}, which is much shorter than 
the proposed age of the dSph of $\sim 10\,$Gyr~\cite{FornaxDES,FornaxAge2}.
If the clusters are in fact still infalling, it seems extremely unlikely
that we would happen to observe them all \textit{just} before they fall into the center of Fornax.  This issue has come to be known as the 
\textit{timing problem} \citep{Tremaine76,Tremaine75}.

There have been multiple proposed solutions to this problem, such as
tidal effects of the Milky Way, a population of massive black 
holes which act to dynamically heat the clusters \citep{Oh2000}, 
non-standard dark matter models (such as we investigate here) 
\cite{Boldrini18}, alterations to the dark matter distribution in the 
Fornax dSph \cite{Cole12}, and more complicated treatments of the 
dynamical friction problem beyond a Chandrasekhar-type formula 
\citep{1943ApJ....97..255C,Goerdt06,Boldrini18,Kaur18}. We mention these 
arguments for the interested reader, but will not reiterate them here. 
Instead, we will focus on the extent to which FDM may be able to 
independently solve this problem.

Following the approach of \cite{HOTW}, we will estimate the infall 
time of each globular cluster due to dynamical friction, $\tau_{\rm DF}$,
as the angular momentum of the globular cluster's orbit divided by the 
torque due to dynamical friction. Assuming that each globular cluster is 
on a circular orbit, this time is given by
\begin{equation}
    \label{eq:taudf_formal}
    \tau_{\rm DF} = \frac{L}{r|F_{\rm DF}|} 
    = \frac{\vrel^3}{4\pi\rhobar G^2 M C_{\rm rel}} \, ,
\end{equation}
where $\vrel$ is the circular orbit velocity, $\rhobar$ is the local density of dark matter at the position of the given globular
cluster, $M$ is the mass of the globular cluster, $G$ is Newton's constant, 
and $C_{\rm rel}$ is as defined in Eq.~\ref{eq:Cdef}.

\begin{figure}
    \centering
    \includegraphics{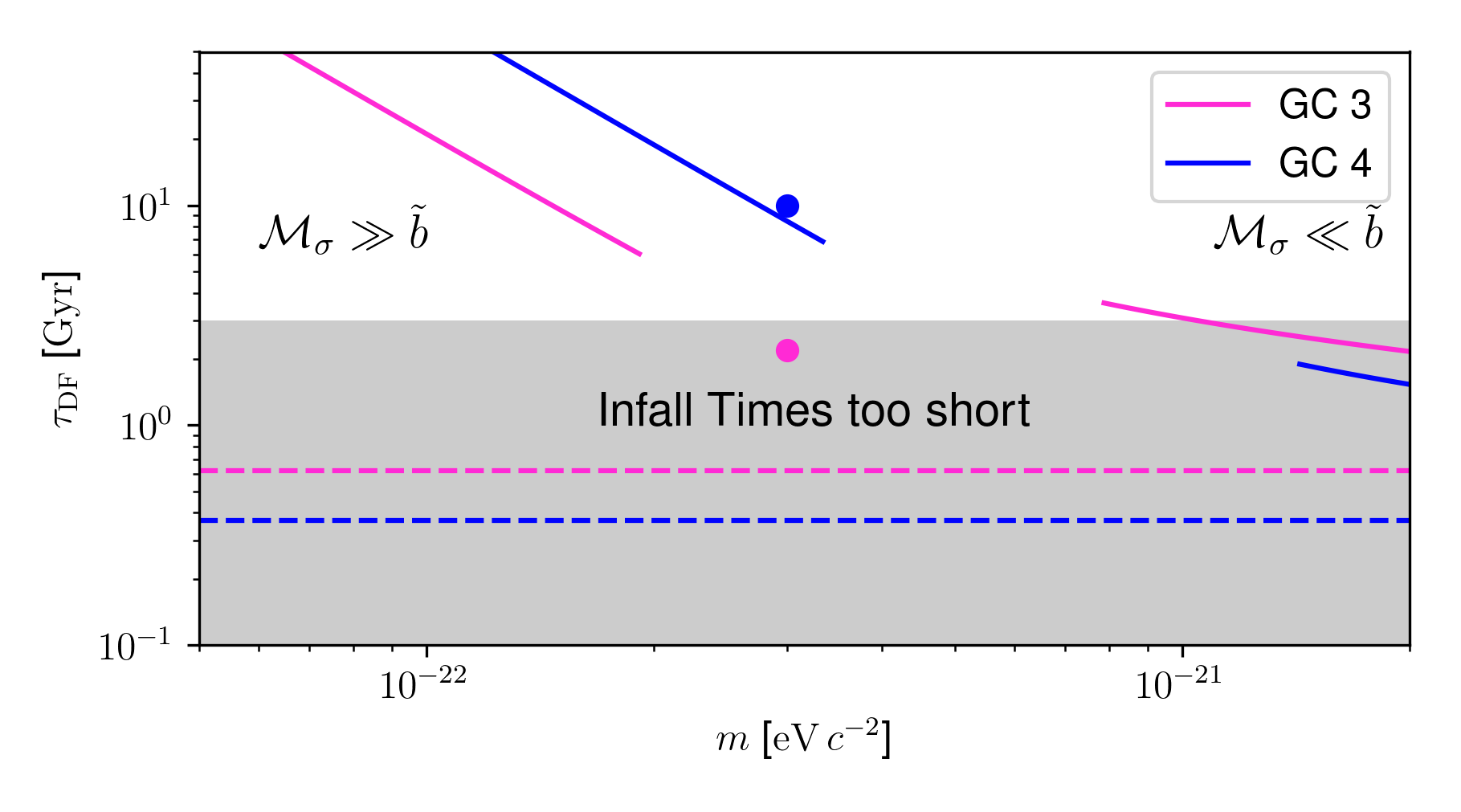}
    \caption{Infall times as a function of FDM particle mass for 
    the two Fornax globular clusters with the shortest infall times, GC 3 and 
    GC 4. The grey shaded region indicates $\tau_{\rm DF}< 3\,$Gyr, roughly the 
    region in which a timing problem exists. The horizontal dashed lines 
    indicate the infall times in a $\Lambda$CDM universe as calculated in 
    \cite{HOTW}, which we can see clearly lies well within the timing-problem 
    domain. The points indicate the infall times in an FDM scenario calculated 
    by \cite{HOTW} for $m=3\times 10^{-22}\,{\rm eV}$ using 
    Eq.~\ref{eq:Crel_lpt}. The solid lines indicate our estimates for the 
    infall times in the FDM scenario where Eq.~\ref{eq:Crel_lpt} is applied 
    at the left-hand side of the plot and Eq.~\ref{eq:Csig_baror} is applied 
    at the right-hand side of the plot.
    The break between the solid lines indicates the region where the velocity 
    dispersion de Broglie wavelength is on the order of the size of the system 
    and individual over- and under-densities make $\tau_{\rm DF}$ quite 
    uncertain/stochastic. Our estimates differ from those of \cite{HOTW} 
    (points) in that we use updated constraints on the dynamic properties of 
    the Fornax dSph.  Additionally, the \cite{HOTW} estimates are simply 
    extrapolations of the point-mass LPT curves (curves shown on the left hand 
    side of the plot) whereas we make sure to apply this theory only in the 
    appropriate regime.
    }
    \label{fig:fornax_gcs_infall}
\end{figure}

To estimate the infall time, we must determine the correct
formula to use to calculate $C_{\rm rel}$. The typical size of the globular clusters in the Fornax dSph system is 
about $\ell \sim 2 \, {\rm pc}$, they have a typical mass of about 
$M \sim 2 \times 10^5\, M_{\odot}$, and a typical orbital velocity of $\sim10\,{\rm km}\, {\rm s}^{-1}$ (assuming an isotropic velocity distribution)~\cite{Cole12,Mackey03,Mateo91}.  Though the distances from the center of the 
Fornax dSph vary over the collection of globular clusters, we will take the 
typical globular cluster to be located at the core radius of the King profile 
fit to the Fornax dSph, which is $r\approx 668\, {\rm pc}$ \cite{Walker09}. The 
density of dark matter at this radius has been estimated to be 
$\rhobar \approx 2\times 10^7\, M_{\odot} \, {\rm kpc}^{-3}$ \cite{Read19} and 
the velocity dispersion of the dark matter at this radius (estimated from 
the velocity dispersion of the stars) is 
$\sigma \approx 10 \, {\rm km}\, {\rm s}^{-1}$~\cite{Walker09}.

With these numbers in mind, we can calculate the dimensionless parameters of 
interest:
\begin{equation}
    \tilde{\ell} \approx 1.04 \times 10^{-3} 
    \left( \frac{m}{10^{-22}\, {\rm eV}}\right) \ \ , \ \ 
    \tilde{b} \approx 3.47  \times 10^{-1}
    \left( \frac{m}{10^{-22}\, {\rm eV}}\right) \ \ 
\end{equation}
\begin{equation}
    \mathcal{M}_Q \approx 90 \left( \frac{m}{10^{-22}\, {\rm eV}}\right)^{-1} 
    \ \ , \ \ 
    \mathcal{M}_{\sigma} \approx 1 \, .
\end{equation}
We can immediately see that we are almost always in the LPT regime ($\mathcal{M}_Q \gg 1$) and that the globular clusters
can be accurately treated as point masses 
($\tilde{b}\gg \tilde{\ell}$).  At low FDM mass, 
the de Broglie wavelength of the velocity dispersion is much greater than the 
size of the system $\lambdabar_{\sigma}\gg b$ (or equivalently, 
$\mathcal{M}_{\sigma} \gg \tilde{b}$) and we may treat the background as constant 
density, applying Eq.~\ref{eq:Crel_lpt} for $C_{\rm rel}$ in 
Eq.~\ref{eq:taudf_formal}. At high FDM mass, we are in the opposite regime, 
$\mathcal{M}_{\sigma} \ll \tilde{b}$, and the inhomogeneities caused by the 
velocity dispersion are very small compared to the size of the system, so  
we may accurately apply Eq.~\ref{eq:Csig_baror}. However, between these 
two regimes, the precise dynamical friction force will be strongly affected by 
the presence or absence of single over-densities/under-densities in the FDM 
caused by the velocity dispersion. In this regime, the infall time is uncertain.

We estimate the infall times for the two Fornax globular clusters with 
the shortest infall times (for which the timing problem is most severe). We 
use information on the mass $M$, projected radial distance from the center 
of the Fornax dSph $r_{\perp}$, and core sizes $\ell$ taken from \cite{Cole12,Mackey03,Walker09}. 
These are the same parameters as used in \cite{HOTW}. Following the procedure 
of \cite{HOTW} further, we take the true radial distance of the globular clusters from the 
center of the Fornax dSph to be $r=2r_{\perp}/\sqrt{3}$ and then take the 
velocity dispersion curve to be roughly flat at 
$\sigma \approx 10\, {\rm km}\, {\rm s}^{-1}$~\cite{Cole12,Read19}. We also 
take the globular clusters to be on circular orbits, with velocities 
determined as $\vrel = \sqrt{GM_{\rm encl}(r)/r}$, where $M_{\rm encl}(r)$ is 
the mass of the Fornax dSph contained within the radius $r$.
However, 
unlike \cite{HOTW}, we use updated fits to the mass and density profiles of 
the Fornax dSph taken from \cite{Read19} and we only apply Eq.~\ref{eq:Crel_lpt} 
when $\mathcal{M}_{\sigma}> 2 \tilde{b}$, using Eq.~\ref{eq:Csig_baror} when 
$\mathcal{M}_{\sigma}<\tilde{b}/2$. As argued above, in the regime 
$\tilde{b}/2<\mathcal{M}_{\sigma}<2\tilde{b}$, the dynamical friction force is 
made uncertain by the presence of individual over-densities in the FDM, so we
do not make a prediction for the infall time.

We show the results of this calculation in Fig.~\ref{fig:fornax_gcs_infall}. 
It is clear that for FDM masses $m \gtrsim 10^{-21}\, {\rm eV}$, FDM no longer 
resolves the timing problem. It should also be noted that as $m$ becomes larger 
we will move first into the non-linear regime ($\mathcal{M}_Q \sim 1$) where 
our LPT calculations do not apply and one must be careful applying analytic 
arguments. As $m$ grows larger, we will move in to the classical regime 
($m \to \infty$ or $\mathcal{M}_Q \ll 1$) where we may apply the classical 
Chandrasekhar formula. One must be aware of these different regimes when 
evaluating whether dynamical friction changes significantly in the FDM picture.

\subsection{The Sagittarius Stream}
\label{subsec:sgr}

The Sagittarius (Sgr) stream is a massive stellar stream that consists of the 
tidal debris of one of the Milky Way's most massive recent accretion events~
\cite{Ibata94,Johnston95,Helmi04,LJM05,Belokurov06,LM10,Purcell11,Koposov12,LM16,DL17,Hernitschek17,Fardal19}. A recent review of the literature is given in \cite{LM16} with 
important, more recent, contributions in \cite{DL17,Hernitschek17,Fardal19}.
Dynamical modeling of the Sgr orbit can constrain the shape of the Galactic potential, and 
therefore the structure, of our Galaxy \cite{Johnston95,LM10,Purcell11,DL17,Fardal19}. 

It has long been known that Sgr's orbit depends sensitively on the effects of 
dynamical friction \cite{JB00}, making various families of orbital parameters 
viable for different ranges of initial masses for Sgr. 
Here, we simply make an estimate of the dimensionless parameters 
which determine the applicable theory for $C_{\rm rel}$, as a guide for future work investigating the infall of Sgr in an FDM universe.

To do this, we follow the orbital model of \cite{DL17}, taking the initial 
scale-size of Sgr (upon first infall) as $\ell = 25\, {\rm kpc}$, the 
initial mass as $M_{\rm Sgr} = 1.3 \times 10^{10} \, M_{\odot}$, the initial 
distance from the center of the Milky Way as $d_{\rm init} = 125 \, {\rm kpc}$,
and the initial velocity relative to the center of the Milky Way as 
$v_{\rm init} = 72.6\, {\rm km}\, {\rm s}^{-1}$. Using these parameters, we infer
\begin{equation}
    \mathcal{M}_Q = 2.49 \times 10^{-2} 
    \left( \frac{m}{10^{-22}\, {\rm eV}}\right)^{-1} \ \ , \ \ 
    \tilde{\ell} = 94.7 \left( \frac{m}{10^{-22}\, {\rm eV}}\right) \ \ , \ \ 
    \tilde{b} = 473 \left( \frac{m}{10^{-22}\, {\rm eV}}\right) \, .
\end{equation}
We can see that the large mass of the Sgr dwarf puts this 
problem strongly outside the regime of LPT,  but most likely into the 
regime where classical arguments should still be viable ($\mathcal{M}_Q \ll 1$). Additionally, 
the size of the Sgr dwarf is large enough to play an important role in 
reducing the dynamical friction significantly.

Recent studies enabled by the data provided by the \textit{Gaia} satellite 
\cite{GAIA,DR2} have allowed authors to look at the shape of the velocity 
ellipsoid of the Milky Way's stellar halo out to large Galactocentric 
radii \cite{Bird18,Lancaster19a}. This allows us to estimate the final 
dimensionless parameter of interest, the classical Mach number 
$\mathcal{M}_{\sigma}$. Taking the one-dimensional velocity dispersion 
at Galactocentric radii of $r=125\, {\rm kpc}$ to be 
$\sigma \approx 80 \, {\rm km} \, {\rm s}^{-1}$ \cite{Bird18}, we see that 
$\mathcal{M}_{\sigma} \approx 1$, meaning that the finite velocity 
dispersion should be important, especially as the system is large.   The parameter estimates above indicate that for most values of $m$ that are 
of interest, classical dynamical friction arguments should still accurately 
describe the orbit of Sagittarius.

\subsection{The Magellanic Clouds}
\label{subsec:MCs}

The Magellanic Clouds, consisting of the Large Magellanic Cloud (LMC) and 
the Small Magellanic Cloud (SMC), are some of the most massive satellites 
of the Milky Way and have been studied in depth in the literature 
\cite{Besla07,Kalli13,Erkal19,GC19,Nidever19,Choi19}.
Dynamical friction has a very important role to play in the orbit of the 
Magellanic Clouds around the Milky Way, as was recently shown in \cite{GC19} 
using high-resolution $N$-body simulations in a universe with `traditional' 
CDM. In these simulations, the authors of \cite{GC19} were able to show that the dark matter 
wake created by the infall of the LMC should have observable effects on the 
outer parts of the stellar halo. 

Since the form of the DM wake formed in the FDM paradigm is so drastically 
different, we would of course expect the observable effects 
to change in the FDM scenario. Similarly as for Sgr, we aim here to estimate 
the values of the dimensionless parameters of interest to determine in what 
regime it lies. To this end, we take parameters for the LMC from \cite{GC19}, 
using a scale 
size of $\ell \approx 20\,  {\rm kpc}$, a total mass of 
$M_{\rm LMC} \approx 2 \times 10^{11}\, M_{\odot}$, and a total orbital 
velocity of $\vrel \approx 325 \, {\rm km} \, {\rm s}^{-1}$ gives us 
the dimensionless parameters of interest:
\begin{equation}
    \mathcal{M}_Q = 7.24 \times 10^{-3} 
    \left( \frac{m}{10^{-22}\, {\rm eV}}\right)^{-1} \ \ , \ \ 
    \tilde{\ell} = 340 \left( \frac{m}{10^{-22}\, {\rm eV}}\right) \ \ , \ \ 
    \tilde{b} = 848 \left( \frac{m}{10^{-22}\, {\rm eV}}\right) \, .
\end{equation}
As we can see, this system, like Sgr, lies deeply in the non-linear regime, 
and the finite size of the system is large enough to play a significant 
role in altering the dynamical friction.  The LMC system is in a regime 
where the classical Chandrasekhar formula will well-approximate the net 
dynamical friction on the object. However, the de Broglie wavelength of the relative motion
of the system is not negligible: $\lambdabar \sim 60\, {\rm pc}$, and the gravitational wake will have transient structures on this length scale, which may have observable effects.

\begin{table}
    \centering
    \begin{tabular}{|c|c|c|c|c|}\hline
        \textbf{Physical Setup}    & Exact PM 
                          & LPT PM
                          & LPT EM
                          & LPT VD, PM\\ \hline
        \textbf{Relevant equation} & \ref{eq:df_exact}   
                          & \ref{eq:df_point_source_lpt_int}, \ref{eq:Crel_lpt}
                          & \ref{eq:df_extended_lpt_int}
                          & \ref{eq:Csig_baror}\\ \hline
                          
    \end{tabular}
    \caption{Here we provide a reference to the relevant equations for 
    dynamical friction coefficients in a given physical scenario. In the above 
    we make the following abbreviations: point mass (PM), extended mass (EM), 
    velocity dispersed (VD), and (as throughout the text) linear perturbation 
    theory (LPT).}
    \label{tab:eqs_ref}
\end{table}{}

\section{Conclusions}
\label{sec:Conclusions}

In this work, we have investigated the phenomenon of dynamical friction acting on a satellite moving through an FDM background. We provided a detailed exploration of the analytic theory associated with this 
phenomenon, in a formalism which allows the calculation of the spatial dependence of the overdensity created by the massive satellite. After briefly reviewing the case of a point-source satellite in a background with no velocity dispersion and checking our results against the known 
exact solution, we moved to linear perturbutation theory (LPT), where we 
derived results for \emph{(i)}~point masses, \emph{(ii)}~finite-size satellites with a Plummer 
density profile, and \emph{(iii)}~a point-mass satellite in a velocity-dispersed background of FDM. 

To test our analytic results and to explore more non-linear regimes, 
we ran a large suite of fully non-linear simulations of the formation of the 
gravitational wake in an FDM medium. Comparing these simulations to 
our analytic results, we found excellent agreement in the linear regime, where our 
analytic results should be valid. Importantly, we determined that the LPT tends 
to overestimate the dynamical friction force when applied in the non-linear regime, both for the case of zero dispersion and finite dispersion.

Finally, having extensively validated and tested our analytic results, we applied them to the investigation of 
various astrophysical problems of interest: \emph{(i)}~the globular clusters in the 
Fornax dSph galaxy, \emph{(ii)}~the Sagittarius stream, 
and \emph{(iii)}~the Magellanic Clouds.  Our application to the Fornax globular clusters 
allowed us to quantitatively show in what regime FDM solves the so-called timing problem. In the case of Sagittarius and the Magellanic Clouds, 
we determined that the net drag of dynamical friction on these bodies 
is most likely well within the regime where classical arguments are 
applicable.  That said, transient wave-like effects in the gravitational wake of these 
objects could be significant enough to have observable consequences. 
Validating this would, however, require detailed simulations 
of the objects in question.

We summarize our main conclusions as follows:
\begin{itemize}
    \item FDM particles with masses $m \gtrsim 10^{-21}\, {\rm eV}$ 
    no longer solve the so-called `timing problem' of the Fornax globular 
    clusters.
    \item There are three distinct regimes for dynamical friction in 
	FDM: (1) the de Broglie wavelength is large and the wake is set by the 
	quantum pressure, well described by LPT; (2) the background fluid has velocity dispersion, 
	de Broglie wavelength is small, and the wake behaves similarly to a 
	collisionless Chandrasekhar wake; and (3) the length scales of the wake and the de Broglie wavelength of the velocity dispersion are comparable, and the wake has a stochastic character arising from
	interference crests of the background.  
    \item LPT calculations overestimate the dynamical 
    friction on a body when applied in the non-linear regime. 
    \item Time-independent LPT is an excellent 
    approximation to the true time-dependent answer, as long as it is 
    applied in the correct way while in the linear regime 
    ($\mathcal{M}_Q\gtrsim 1$).
    \item The dynamical friction force is relatively insensitive to the shape of the satellite's density distribution as long as the `system size' $b$ is much greater than the satellite's size $\ell$. Instead, the parameter that matters the most for the overall dynamical friction force is the satellite's size itself, $\ell$.
\end{itemize}

While FDM is an intriguing model of dark matter with the potential to resolve outstanding issues with small-scale structure, its unique phenomenology, with wave-like effects manifesting on galactic scales, offers numerous opportunities to constrain the model with astrophysical systems. Some constraints, such as those derived from pulsar timing \cite{Khmelnitsky:2013lxt,Porayko:2018sfa}, are quite robust but are limited by instrumental sensitivity. On the other hand, the changes that the FDM model implies for dynamical friction provide several test beds for the model itself. Given the range of satellite masses probed by our application to classic dynamical friction problems (Fornax globular clusters, Sgr, LMC) it seems that some of the best tests of these effects could come in the 
intermediate mass regime where $\mathcal{M}_Q \sim \mathcal{O}(1/2\pi)$, which is outside the regime of both our analytic theory and that of classical treatments. As seen from Eqs.~\ref{eq:MQ} and \ref{eq:IntermediateMassEstimate}, for $m \sim 10^{-22} \ {\rm eV}$ and relative velocities $v_{\rm rel} \sim 100 \ {\rm km/s}$, these effects will be seen for satellite masses of $M \sim 10^9 M_{\odot}$. Some potential candidate objects could be nuclei of galaxies during mergers or SMBHs on the higher mass end, also during galaxy mergers.
Thus, dedicated simulations of infalling intermediate-mass satellites are crucial, and we look forward to the development of such simulations to shed light on the nature of dark matter in our Galaxy.

\acknowledgments

The authors would like to thank  Mustafa Amin, Lasha Berezhiani, Pierre-Henri Chavanis, Wyn Evans, Beno\^{i}t Famaey, Rodrigo Ibata, Justin Khoury, Sergey Koposov, Jens Niemeyer, Eve Ostriker,  Justin Read, Vassilios Tsiolis, Matthew Walker, and Wenrui Xu for helpful comments and conversations.
We would especially like to thank Scott Tremaine for his detailed reading and thoughtful comments on this work.
P.M. is supported by NASA through the Einstein Postdoctoral Fellowship grant number PF7-180164 awarded by the Chandra X-ray Center, which is operated by the Smithsonian Astrophysical Observatory for NASA under contract NAS8-03060.
M.L. is supported by the DOE under Award Number
DESC0007968 and the
Cottrell Scholar Program through the Research Corporation for Science Advancement.
Some of the computations in this paper were run on the Odyssey cluster supported by the FAS Division of Science, Research Computing Group at Harvard University. This work was supported in part by the Kavli Institute for Cosmological Physics at the University of Chicago through an endowment from the Kavli Foundation and its founder Fred Kavli.

\appendix
\section{Integrals for Linear Perturbation Theory Results}
\label{app:ints_lpt}

We proceed here with the solution to the form of the overdensity 
(or wake) created as a point mass travels through an initially 
uniform condensate in the limit of infinite time. We continue 
where we left off with Eq.~\ref{eq:alpha_sol2}. To simplify the notation, we will drop all tildes for dimensionless quantities, and in this Appendix all distances and wave vectors will be in units of $\lambdabar$ or $1/\lambdabar$ as appropriate.

We note that we may perform the integral over $k_z$ with contour 
integration, which allows us to temporarily ignore the factor 
of $k_R J_0(k_R R)$. For simplicity we will now define the 
dimensionless integrand
\begin{equation}
    \label{eq:Iint1}
    I(z,k_R) = \int_{-\infty}^{\infty} \diff k_z
    \frac{e^{ik_z z}}{k^4 - 4k_z^2} \, ,
\end{equation}
This integral will be evaluated by contour integration, so 
it is useful to specify the poles of the integrand. As 
$k^2 = k_R^2 + k_z^2$, one can show that the denominator of the 
integrand above can be factorized as
\begin{equation}
    \label{eq:pole_factorize}
    k^4 -4k_z^2 = (k_z^2 - \chi_+^2)(k_z^2 - \chi_-^2) \, ,
\end{equation}
where $\chi_{\pm} = 1 \pm \sqrt{1 - k_R^2}$. The poles 
of the integrand are then $k_z = +\chi_+,-\chi_+,+\chi_-,-\chi_-$,
which can either lie on or off of the real line depending on whether 
$k_R> 1$ or $k_R<1$; we will deal with each of these cases individually.
First we make the poles explicit by rewriting the integrand as
\begin{equation}
    \label{eq:Iint2}
    I(z,k_R) = \int_{-\infty}^{\infty} \diff k_z
    \frac{e^{ik_z z}}{(k_z-\chi_-)(k_z+\chi_-)(k_z-\chi_+)(k_z+\chi_+)} \, .
\end{equation}

\begin{itemize}
   \item[\textbf{Case 1:}] $k_R>1$
    
    In this case, all of the poles of the integrand lie off of the 
    real line. We show the poles and two possible contours in the complex plane of $k_z$ below,
    where the original integral $I(z,k_R)$ is along the real axis.
    \begin{center}
    \begin{tikzpicture}[thick,
    decoration={
    markings,
    mark=at position .5 with {\arrow{latex}}},
    every pin/.append style = {pin distance=10mm, pin edge={<-,black}}
             ]
    
    
    \draw [<->] (-4,0) -- (4,0) node [above right]  {$\re(k_z)$};
    \draw [<->] (0,-4) -- (0,4) node [below left=-1pt] {$\im(k_z)$};
    
    \draw[fill] (-2,1) circle (1pt) node [right] {$- \chi_{-}$};
    \draw[fill] (-2,-1) circle (1pt) node [right] {$- \chi_{+}$};
    \draw[fill] (2,1) circle (1pt) node [right] {$ \chi_{+}$};
    \draw[fill] (2,-1) circle (1pt) node [right] {$\chi_{-}$};
    
    \draw[DarkBlue,postaction={decorate}, very thick]   ( -4,0) -- ( 0, 0);
    \draw[DarkBlue,postaction={decorate}, very thick]   ( 0,0) -- ( 4, 0) node [below right] {$C$};
    \draw[DarkBlue,postaction={decorate},dashed, very thick]  (4,0) arc (0:90:4 and 3) node [above right] {$z > 0$};
    \draw[DarkBlue,postaction={decorate},dashed, very thick] (0,3) arc (90:180:4 and 3) ;
    \draw[DarkBlue,postaction={decorate},dashed, very thick] (0,-3) arc (270:180:4 and 3);
    \draw[DarkBlue,postaction={decorate},dashed, very thick] (4,0) arc (360:270:4 and 3) node [below left] {$z < 0$};
    \end{tikzpicture}
    \end{center}
    We see upon examining the exponential term in Eq.~\ref{eq:Iint2}
    that we should close the contour in the upper-half of the 
    complex plane when $z>0$ (making the contribution to the integral 
    that is off of the real line exponentially small), and we thereby pick up the 
    residues at $k_z = -\chi_-, \chi_+$. Similarly, when $z<0$ we close 
    in the lower-half plane and pick up the residues of the other two poles.
    
    Using the definitions of $\chi_{\pm}$ and Cauchy's integral 
    formula we can then easily evaluate the integral \ref{eq:Iint2}.
    In principle, there should be distinct cases for $z$ positive and negative, but it turns out that the 
    solution can be succinctly written as
    \begin{equation}
        \label{eq:Iint_krgt1}
        I(z,k_R>1) = \frac{\pi e^{-\sqrt{k_R^2 - 1}|z|}}{2\sqrt{k_R^2 -1}k_R^2}
        \left(\cos(z) + \sqrt{k_R^2 - 1}\sin(|z|) \right).
    \end{equation}
    However, since this formula is even in $z$, this part of the 
    integrand will not actually contribute to any dynamical friction 
    forces, as the overdensity will be equal in front of and behind the perturber.

    \item[\textbf{Case 2:}] $k_R<1$
    
    In this case, the poles all lie on the real line and the choice of contour will depend on a prescription for avoiding the poles. Below, we show our choice of pole prescription along the real line along with the two choices for closing the contour depending on the sign of $z$; we explain the rationale behind our pole prescription below.
    
    \begin{center}
    \begin{tikzpicture}[thick,
    decoration={
    markings,
    mark=at position .5 with {\arrow{latex}}},
    every pin/.append style = {pin distance=21mm, pin edge={<-,black}}
             ]
    
    
    \draw [<->] (-4.2,0) -- (4.2,0) node [above right]  {$\re(k_z)$};
    \draw [<->] (0,-4) -- (0,4) node [below left=-1pt] {$\im(k_z)$};
    
    \draw[fill] (-1,0) circle (1pt) node [below] {$- \chi_{-}$};
    \draw[fill] (-3,0) circle (1pt) node [above] {$- \chi_{+}$};
    \draw[fill] (3,0) circle (1pt) node [above] {$ \chi_{+}$};
    \draw[fill] (1,0) circle (1pt) node [below] {$\chi_{-}$};
    
    \draw[DarkBlue,postaction={decorate}, very thick]   ( -4,0) -- ( -3.5, 0);
    \draw[DarkBlue,postaction={decorate}, very thick] (-3.5,0) arc (180:360:0.5) ;
    \draw[DarkBlue,postaction={decorate}, very thick]   ( -2.5,0) -- ( -1.5, 0);
    \draw[DarkBlue,postaction={decorate}, very thick] (-1.5,0) arc (180:0:0.5) ;
    \draw[DarkBlue,postaction={decorate}, very thick]   ( -.5,0) -- ( 0.5, 0);
    \draw[DarkBlue,postaction={decorate}, very thick] (.5,0) arc (180:0:0.5) ;
    \draw[DarkBlue,postaction={decorate}, very thick]   ( 1.5,0) -- ( 2.5, 0);
    \draw[DarkBlue,postaction={decorate}, very thick] (2.5,0) arc (180:360:0.5) ;
    \draw[DarkBlue,postaction={decorate}, very thick]   ( 3.5,0) -- ( 4, 0) node [below right] {$C$};
    \draw[DarkBlue,postaction={decorate},dashed, very thick]  (4,0) arc (0:90:4 and 3) node [above right] {$z > 0$};
    \draw[DarkBlue,postaction={decorate},dashed, very thick] (0,3) arc (90:180:4 and 3) ;
    \draw[DarkBlue,postaction={decorate},dashed, very thick] (0,-3) arc (270:180:4 and 3);
    \draw[DarkBlue,postaction={decorate},dashed, very thick] (4,0) arc (360:270:4 and 3) node [below left] {$z < 0$};
    \end{tikzpicture}
    \end{center}
    We know that the result of the integral must be real, as it is directly proportional to $\alpha$ via a real 
    constant and $\alpha$ must be real.
    Returning to the integral in Eq.~\ref{eq:Iint2}, we can see that 
    if all of the poles are real, then the denominators of all of 
    the residues will be real. This means that, in order to have a 
    chance of the overall integral being real, we need the imaginary 
    part of the exponential term to cancel with another term.  This 
    only happens in the case that $\pm \chi_+$ are both on the same side
    of the contour, and similarly for $\pm \chi_-$. This fact leaves us 
    with four options for our contour. Two of these four choices, 
    the ones where all poles are placed on one half of the complex 
    plane, leave the contribution from this case to be zero in one 
    half of the plane. Though previous authors have motivated that 
    this should be the case \citep{Lora12}, the non-linear
    dispersion relation given in Eq.~\ref{eq:quantum_disp} 
    points to the fact that a causality argument should not 
    motivate this choice, as only linear dispersion relations 
    have finite propagation speeds. Restricting to the case where the 
    $\pm \chi_-$ and $\pm \chi_+$ terms lie on separate 
    sides of the contour, we are left with two cases, 
    corresponding to which terms we put on which side 
    (\emph{e.g.}, $\pm \chi_-$ terms on the upper or lower half of 
    the contour). Since our linear-theory, time-independent 
    treatment has no notion of time, it is time-reversible, and 
    indeed these two choices of contour correspond to a choice 
    of the arrow of time (relative to the velocity vector)~\citep{watanabe2014integral}. Thus, we find that the correct choice of contour for forward time is the one given above.
    
    If we choose the contour given in the figure above, then for 
    $z>0$, we have
    \begin{equation}
        \label{eq:Iint1_krlt1}
        I(z>0,k_R>1) = - \frac{\pi \sin \left(\chi_+ z \right)}{2\chi_+ \sqrt{1-k_R^2}} \, ,
    \end{equation}
and for $z<0$ we find
    \begin{equation}
        \label{eq:Iint2_krlt1}
        I(z<0,k_R>1) = - \frac{\pi \sin \left(\chi_- z \right)}{2\chi_- \sqrt{1-k_R^2}} \, .
    \end{equation}
\end{itemize}

Now that we have evaluated Eq.~\ref{eq:Iint1} in all 
relevant cases, we may return to the statement of the integral 
that we have for $\alpha$ in Eq.~\ref{eq:alpha_sol2}. Following \cite{Lora12}, and noting that the form of the overdensity depends on the cases $k_R >1$ or $k_R<1$, we define
\begin{equation}
    \label{eq:alpha0I}
    \alpha_I (\mathbf{r}) \equiv \frac{16\pi}{(2\pi)^2 \mathcal{M}_Q}
    \int_0^{1} \diff k_R \int_{-\infty}^{\infty} \diff k_z 
    \frac{k_R J_0(k_R R)e^{ik_z z}}{k^4 -4k_z^2}
\end{equation}
and
\begin{equation}
    \label{eq:alpha0II}
    \alpha_{II} (\mathbf{r}) \equiv \frac{16\pi}{(2\pi)^2 \mathcal{M}_Q}
    \int_1^{\infty} \diff k_R \int_{-\infty}^{\infty} \diff k_z 
    \frac{k_R J_0(k_R R)e^{ik_z z}}{k^4 -4k_z^2} \, ,
\end{equation}
so that we have $\alpha = \alpha_I + \alpha_{II}$. As we have solved 
for the integrals over $k_z$ in either case above, we can state these 
integrals in a more simplified way as
\begin{equation}
    \label{eq:alpha0I2_zgt0}
    \alpha_{I}(\mathbf{r}; z>0) = -\frac{2}{\mathcal{M}_Q} 
    \int_{0}^{1} \diff k_R k_R J_0(k_R R) 
    \frac{\sin\left( \chi_+ z\right)}{\chi_+\sqrt{1-k_R^2}}
\end{equation}
\begin{equation}
    \label{eq:alpha0I2_zlt0}
    \alpha_{I}(\mathbf{r}; z<0) = -\frac{2}{\mathcal{M}_Q} 
    \int_{0}^{1} \diff k_R k_R J_0(k_R R) 
    \frac{\sin\left( \chi_- z\right)}{\chi_-\sqrt{1-k_R^2}} \, .
\end{equation}
We now make a change of variables in Eq.~\ref{eq:alpha0I2_zgt0} and 
\ref{eq:alpha0I2_zlt0} so that the integrals are over $\chi_{\pm}$ 
instead of over $k_R$:
\begin{equation}
    \label{eq:hange_of_variables}
    \chi_{\pm} = 1 \pm \sqrt{1-k_R^2} \ \  \ \ 
    \diff \chi_{\pm} = \mp \frac{k_R\,\diff k_R}{\sqrt{1-k_R^2}} \ \  \ \ 
    k_R = \sqrt{2\chi_{\pm} - \chi_{\pm}^2}
\end{equation}
\begin{equation}
    \label{eq:change_of_bounds}
    k_R = 0 \to \chi_+ = 2 \, , \ \chi_- = 0 \ \  \ \ 
    k_R = 1 \to \chi_+ = 1 \, , \ \chi_- = 1 \, .
\end{equation}
Implementing the change of variables above in 
Eqs.~\ref{eq:alpha0I2_zgt0} and \ref{eq:alpha0I2_zlt0} 
($\chi_+$ for \ref{eq:alpha0I2_zgt0} and $\chi_-$ for 
\ref{eq:alpha0I2_zlt0}) and relabeling the dummy 
integration variables as $x$ in both cases, we have
\begin{equation}
    \label{eq:alpha0I3_zgt0}
    \alpha_{I}(\mathbf{r}; z>0) = -\frac{2}{\mathcal{M}_Q} 
    \int_{1}^{2} \diff x J_0\left(\sqrt{2x - x^2} R\right) 
    \frac{\sin\left( x z\right)}{x}
\end{equation}
\begin{equation}
    \label{eq:alpha0I3_zlt0}
    \alpha_{I}(\mathbf{r}; z<0) = -\frac{2}{\mathcal{M}_Q} 
    \int_{0}^{1} \diff x J_0\left(\sqrt{2x - x^2} R\right) 
    \frac{\sin\left( x z\right)}{x} \, .
\end{equation}
At this point, we can combine the cases $z < 0$ and $z > 0$ by integrating from 0 to 2 in $x$. Plugging this into the general expression for the dynamical friction,  Eq.~\ref{eq:df_force}, gives Eq.~\ref{eq:df_point_source_lpt_int}. It is
clear that $\alpha_I$ is not symmetric in $z$ and 
therefore will contribute to the dynamical friction.

For completeness, we also evaluate $\alpha_{II}$:
\begin{equation}
    \label{eq:alpha0II2}
    \alpha_{II}(\mathbf{r}) = \frac{2}{\mathcal{M}_Q} 
    \int_{1}^{\infty} \diff k_R k_R J_0(k_R R) 
    \frac{e^{-\sqrt{k_R^2 -1}|z|}}{\sqrt{k_R^2 - 1}k_R^2}
    \left(\cos(z) + \sqrt{k_R^2 - 1} \sin(|z|) \right) \, .
\end{equation}
This can be further simplified by a change of coordinates to 
$x\equiv\sqrt{k_R^2 -1}$ so that the integral becomes:
\begin{equation}
    \label{eq:alpha0II3}
    \alpha_{II}(\mathbf{r}) = \frac{2}{\mathcal{M}_Q} 
    \int_{0}^{\infty} \diff x 
    \frac{J_0 \left( \sqrt{x^2 + 1}R\right)e^{-x|z|}}{x^2 + 1}
    \left(\cos(z) + x \sin(|z|) \right).
\end{equation}
As opposed to the previous term, this term is clearly symmetric in 
$z$ and will therefore not contribute to the dynamical friction force.

\section{Analytical Calculation for Truncated Isothermal Sphere}
\label{app:dist}
The density profile considered for most of this paper, the Plummer sphere, provides only one example of a possible matter distribution of the satellite.  While the expression for the dimensionless coefficient of dynamical friction $C_{\textrm{rel}}$ can only be found analytically in a small number of cases, it is still instructive to probe the end of the density spectrum opposite to the Plummer sphere in order to determine the impact of the satellite mass distribution on dynamical friction.  We therefore calculate $C_{\textrm{rel}}$ for a more concentrated model, namely an isothermal sphere with an exponential cutoff, below.  As in Appendix~\ref{app:ints_lpt}, we will drop all tildes for dimensionless quantities in order to simplify the notation, and all distances and wave vectors appearing in this Appendix will be in units of $\lambdabar$ or $1/\lambdabar$ as appropriate.
\par
The truncated isothermal sphere potential we use is given by
\begin{equation}
    \nabla^2 U_I=\frac{GM}{r^2\ell}e^{-r/\ell}\, ,
\end{equation}
where $G$ is the gravitational constant, $M$ is the mass of the satellite, $\ell$ is a characteristic satellite size, and $r$ is the radial distance from the center of the satellite.  Solving Poisson's equation above yields a potential of the form
\begin{equation}
    U_I=GM\left(\frac{e^{-r/\ell}-1}{r}+\frac{\textrm{Ei}\left(-\frac{r}{\ell}\right)}{\ell}\right)\, ,
\end{equation}
where $\textrm{Ei}\left(-\frac{r}{\ell}\right)$ is a one-argument exponential integral.  To find the dynamical friction for this density profile, we begin with the analog of Eq.~\ref{eq:four_tind_extended_psoure_lpt}, but instead of using the potential for the Plummer sphere, we use the potential for the isothermal sphere:
\begin{equation}
    \left({k}^4-4{k}_z^2\right)\hat{\alpha}=\frac{16\pi\lambdabar^3}{\mathcal{M}_Q}\frac{\tan^{-1}\left({k}{\ell}\right)}{{k}{\ell}}\, .
\end{equation}
The analog of Eq.~\ref{eq:alpha_sol2_extended} can be obtained by transforming to dimensionless variables and writing the integral
\begin{equation}
    \alpha(\mathbf{{r}})=\frac{16\pi}{(2\pi)^2\mathcal{M}_Q{\ell}}\int_0^\infty \textrm{d}{k}_R\int_{-\infty}^\infty \textrm{d}{k}_z \frac{{k}_R \tan^{-1}\left({k}{\ell}\right)J_0({k}_R{R})e^{i{k}_z{z}}}{{k}({k}^4-4{k}_z^2)}\, ;
\end{equation}
performing the contour integration over $k_z$ as detailed in Appendix~\ref{app:ints_lpt} and making the relevant change of variables yields 
\begin{equation}
    \alpha(\mathbf{{r}})=-\frac{2}{\mathcal{M}_Q{\ell}}\int \textrm{d}x  \tan^{-1}\left(\sqrt{2x}{\ell}\right)J_0(\sqrt{2x-x^2}{R})\frac{\sin(x{z})}{\sqrt{2}x^{3/2}}\, .\label{eq:dimless_alpha_isothermal}
\end{equation}
Finally, the force is given by Eq.~\ref{eq:df_def}, which in this case is
\begin{equation}
F_z=GM \lambdabar \rhobar\int \textrm{d}{R}\,\textrm{d}{z}\,\textrm{d}\theta \frac{\alpha({R}, {z}){R}{z}}{\left({z}^2+{R}^2\right)^{3/2}}\left(e^{-\frac{\sqrt{R^2 + {z}^2}}{\ell}}-1\right)\, ,
\end{equation}
where we have included the appropriate dimensional prefactors but all quantities in the integrand are dimensionless in units of $\lambdabar$.
Plugging Eq.~\ref{eq:dimless_alpha_isothermal} into this expression gives
\begin{multline}
    F_z=-\frac{4\pi GM \lambdabar\rhobar}{\sqrt{2}\mathcal{M}_Q{\ell}}\int\displaylimits_0^{b}\!\! \diff R \!\!\!\!\!\!
    \int\displaylimits^{\sqrt{b^2 - R^2}}_0 \!\!\!\!\!\!\!\diff z \int\displaylimits_0^2 \!\! \diff x \, \tan^{-1}\left(\sqrt{2x}{\ell}\right) \\
    \frac{{R}{z}\,J_0(\sqrt{2x-x^2}{R})\,\sin(x{z})}{x^{3/2}\left({z}^2+{R}^2\right)^{3/2}}
    \left(e^{-\frac{\sqrt{R^2 + {z}^2}}{\ell}}-1\right)\, ,
\end{multline}
or
\begin{equation}
    C_{\textrm{rel}}=-\frac{1}{\sqrt{2}{\ell}}\int\displaylimits_0^{b}\!\! \diff R\!\!\!\!\!\!
    \int\displaylimits^{\sqrt{b^2 - R^2}}_0 \!\!\!\!\!\!\!\diff z \int\displaylimits_0^2 \!\! \diff x \, \tan^{-1}\left(\sqrt{2x}{\ell}\right)\frac{{R}{z}\,J_0(\sqrt{2x-x^2}{R})\,\sin(x{z})}{x^{3/2}\left({z}^2+{R}^2\right)^{3/2}}
    \left(e^{-\frac{\sqrt{R^2 + {z}^2}}{\ell}}-1\right)\, .\label{eq:isotherm_crel}
\end{equation}
One can confirm that this solution reduces to that of a point mass in the limit $\ell\rightarrow0$.

\begin{figure}[t]
\centering
\includegraphics{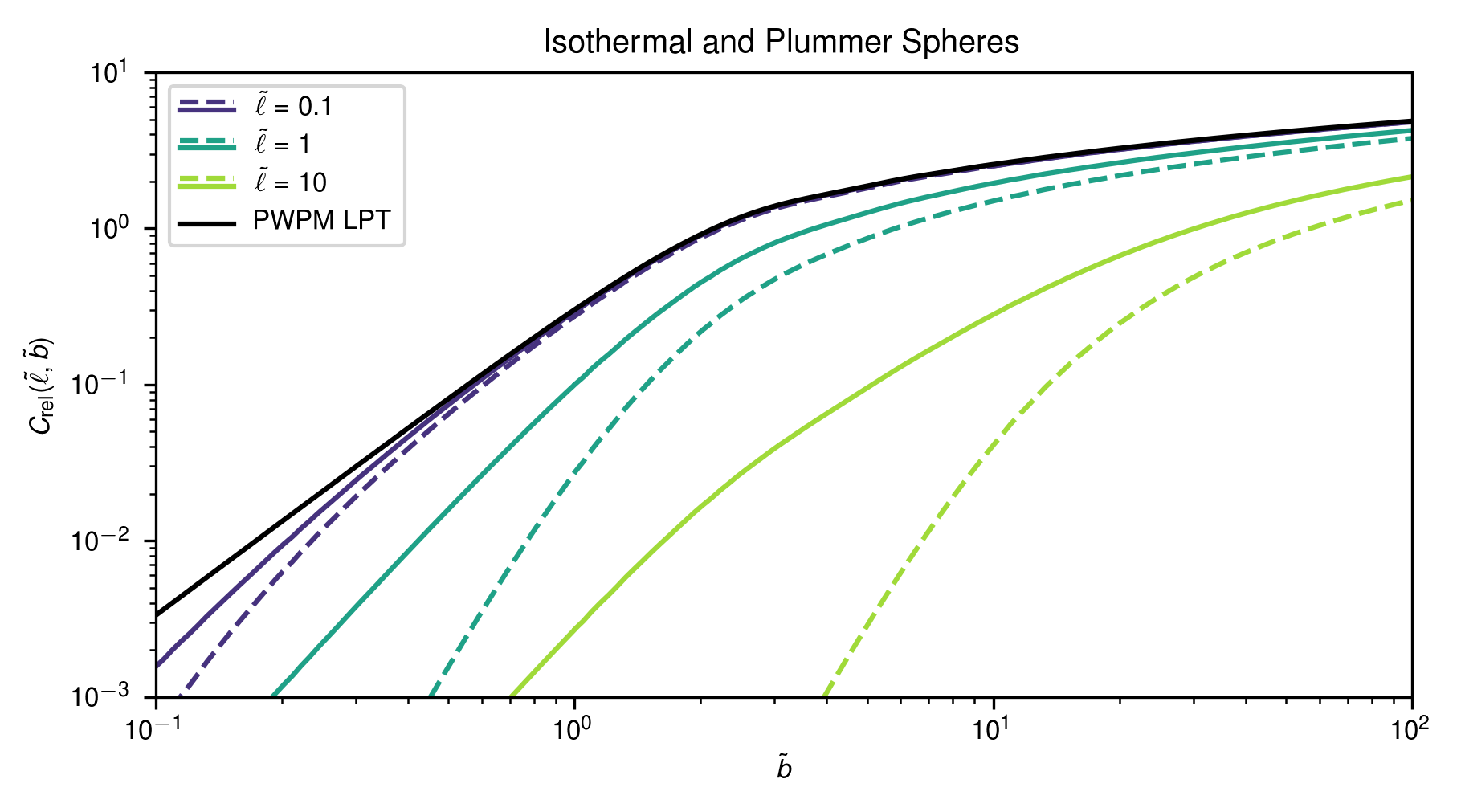}
\caption{A comparison of the curves generated from plotting Eq.~\ref{eq:isotherm_crel} (shown in solid lines) and its analog for the case of a Plummer sphere (Eq.~\ref{eq:df_extended_lpt_int}, shown in dashed lines) for three different satellite sizes, as a function of the dimensionless cutoff parameter $\tilde{b}$.  The black line is the analagous curve for the perturbative solution in the case of a plane wave, point mass, linear perturbation theory (PWPM) solution.
Note the qualitative similarities among all pairs of curves; where $b\gtrsim 100\ell$, the curves tend to agree for both the truncated isothermal and Plummer spheres, suggesting the dynamical friction is relatively insensitive to the matter distribution of the satellite in this regime.}
\label{fig:isotherm_and_plummer_comparison}
\end{figure}

\par
The resulting curves from plotting Eq.~\ref{eq:isotherm_crel} are depicted in Fig.~\ref{fig:isotherm_and_plummer_comparison}, alongside the analogous curves for the case of a Plummer sphere (Eq.~\ref{eq:df_extended_lpt_int}) for comparison.  Despite residing at opposite ends of the spectrum of profile densities, the curves are qualitatively very similar.  This demonstrates that the dynamical friction from FDM is instead more sensitive to other parameters describing the satellite, such as its size, rather than the precise mass distribution within the satellite.

\section{Techniques for Numerical Simulations}
\label{app:num}

Here, we outline the equations solved in our numerical simulations.
The Schr\"odinger system is evolved with the unitary spectral method of \cite{Mocz17}.
Our code solves Eq.~\ref{eqn:S1} with a Plummer profile satellite.
The code solves the equations in dimensionless form, normalizing against the length-scale $L_Q$ (this is in contrast to normalization by $\lambdabar$ in the analytic calculations of the paper).
We will indicate this by placing a breve ($\,\breve{\,}\,$) above dimensionless variables. Our dimensionless system of units for the simulation are:
\begin{equation}
\breve{x} \equiv \frac{x}{L_{\rm Q}}
\end{equation}
\begin{equation}
\breve{t} \equiv \frac{t}{t_{\rm Q}}
\end{equation}
\begin{equation}
\breve{\psi} \equiv \frac{\psi}{\sqrt{\rhobar}},
\end{equation}
where 
\begin{equation}
t_{\rm Q} \equiv \frac{m}{\hbar}L_{\rm Q}^2 
\end{equation}
is the characteristic quantum-wave timescale.
The satellite has dimensionless object size $\breve{\ell}=\ell/L_{\rm Q}$.

The Schr\"odinger equation thus becomes
\begin{equation}
i\frac{\partial}{\partial \breve{t}} \breve{\psi}
= 
\left[-\frac{\breve{\nabla}^2}{2} -\frac{1}{\sqrt{\breve{r}^2+\breve{\ell}^2}}   \right]  \breve{\psi} 
\end{equation}
with initial condition:
\begin{equation}
\breve{\psi}_0 = e^{i \mathcal{M}_{\rm Q}  \breve{z} }.
\end{equation}
The dimensionless satellite profile is:
\begin{equation}
\breve{\rho}_{\rm p} = \frac{3\breve{\ell}^2}{4\pi(\breve{\ell}^2+\breve{r}^2)^{5/2}}
\times \left(\frac{M_{\rm p}/L_{\rm Q}^3}{\rhobar}\right).
\end{equation}

The solution is updated via a second-order spectral method: each timestep can be broken into a half-step `kick' from the potential, carried out in real space, followed by a full-step `drift' from the kinetic operator carried out in Fourier space, followed by another half-step `kick', as follows:
\begin{equation}
\breve{\psi} \leftarrow \exp\left[-i \frac{\Delta \breve{t}}{2}\times\frac{-1}{\sqrt{\breve{r}^2+\breve{\ell}^2}}\right]\breve{\psi}
\end{equation}
\begin{equation}
\breve{\psi} \leftarrow 
{\rm ifft} \left[ \exp\left[ -i \Delta \breve{t} \times \breve{k}^2/2  \right] {\rm fft}\left[ \breve{\psi} \right] \right]
\end{equation}
\begin{equation}
\breve{\psi} \leftarrow \exp\left[-i \frac{\Delta \breve{t}}{2}\frac{-1}{\sqrt{\breve{r}^2+\breve{\ell}^2}}\right]\breve{\psi} 
\end{equation}
where ${\rm fft}$ and ${\rm ifft}$ are numerical fast Fourier transform and inverse fast Fourier transform operations.

The drag coefficient can be calculated from the resulting acceleration field of the wavefuncion.
The dimensionless acceleration field is:
\begin{equation}
\breve{\mathbf{a}} = -\nabla \breve{V} \cdot \left(  \frac{4\pi G \rhobar L_{\rm Q}^4 m^2}{\hbar^2} \right),
\end{equation}
where  $\breve{V}$ is the dimensionless self-potential, $V \equiv \breve{V} \cdot  (4\pi L_{\rm Q}^2 G \rhobar)$, 
given by Poisson's equation
\begin{equation}
\nabla^2 \breve{V} =  \left( |\breve{\psi}|^2- 1 \right).
\end{equation}
This can be solved numerically in Fourier space:
\begin{equation}
\breve{V} = 
{\rm ifft}\left[
-{\rm fft}\left[
\left( |\breve{\psi}|^2- 1 \right)
\right]/ \breve{k}^2
\right].
\end{equation}
Finally, the drag coefficient is
\begin{equation}
C_{\rm Q} = \frac{1}{4\pi}\int \breve{\rho}_{\rm p} \breve{a}_{\breve{z}} \,d\mathbf{\breve{x}}^3
=  -\int 
\frac{3\breve{\ell}^2  \partial_{\breve{z}}\breve{V} }{4\pi(\breve{\ell}^2+\breve{r}^2)^{5/2}} 
\,d\mathbf{\breve{x}}^3.
\end{equation}
We carry out this integral numerically on the discretized domain.

Note that there is a general scaling symmetry of the equations:
\begin{equation}
\{ x,t,\rho,m,M_{\rm p}\} \rightarrow \{ ax,bt,c\rho,a^{-2}bm,a^3b^{-2}M_{\rm p}\} \, ,
\end{equation}
which can be used to transform our solutions to different satellite masses and FDM particle masses.

\bibliography{mybib}

\providecommand{\href}[2]{#2}\begingroup\raggedright\begin{thebibliography}{100}

\bibitem{WMAP03}
D.~N. {Spergel}, L.~{Verde}, H.~V. {Peiris}, E.~{Komatsu}, M.~R. {Nolta}, C.~L.
  {Bennett} et~al., \emph{{First-Year Wilkinson Microwave Anisotropy Probe
  (WMAP) Observations: Determination of Cosmological Parameters}},
  \href{https://doi.org/10.1086/377226}{\emph{\apjs} {\bfseries 148} (2003)
  175} [\href{https://arxiv.org/abs/astro-ph/0302209}{{\ttfamily
  astro-ph/0302209}}].

\bibitem{Planck18}
{Planck Collaboration}, N.~{Aghanim}, Y.~{Akrami}, M.~{Ashdown}, J.~{Aumont},
  C.~{Baccigalupi} et~al., \emph{{Planck 2018 results. VI. Cosmological
  parameters}}, {\emph{arXiv e-prints} (2018) arXiv:1807.06209}
  [\href{https://arxiv.org/abs/1807.06209}{{\ttfamily 1807.06209}}].

\bibitem{BAODR12}
S.~{Alam}, M.~{Ata}, S.~{Bailey}, F.~{Beutler}, D.~{Bizyaev}, J.~A. {Blazek}
  et~al., \emph{{The clustering of galaxies in the completed SDSS-III Baryon
  Oscillation Spectroscopic Survey: cosmological analysis of the DR12 galaxy
  sample}}, \href{https://doi.org/10.1093/mnras/stx721}{\emph{\mnras}
  {\bfseries 470} (2017) 2617}
  [\href{https://arxiv.org/abs/1607.03155}{{\ttfamily 1607.03155}}].

\bibitem{2018MNRAS.475..676S}
V.~{Springel}, R.~{Pakmor}, A.~{Pillepich}, R.~{Weinberger}, D.~{Nelson},
  L.~{Hernquist} et~al., \emph{{First results from the IllustrisTNG
  simulations: matter and galaxy clustering}},
  \href{https://doi.org/10.1093/mnras/stx3304}{\emph{\mnras} {\bfseries 475}
  (2018) 676} [\href{https://arxiv.org/abs/1707.03397}{{\ttfamily
  1707.03397}}].

\bibitem{2018RPPh...81f6201R}
L.~{Roszkowski}, E.~M. {Sessolo} and S.~{Trojanowski}, \emph{{WIMP dark matter
  candidates and searches--current status and future prospects}},
  \href{https://doi.org/10.1088/1361-6633/aab913}{\emph{Reports on Progress in
  Physics} {\bfseries 81} (2018) 066201}
  [\href{https://arxiv.org/abs/1707.06277}{{\ttfamily 1707.06277}}].

\bibitem{MBKJBRev17}
J.~S. {Bullock} and M.~{Boylan-Kolchin}, \emph{{Small-Scale Challenges to the
  {\ensuremath{\Lambda}}CDM Paradigm}},
  \href{https://doi.org/10.1146/annurev-astro-091916-055313}{\emph{Annual
  Review of Astronomy and Astrophysics} {\bfseries 55} (2017) 343}
  [\href{https://arxiv.org/abs/1707.04256}{{\ttfamily 1707.04256}}].

\bibitem{PontzenGovernato12}
A.~{Pontzen} and F.~{Governato}, \emph{{How supernova feedback turns dark
  matter cusps into cores}},
  \href{https://doi.org/10.1111/j.1365-2966.2012.20571.x}{\emph{\mnras}
  {\bfseries 421} (2012) 3464}
  [\href{https://arxiv.org/abs/1106.0499}{{\ttfamily 1106.0499}}].

\bibitem{Read16}
J.~I. {Read}, O.~{Agertz} and M.~L.~M. {Collins}, \emph{{Dark matter cores all
  the way down}}, \href{https://doi.org/10.1093/mnras/stw713}{\emph{\mnras}
  {\bfseries 459} (2016) 2573}
  [\href{https://arxiv.org/abs/1508.04143}{{\ttfamily 1508.04143}}].

\bibitem{Zolotov12}
A.~{Zolotov}, A.~M. {Brooks}, B.~{Willman}, F.~{Governato}, A.~{Pontzen},
  C.~{Christensen} et~al., \emph{{Baryons Matter: Why Luminous Satellite
  Galaxies have Reduced Central Masses}},
  \href{https://doi.org/10.1088/0004-637X/761/1/71}{\emph{\apj} {\bfseries 761}
  (2012) 71} [\href{https://arxiv.org/abs/1207.0007}{{\ttfamily 1207.0007}}].

\bibitem{Wetzel16}
A.~R. {Wetzel}, P.~F. {Hopkins}, J.-h. {Kim}, C.-A. {Faucher-Gigu{\`e}re},
  D.~{Kere{\v{s}}} and E.~{Quataert}, \emph{{Reconciling Dwarf Galaxies with
  {\ensuremath{\Lambda}}CDM Cosmology: Simulating a Realistic Population of
  Satellites around a Milky Way-mass Galaxy}},
  \href{https://doi.org/10.3847/2041-8205/827/2/L23}{\emph{\apj} {\bfseries
  827} (2016) L23} [\href{https://arxiv.org/abs/1602.05957}{{\ttfamily
  1602.05957}}].

\bibitem{Hopkins18}
P.~F. {Hopkins}, A.~{Wetzel}, D.~{Kere{\v{s}}}, C.-A. {Faucher-Gigu{\`e}re},
  E.~{Quataert}, M.~{Boylan-Kolchin} et~al., \emph{{FIRE-2 simulations: physics
  versus numerics in galaxy formation}},
  \href{https://doi.org/10.1093/mnras/sty1690}{\emph{\mnras} {\bfseries 480}
  (2018) 800} [\href{https://arxiv.org/abs/1702.06148}{{\ttfamily
  1702.06148}}].

\bibitem{Dutton16}
A.~A. {Dutton}, A.~V. {Macci{\`o}}, J.~{Frings}, L.~{Wang}, G.~S. {Stinson},
  C.~{Penzo} et~al., \emph{{NIHAO V: too big does not fail - reconciling the
  conflict between {\ensuremath{\Lambda}}CDM predictions and the circular
  velocities of nearby field galaxies}},
  \href{https://doi.org/10.1093/mnrasl/slv193}{\emph{\mnras} {\bfseries 457}
  (2016) L74} [\href{https://arxiv.org/abs/1512.00453}{{\ttfamily
  1512.00453}}].

\bibitem{HOTW}
L.~{Hui}, J.~P. {Ostriker}, S.~{Tremaine} and E.~{Witten}, \emph{{Ultralight
  scalars as cosmological dark matter}},
  \href{https://doi.org/10.1103/PhysRevD.95.043541}{\emph{\prd} {\bfseries 95}
  (2017) 043541} [\href{https://arxiv.org/abs/1610.08297}{{\ttfamily
  1610.08297}}].

\bibitem{Goodman00}
J.~{Goodman}, \emph{{Repulsive dark matter}},
  \href{https://doi.org/10.1016/S1384-1076(00)00015-4}{\emph{New Astronomy}
  {\bfseries 5} (2000) 103}
  [\href{https://arxiv.org/abs/astro-ph/0003018}{{\ttfamily
  astro-ph/0003018}}].

\bibitem{Hu00fdm}
W.~{Hu}, R.~{Barkana} and A.~{Gruzinov}, \emph{{Fuzzy Cold Dark Matter: The
  Wave Properties of Ultralight Particles}},
  \href{https://doi.org/10.1103/PhysRevLett.85.1158}{\emph{\prl} {\bfseries 85}
  (2000) 1158} [\href{https://arxiv.org/abs/astro-ph/0003365}{{\ttfamily
  astro-ph/0003365}}].

\bibitem{SIDM01}
B.~D. {Wandelt}, R.~{Dave}, G.~R. {Farrar}, P.~C. {McGuire}, D.~N. {Spergel}
  and P.~J. {Steinhardt}, \emph{{Self-Interacting Dark Matter}},  in
  \emph{Sources and Detection of Dark Matter and Dark Energy in the Universe}
  (D.~B. {Cline}, ed.), p.~263, Jan, 2001,
  \href{https://arxiv.org/abs/astro-ph/0006344}{{\ttfamily astro-ph/0006344}}.

\bibitem{Berezhiani15}
L.~{Berezhiani} and J.~{Khoury}, \emph{{Theory of dark matter superfluidity}},
  \href{https://doi.org/10.1103/PhysRevD.92.103510}{\emph{\prd} {\bfseries 92}
  (2015) 103510} [\href{https://arxiv.org/abs/1507.01019}{{\ttfamily
  1507.01019}}].

\bibitem{Schive14}
H.-Y. {Schive}, T.~{Chiueh} and T.~{Broadhurst}, \emph{{Cosmic structure as the
  quantum interference of a coherent dark wave}},
  \href{https://doi.org/10.1038/nphys2996}{\emph{Nature Physics} {\bfseries 10}
  (2014) 496} [\href{https://arxiv.org/abs/1406.6586}{{\ttfamily 1406.6586}}].

\bibitem{Schive14b}
H.-Y. {Schive}, M.-H. {Liao}, T.-P. {Woo}, S.-K. {Wong}, T.~{Chiueh},
  T.~{Broadhurst} et~al., \emph{{Understanding the Core-Halo Relation of
  Quantum Wave Dark Matter from 3D Simulations}},
  \href{https://doi.org/10.1103/PhysRevLett.113.261302}{\emph{\prl} {\bfseries
  113} (2014) 261302} [\href{https://arxiv.org/abs/1407.7762}{{\ttfamily
  1407.7762}}].

\bibitem{Mocz17}
P.~{Mocz}, M.~{Vogelsberger}, V.~H. {Robles}, J.~{Zavala}, M.~{Boylan-Kolchin},
  A.~{Fialkov} et~al., \emph{{Galaxy formation with BECDM - I. Turbulence and
  relaxation of idealized haloes}},
  \href{https://doi.org/10.1093/mnras/stx1887}{\emph{\mnras} {\bfseries 471}
  (2017) 4559} [\href{https://arxiv.org/abs/1705.05845}{{\ttfamily
  1705.05845}}].

\bibitem{BarOr18}
B.~{Bar-Or}, J.-B. {Fouvry} and S.~{Tremaine}, \emph{{Relaxation in a Fuzzy
  Dark Matter Halo}}, {\emph{ArXiv e-prints} (2018) }
  [\href{https://arxiv.org/abs/1809.07673}{{\ttfamily 1809.07673}}].

\bibitem{Church19}
B.~V. {Church}, P.~{Mocz} and J.~P. {Ostriker}, \emph{{Heating of Milky Way
  disc stars by dark matter fluctuations in cold dark matter and fuzzy dark
  matter paradigms}}, \href{https://doi.org/10.1093/mnras/stz534}{\emph{\mnras}
  {\bfseries 485} (2019) 2861}
  [\href{https://arxiv.org/abs/1809.04744}{{\ttfamily 1809.04744}}].

\bibitem{2019PhRvL.123n1301M}
P.~{Mocz}, A.~{Fialkov}, M.~{Vogelsberger}, F.~{Becerra}, M.~A. {Amin},
  S.~{Bose} et~al., \emph{{First Star-Forming Structures in Fuzzy Cosmic
  Filaments}},
  \href{https://doi.org/10.1103/PhysRevLett.123.141301}{\emph{\prl} {\bfseries
  123} (2019) 141301} [\href{https://arxiv.org/abs/1910.01653}{{\ttfamily
  1910.01653}}].

\bibitem{MarshSilk14}
D.~J.~E. {Marsh} and J.~{Silk}, \emph{{A model for halo formation with axion
  mixed dark matter}},
  \href{https://doi.org/10.1093/mnras/stt2079}{\emph{\mnras} {\bfseries 437}
  (2014) 2652} [\href{https://arxiv.org/abs/1307.1705}{{\ttfamily 1307.1705}}].

\bibitem{Tremaine75}
S.~D. {Tremaine}, J.~P. {Ostriker} and J.~{Spitzer}, L., \emph{{The formation
  of the nuclei of galaxies. I. M31.}},
  \href{https://doi.org/10.1086/153422}{\emph{\apj} {\bfseries 196} (1975)
  407}.

\bibitem{Tremaine76}
S.~D. {Tremaine}, \emph{{The formation of the nuclei of galaxies. II. The local
  group.}}, \href{https://doi.org/10.1086/154085}{\emph{\apj} {\bfseries 203}
  (1976) 345}.

\bibitem{TW84}
S.~{Tremaine} and M.~D. {Weinberg}, \emph{{Dynamical friction in spherical
  systems.}}, \href{https://doi.org/10.1093/mnras/209.4.729}{\emph{\mnras}
  {\bfseries 209} (1984) 729}.

\bibitem{Weinberg85}
M.~D. {Weinberg}, \emph{{Evolution of barred galaxies by dynamical friction.}},
  \href{https://doi.org/10.1093/mnras/213.3.451}{\emph{\mnras} {\bfseries 213}
  (1985) 451}.

\bibitem{BBR80}
M.~C. {Begelman}, R.~D. {Blandford} and M.~J. {Rees}, \emph{{Massive black hole
  binaries in active galactic nuclei}},
  \href{https://doi.org/10.1038/287307a0}{\emph{\nat} {\bfseries 287} (1980)
  307}.

\bibitem{Yu02}
Q.~{Yu}, \emph{{Evolution of massive binary black holes}},
  \href{https://doi.org/10.1046/j.1365-8711.2002.05242.x}{\emph{\mnras}
  {\bfseries 331} (2002) 935}
  [\href{https://arxiv.org/abs/astro-ph/0109530}{{\ttfamily
  astro-ph/0109530}}].

\bibitem{Berezhiani:2019pzd}
L.~Berezhiani, B.~Elder and J.~Khoury, \emph{{Dynamical Friction in
  Superfluids}},  \href{https://arxiv.org/abs/1905.09297}{{\ttfamily
  1905.09297}}.

\bibitem{Safarzadeh2019}
M.~{Safarzadeh} and D.~N. {Spergel}, \emph{{Ultra-light Dark Matter is
  Incompatible with the Milky Way's Dwarf Satellites}}, {\emph{arXiv e-prints}
  (2019) arXiv:1906.11848} [\href{https://arxiv.org/abs/1906.11848}{{\ttfamily
  1906.11848}}].

\bibitem{1943ApJ....97..255C}
S.~{Chandrasekhar}, \emph{{Dynamical Friction. I. General Considerations: the
  Coefficient of Dynamical Friction.}},
  \href{https://doi.org/10.1086/144517}{\emph{\apj} {\bfseries 97} (1943) 255}.

\bibitem{Marochnik68}
L.~S. {Marochnik}, \emph{{A Test Star in a Stellar System.}}, {\emph{\sovast}
  {\bfseries 11} (1968) 873}.

\bibitem{Kalnajs72}
A.~J. {Kalnajs}, \emph{{Polarization Clouds and Dynamical Friction}},  in
  \emph{IAU Colloq. 10: Gravitational N-Body Problem} (M.~{Lecar}, ed.),
  vol.~31 of \emph{Astrophysics and Space Science Library}, p.~13, Jan, 1972,
  \href{https://doi.org/10.1007/978-94-010-2870-7_2}{DOI}.

\bibitem{Ostriker99}
E.~C. {Ostriker}, \emph{{Dynamical Friction in a Gaseous Medium}},
  \href{https://doi.org/10.1086/306858}{\emph{\apj} {\bfseries 513} (1999) 252}
  [\href{https://arxiv.org/abs/astro-ph/9810324}{{\ttfamily
  astro-ph/9810324}}].

\bibitem{2008gady.book.....B}
J.~{Binney} and S.~{Tremaine}, \emph{{Galactic Dynamics: Second Edition}}.
  Princeton University Press, 2008.

\bibitem{Weinberg86}
M.~D. {Weinberg}, \emph{{Orbital Decay of Satellite Galaxies in Spherical
  Systems}}, \href{https://doi.org/10.1086/163785}{\emph{\apj} {\bfseries 300}
  (1986) 93}.

\bibitem{GHPW15}
A.~H. {Guth}, M.~P. {Hertzberg} and C.~{Prescod-Weinstein}, \emph{{Do dark
  matter axions form a condensate with long-range correlation?}},
  \href{https://doi.org/10.1103/PhysRevD.92.103513}{\emph{\prd} {\bfseries 92}
  (2015) 103513} [\href{https://arxiv.org/abs/1412.5930}{{\ttfamily
  1412.5930}}].

\bibitem{Madelung27}
E.~{Madelung}, \emph{{Quantentheorie in hydrodynamischer Form}},
  \href{https://doi.org/10.1007/BF01400372}{\emph{Zeitschrift fur Physik}
  {\bfseries 40} (1927) 322}.

\bibitem{WidrowKaiser93}
L.~M. {Widrow} and N.~{Kaiser}, \emph{{Using the Schroedinger Equation to
  Simulate Collisionless Matter}},
  \href{https://doi.org/10.1086/187073}{\emph{\apj} {\bfseries 416} (1993)
  L71}.

\bibitem{2018PhRvD..97h3519M}
P.~{Mocz}, L.~{Lancaster}, A.~{Fialkov}, F.~{Becerra} and P.-H. {Chavanis},
  \emph{{Schr{\"o}dinger-Poisson-Vlasov-Poisson correspondence}},
  \href{https://doi.org/10.1103/PhysRevD.97.083519}{\emph{\prd} {\bfseries 97}
  (2018) 083519} [\href{https://arxiv.org/abs/1801.03507}{{\ttfamily
  1801.03507}}].

\bibitem{Li19}
X.~{Li}, L.~{Hui} and G.~L. {Bryan}, \emph{{Numerical and perturbative
  computations of the fuzzy dark matter model}},
  \href{https://doi.org/10.1103/PhysRevD.99.063509}{\emph{\prd} {\bfseries 99}
  (2019) 063509} [\href{https://arxiv.org/abs/1810.01915}{{\ttfamily
  1810.01915}}].

\bibitem{Lora12}
V.~{Lora}, J.~{Maga{\~n}a}, A.~{Bernal}, F.~J. {S{\'a}nchez-Salcedo} and E.~K.
  {Grebel}, \emph{{On the mass of ultra-light bosonic dark matter from galactic
  dynamics}}, \href{https://doi.org/10.1088/1475-7516/2012/02/011}{\emph{\jcap}
  {\bfseries 2} (2012) 011} [\href{https://arxiv.org/abs/1110.2684}{{\ttfamily
  1110.2684}}].

\bibitem{roetman1967biharmonic}
E.~Roetman, \emph{On the biharmonic wave equation}, {\emph{Pacific Journal of
  Mathematics} {\bfseries 22} (1967) 139}.

\bibitem{timoshenko1953history}
S.~Timoshenko, \emph{History of strength of materials mcgraw-hill book
  company}, {\emph{Inc., New York/Toronto/London} (1953) }.

\bibitem{bethe2013quantum}
H.~Bethe and E.~Salpeter, \emph{Quantum Mechanics of One- and Two-Electron
  Atoms}. Springer Berlin Heidelberg, 2013.

\bibitem{AtomicCollisions}
N.~F. {Mott} and H.~S.~W. {Massey}, \emph{{The theory of atomic collisions}}.
  Oxford University Press, 1949.

\bibitem{2019arXiv190804790Y}
E.~{Yarnell Davies} and P.~{Mocz}, \emph{{Fuzzy Dark Matter Soliton Cores
  around Supermassive Black Holes}}, {\emph{arXiv e-prints} (2019)
  arXiv:1908.04790} [\href{https://arxiv.org/abs/1908.04790}{{\ttfamily
  1908.04790}}].

\bibitem{Plummer}
H.~C. {Plummer}, \emph{{On the problem of distribution in globular star
  clusters}}, \href{https://doi.org/10.1093/mnras/71.5.460}{\emph{\mnras}
  {\bfseries 71} (1911) 460}.

\bibitem{NFW97}
J.~F. {Navarro}, C.~S. {Frenk} and S.~D.~M. {White}, \emph{{A Universal Density
  Profile from Hierarchical Clustering}},
  \href{https://doi.org/10.1086/304888}{\emph{\apj} {\bfseries 490} (1997) 493}
  [\href{https://arxiv.org/abs/astro-ph/9611107}{{\ttfamily
  astro-ph/9611107}}].

\bibitem{WhiteRees78}
S.~D.~M. {White} and M.~J. {Rees}, \emph{{Core condensation in heavy halos: a
  two-stage theory for galaxy formation and clustering.}},
  \href{https://doi.org/10.1093/mnras/183.3.341}{\emph{\mnras} {\bfseries 183}
  (1978) 341}.

\bibitem{Foster:2017hbq}
J.~W. Foster, N.~L. Rodd and B.~R. Safdi, \emph{{Revealing the Dark Matter Halo
  with Axion Direct Detection}},
  \href{https://doi.org/10.1103/PhysRevD.97.123006}{\emph{Phys. Rev.}
  {\bfseries D97} (2018) 123006}
  [\href{https://arxiv.org/abs/1711.10489}{{\ttfamily 1711.10489}}].

\bibitem{Coleman08}
M.~G. {Coleman} and J.~T.~A. {de Jong}, \emph{{A Deep Survey of the Fornax
  dSph. I. Star Formation History}},
  \href{https://doi.org/10.1086/589992}{\emph{\apj} {\bfseries 685} (2008) 933}
  [\href{https://arxiv.org/abs/0805.1365}{{\ttfamily 0805.1365}}].

\bibitem{Poretti08}
E.~{Poretti}, G.~{Clementini}, E.~V. {Held}, C.~{Greco}, M.~{Mateo},
  L.~{Dell'Arciprete} et~al., \emph{{Variable Stars in the Fornax dSph Galaxy.
  II. Pulsating Stars below the Horizontal Branch}},
  \href{https://doi.org/10.1086/591241}{\emph{\apj} {\bfseries 685} (2008) 947}
  [\href{https://arxiv.org/abs/0806.4453}{{\ttfamily 0806.4453}}].

\bibitem{Larsen12}
S.~S. {Larsen}, J.~{Strader} and J.~P. {Brodie}, \emph{{Constraints on mass
  loss and self-enrichment scenarios for the globular clusters of the Fornax
  dSph}}, \href{https://doi.org/10.1051/0004-6361/201219897}{\emph{\aap}
  {\bfseries 544} (2012) L14}
  [\href{https://arxiv.org/abs/1207.5792}{{\ttfamily 1207.5792}}].

\bibitem{Cole12}
D.~R. {Cole}, W.~{Dehnen}, J.~I. {Read} and M.~I. {Wilkinson}, \emph{{The mass
  distribution of the Fornax dSph: constraints from its globular cluster
  distribution}},
  \href{https://doi.org/10.1111/j.1365-2966.2012.21885.x}{\emph{\mnras}
  {\bfseries 426} (2012) 601}
  [\href{https://arxiv.org/abs/1205.6327}{{\ttfamily 1205.6327}}].

\bibitem{DeBoer16}
T.~J.~L. {de Boer} and M.~{Fraser}, \emph{{Four and one more: The formation
  history and total mass of globular clusters in the Fornax dSph}},
  \href{https://doi.org/10.1051/0004-6361/201527580}{\emph{\aap} {\bfseries
  590} (2016) A35} [\href{https://arxiv.org/abs/1510.05642}{{\ttfamily
  1510.05642}}].

\bibitem{DelPino17}
A.~{del Pino}, A.~{Aparicio}, S.~L. {Hidalgo} and E.~L. {{\L}okas},
  \emph{{Rotating stellar populations in the Fornax dSph galaxy}},
  \href{https://doi.org/10.1093/mnras/stw3016}{\emph{\mnras} {\bfseries 465}
  (2017) 3708} [\href{https://arxiv.org/abs/1605.09414}{{\ttfamily
  1605.09414}}].

\bibitem{Reid19Fornax}
J.~I. {Read}, M.~G. {Walker} and P.~{Steger}, \emph{{Dark matter heats up in
  dwarf galaxies}}, \href{https://doi.org/10.1093/mnras/sty3404}{\emph{\mnras}
  {\bfseries 484} (2019) 1401}
  [\href{https://arxiv.org/abs/1808.06634}{{\ttfamily 1808.06634}}].

\bibitem{Kowalczyk19}
K.~{Kowalczyk}, A.~{del Pino}, E.~L. {{\L}okas} and M.~{Valluri},
  \emph{{Schwarzschild dynamical model of the Fornax dwarf spheroidal galaxy}},
  \href{https://doi.org/10.1093/mnras/sty3100}{\emph{\mnras} {\bfseries 482}
  (2019) 5241} [\href{https://arxiv.org/abs/1807.07852}{{\ttfamily
  1807.07852}}].

\bibitem{Boldrini18}
P.~{Boldrini}, R.~{Mohayaee} and J.~{Silk}, \emph{{Does Fornax have a cored
  halo? Implications for the nature of dark matter}}, {\emph{arXiv e-prints}
  (2018) arXiv:1806.09591} [\href{https://arxiv.org/abs/1806.09591}{{\ttfamily
  1806.09591}}].

\bibitem{WangMeiDES18}
M.-Y. {Wang}, T.~{de Boer}, A.~{Pieres}, T.~S. {Li}, A.~{Drlica-Wagner}, S.~E.
  {Koposov} et~al., \emph{{The morphology and structure of stellar populations
  in the Fornax dwarf spheroidal galaxy from Dark Energy Survey Data}},
  {\emph{arXiv e-prints} (2018) arXiv:1809.07801}
  [\href{https://arxiv.org/abs/1809.07801}{{\ttfamily 1809.07801}}].

\bibitem{Aaronson83}
M.~{Aaronson}, \emph{{Accurate radial velocities for carbon stars in Draco and
  Ursa Minor :the first hint of a dwarf spheroidal mass-to-light ratio.}},
  \href{https://doi.org/10.1086/183969}{\emph{\apj} {\bfseries 266} (1983)
  L11}.

\bibitem{Mateo98}
M.~L. {Mateo}, \emph{{Dwarf Galaxies of the Local Group}},
  \href{https://doi.org/10.1146/annurev.astro.36.1.435}{\emph{Annual Review of
  Astronomy and Astrophysics} {\bfseries 36} (1998) 435}
  [\href{https://arxiv.org/abs/astro-ph/9810070}{{\ttfamily
  astro-ph/9810070}}].

\bibitem{Gilmore07}
G.~{Gilmore}, M.~{Wilkinson}, J.~{Kleyna}, A.~{Koch}, W.~{Evans}, R.~F.~G.
  {Wyse} et~al., \emph{{Observed Properties of Dark Matter: dynamical studies
  of dSph galaxies}},
  \href{https://doi.org/10.1016/j.nuclphysbps.2007.08.143}{\emph{Nuclear
  Physics B Proceedings Supplements} {\bfseries 173} (2007) 15}
  [\href{https://arxiv.org/abs/astro-ph/0608528}{{\ttfamily
  astro-ph/0608528}}].

\bibitem{WP11}
M.~G. {Walker} and J.~{Pe{\~n}arrubia}, \emph{{A Method for Measuring (Slopes
  of) the Mass Profiles of Dwarf Spheroidal Galaxies}},
  \href{https://doi.org/10.1088/0004-637X/742/1/20}{\emph{\apj} {\bfseries 742}
  (2011) 20} [\href{https://arxiv.org/abs/1108.2404}{{\ttfamily 1108.2404}}].

\bibitem{WangMeiClusterSix}
M.-Y. {Wang}, S.~{Koposov}, A.~{Drlica-Wagner}, A.~{Pieres}, T.~{Li}, T.~{de
  Boer} et~al., \emph{{Rediscovery of the Sixth Star Cluster in the Fornax
  Dwarf Spheroidal Galaxy}}, {\emph{arXiv e-prints} (2019) arXiv:1902.04589}
  [\href{https://arxiv.org/abs/1902.04589}{{\ttfamily 1902.04589}}].

\bibitem{Hernandez98}
X.~{Hernandez} and G.~{Gilmore}, \emph{{Dynamical friction in dwarf galaxies}},
  \href{https://doi.org/10.1046/j.1365-8711.1998.01511.x}{\emph{\mnras}
  {\bfseries 297} (1998) 517}
  [\href{https://arxiv.org/abs/astro-ph/9802261}{{\ttfamily
  astro-ph/9802261}}].

\bibitem{Oh2000}
K.~S. {Oh}, D.~N.~C. {Lin} and H.~B. {Richer}, \emph{{Globular Clusters in the
  Fornax Dwarf Spheroidal Galaxy}},
  \href{https://doi.org/10.1086/308477}{\emph{\apj} {\bfseries 531} (2000)
  727}.

\bibitem{Goerdt06}
T.~{Goerdt}, B.~{Moore}, J.~I. {Read}, J.~{Stadel} and M.~{Zemp}, \emph{{Does
  the Fornax dwarf spheroidal have a central cusp or core?}},
  \href{https://doi.org/10.1111/j.1365-2966.2006.10182.x}{\emph{\mnras}
  {\bfseries 368} (2006) 1073}
  [\href{https://arxiv.org/abs/astro-ph/0601404}{{\ttfamily
  astro-ph/0601404}}].

\bibitem{FornaxDES}
M.-Y. {Wang}, T.~{de Boer}, A.~{Pieres}, T.~S. {Li}, A.~{Drlica-Wagner}, S.~E.
  {Koposov} et~al., \emph{{The morphology and structure of stellar populations
  in the Fornax dwarf spheroidal galaxy from Dark Energy Survey Data}},
  {\emph{arXiv e-prints} (2018) arXiv:1809.07801}
  [\href{https://arxiv.org/abs/1809.07801}{{\ttfamily 1809.07801}}].

\bibitem{FornaxAge2}
A.~{del Pino}, S.~L. {Hidalgo}, A.~{Aparicio}, C.~{Gallart}, R.~{Carrera},
  M.~{Monelli} et~al., \emph{{Spatial dependence of the star formation history
  in the central regions of the Fornax dwarf spheroidal galaxy}},
  \href{https://doi.org/10.1093/mnras/stt833}{\emph{\mnras} {\bfseries 433}
  (2013) 1505} [\href{https://arxiv.org/abs/1305.2166}{{\ttfamily 1305.2166}}].

\bibitem{Kaur18}
K.~{Kaur} and S.~{Sridhar}, \emph{{Stalling of Globular Cluster Orbits in Dwarf
  Galaxies}}, \href{https://doi.org/10.3847/1538-4357/aaeacf}{\emph{\apj}
  {\bfseries 868} (2018) 134}
  [\href{https://arxiv.org/abs/1810.00369}{{\ttfamily 1810.00369}}].

\bibitem{Mackey03}
A.~D. {Mackey} and G.~F. {Gilmore}, \emph{{Surface brightness profiles and
  structural parameters for globular clusters in the Fornax and Sagittarius
  dwarf spheroidal galaxies}},
  \href{https://doi.org/10.1046/j.1365-8711.2003.06275.x}{\emph{\mnras}
  {\bfseries 340} (2003) 175}
  [\href{https://arxiv.org/abs/astro-ph/0211396}{{\ttfamily
  astro-ph/0211396}}].

\bibitem{Mateo91}
M.~{Mateo}, E.~{Olszewski}, D.~L. {Welch}, P.~{Fischer} and W.~{Kunkel},
  \emph{{A Kinematic Study of the Fornax Dwarf Spheroid Galaxy}},
  \href{https://doi.org/10.1086/115923}{\emph{\aj} {\bfseries 102} (1991) 914}.

\bibitem{Walker09}
M.~G. {Walker}, M.~{Mateo}, E.~W. {Olszewski}, J.~{Pe{\~n}arrubia}, N.~W.
  {Evans} and G.~{Gilmore}, \emph{{A Universal Mass Profile for Dwarf
  Spheroidal Galaxies?}},
  \href{https://doi.org/10.1088/0004-637X/704/2/1274}{\emph{\apj} {\bfseries
  704} (2009) 1274} [\href{https://arxiv.org/abs/0906.0341}{{\ttfamily
  0906.0341}}].

\bibitem{Read19}
J.~I. {Read}, M.~G. {Walker} and P.~{Steger}, \emph{{Dark matter heats up in
  dwarf galaxies}}, \href{https://doi.org/10.1093/mnras/sty3404}{\emph{\mnras}
  {\bfseries 484} (2019) 1401}
  [\href{https://arxiv.org/abs/1808.06634}{{\ttfamily 1808.06634}}].

\bibitem{Ibata94}
R.~A. {Ibata}, G.~{Gilmore} and M.~J. {Irwin}, \emph{{A dwarf satellite galaxy
  in Sagittarius}}, \href{https://doi.org/10.1038/370194a0}{\emph{\nat}
  {\bfseries 370} (1994) 194}.

\bibitem{Johnston95}
K.~V. {Johnston}, D.~N. {Spergel} and L.~{Hernquist}, \emph{{The Disruption of
  the Sagittarius Dwarf Galaxy}},
  \href{https://doi.org/10.1086/176247}{\emph{\apj} {\bfseries 451} (1995) 598}
  [\href{https://arxiv.org/abs/astro-ph/9502005}{{\ttfamily
  astro-ph/9502005}}].

\bibitem{Helmi04}
A.~{Helmi}, \emph{{Velocity Trends in the Debris of Sagittarius and the Shape
  of the Dark Matter Halo of Our Galaxy}},
  \href{https://doi.org/10.1086/423340}{\emph{\apjl} {\bfseries 610} (2004)
  L97} [\href{https://arxiv.org/abs/astro-ph/0406396}{{\ttfamily
  astro-ph/0406396}}].

\bibitem{LJM05}
D.~R. {Law}, K.~V. {Johnston} and S.~R. {Majewski}, \emph{{A Two Micron All-Sky
  Survey View of the Sagittarius Dwarf Galaxy. IV. Modeling the Sagittarius
  Tidal Tails}}, \href{https://doi.org/10.1086/426779}{\emph{\apj} {\bfseries
  619} (2005) 807} [\href{https://arxiv.org/abs/astro-ph/0407566}{{\ttfamily
  astro-ph/0407566}}].

\bibitem{Belokurov06}
V.~{Belokurov}, D.~B. {Zucker}, N.~W. {Evans}, G.~{Gilmore}, S.~{Vidrih}, D.~M.
  {Bramich} et~al., \emph{{The Field of Streams: Sagittarius and Its
  Siblings}}, \href{https://doi.org/10.1086/504797}{\emph{\apjl} {\bfseries
  642} (2006) L137} [\href{https://arxiv.org/abs/astro-ph/0605025}{{\ttfamily
  astro-ph/0605025}}].

\bibitem{LM10}
D.~R. {Law} and S.~R. {Majewski}, \emph{{The Sagittarius Dwarf Galaxy: A Model
  for Evolution in a Triaxial Milky Way Halo}},
  \href{https://doi.org/10.1088/0004-637X/714/1/229}{\emph{\apj} {\bfseries
  714} (2010) 229} [\href{https://arxiv.org/abs/1003.1132}{{\ttfamily
  1003.1132}}].

\bibitem{Purcell11}
C.~W. {Purcell}, J.~S. {Bullock}, E.~J. {Tollerud}, M.~{Rocha} and
  S.~{Chakrabarti}, \emph{{The Sagittarius impact as an architect of spirality
  and outer rings in the Milky Way}},
  \href{https://doi.org/10.1038/nature10417}{\emph{\nat} {\bfseries 477} (2011)
  301} [\href{https://arxiv.org/abs/1109.2918}{{\ttfamily 1109.2918}}].

\bibitem{Koposov12}
S.~E. {Koposov}, V.~{Belokurov}, N.~W. {Evans}, G.~{Gilmore}, M.~{Gieles},
  M.~J. {Irwin} et~al., \emph{{The Sagittarius Streams in the Southern Galactic
  Hemisphere}}, \href{https://doi.org/10.1088/0004-637X/750/1/80}{\emph{\apj}
  {\bfseries 750} (2012) 80} [\href{https://arxiv.org/abs/1111.7042}{{\ttfamily
  1111.7042}}].

\bibitem{LM16}
D.~R. {Law} and S.~R. {Majewski}, \emph{{The Sagittarius Dwarf Tidal
  Stream(s)}},  in \emph{Tidal Streams in the Local Group and Beyond} (H.~J.
  {Newberg} and J.~L. {Carlin}, eds.), vol.~420, p.~31, Jan, 2016,
  \href{https://doi.org/10.1007/978-3-319-19336-6_2}{DOI}.

\bibitem{DL17}
M.~I.~P. {Dierickx} and A.~{Loeb}, \emph{{Predicted Extension of the
  Sagittarius Stream to the Milky Way Virial Radius}},
  \href{https://doi.org/10.3847/1538-4357/836/1/92}{\emph{\apj} {\bfseries 836}
  (2017) 92} [\href{https://arxiv.org/abs/1611.00089}{{\ttfamily 1611.00089}}].

\bibitem{Hernitschek17}
N.~{Hernitschek}, B.~{Sesar}, H.-W. {Rix}, V.~{Belokurov},
  D.~{Martinez-Delgado}, N.~F. {Martin} et~al., \emph{{The Geometry of the
  Sagittarius Stream from Pan-STARRS1 3{\ensuremath{\pi}} RR Lyrae}},
  \href{https://doi.org/10.3847/1538-4357/aa960c}{\emph{\apj} {\bfseries 850}
  (2017) 96} [\href{https://arxiv.org/abs/1710.09436}{{\ttfamily 1710.09436}}].

\bibitem{Fardal19}
M.~A. {Fardal}, R.~P. {van der Marel}, D.~R. {Law}, S.~T. {Sohn}, B.~{Sesar},
  N.~{Hernitschek} et~al., \emph{{Connecting the Milky Way potential profile to
  the orbital time-scales and spatial structure of the Sagittarius Stream}},
  \href{https://doi.org/10.1093/mnras/sty3428}{\emph{\mnras} {\bfseries 483}
  (2019) 4724} [\href{https://arxiv.org/abs/1804.04995}{{\ttfamily
  1804.04995}}].

\bibitem{JB00}
I.-G. {Jiang} and J.~{Binney}, \emph{{The orbit and mass of the Sagittarius
  dwarf galaxy}},
  \href{https://doi.org/10.1046/j.1365-8711.2000.03311.x}{\emph{\mnras}
  {\bfseries 314} (2000) 468}
  [\href{https://arxiv.org/abs/astro-ph/9908025}{{\ttfamily
  astro-ph/9908025}}].

\bibitem{GAIA}
{Gaia Collaboration}, T.~{Prusti}, J.~H.~J. {de Bruijne}, A.~G.~A. {Brown},
  A.~{Vallenari}, C.~{Babusiaux} et~al., \emph{{The Gaia mission}},
  \href{https://doi.org/10.1051/0004-6361/201629272}{\emph{\aap} {\bfseries
  595} (2016) A1} [\href{https://arxiv.org/abs/1609.04153}{{\ttfamily
  1609.04153}}].

\bibitem{DR2}
{Gaia Collaboration}, A.~G.~A. {Brown}, A.~{Vallenari}, T.~{Prusti}, J.~H.~J.
  {de Bruijne}, C.~{Babusiaux} et~al., \emph{{Gaia Data Release 2. Summary of
  the contents and survey properties}},
  \href{https://doi.org/10.1051/0004-6361/201833051}{\emph{\aap} {\bfseries
  616} (2018) A1}.

\bibitem{Bird18}
S.~A. {Bird}, X.-X. {Xue}, C.~{Liu}, J.~{Shen}, C.~{Flynn} and C.~{Yang},
  \emph{{Anisotropy of the Milky Way's stellar halo using K giants from LAMOST
  and $Gaia$}}, {\emph{ArXiv e-prints} (2018) }
  [\href{https://arxiv.org/abs/1805.04503}{{\ttfamily 1805.04503}}].

\bibitem{Lancaster19a}
L.~{Lancaster}, S.~E. {Koposov}, V.~{Belokurov}, N.~W. {Evans} and A.~J.
  {Deason}, \emph{{The halo's ancient metal-rich progenitor revealed with BHB
  stars}}, \href{https://doi.org/10.1093/mnras/stz853}{\emph{\mnras} {\bfseries
  486} (2019) 378} [\href{https://arxiv.org/abs/1807.04290}{{\ttfamily
  1807.04290}}].

\bibitem{Besla07}
G.~{Besla}, N.~{Kallivayalil}, L.~{Hernquist}, B.~{Robertson}, T.~J. {Cox},
  R.~P. {van der Marel} et~al., \emph{{Are the Magellanic Clouds on Their First
  Passage about the Milky Way?}},
  \href{https://doi.org/10.1086/521385}{\emph{\apj} {\bfseries 668} (2007) 949}
  [\href{https://arxiv.org/abs/astro-ph/0703196}{{\ttfamily
  astro-ph/0703196}}].

\bibitem{Kalli13}
N.~{Kallivayalil}, R.~P. {van der Marel}, G.~{Besla}, J.~{Anderson} and
  C.~{Alcock}, \emph{{Third-epoch Magellanic Cloud Proper Motions. I. Hubble
  Space Telescope/WFC3 Data and Orbit Implications}},
  \href{https://doi.org/10.1088/0004-637X/764/2/161}{\emph{\apj} {\bfseries
  764} (2013) 161} [\href{https://arxiv.org/abs/1301.0832}{{\ttfamily
  1301.0832}}].

\bibitem{Erkal19}
D.~{Erkal}, V.~{Belokurov}, C.~F.~P. {Laporte}, S.~E. {Koposov}, T.~S. {Li},
  C.~J. {Grillmair} et~al., \emph{{The total mass of the Large Magellanic Cloud
  from its perturbation on the Orphan stream}},
  \href{https://doi.org/10.1093/mnras/stz1371}{\emph{\mnras} {\bfseries 487}
  (2019) 2685} [\href{https://arxiv.org/abs/1812.08192}{{\ttfamily
  1812.08192}}].

\bibitem{GC19}
N.~{Garavito-Camargo}, G.~{Besla}, C.~F.~P. {Laporte}, K.~V. {Johnston}, F.~A.
  {G{\'o}mez} and L.~L. {Watkins}, \emph{{Hunting for the Dark Matter Wake
  Induced by the Large Magellanic Cloud}}, {\emph{arXiv e-prints} (2019)
  arXiv:1902.05089} [\href{https://arxiv.org/abs/1902.05089}{{\ttfamily
  1902.05089}}].

\bibitem{Nidever19}
D.~L. {Nidever}, K.~{Olsen}, Y.~{Choi}, T.~J.~L. {de Boer}, R.~D. {Blum}, E.~F.
  {Bell} et~al., \emph{{Exploring the Very Extended Low-surface-brightness
  Stellar Populations of the Large Magellanic Cloud with SMASH}},
  \href{https://doi.org/10.3847/1538-4357/aafaf7}{\emph{\apj} {\bfseries 874}
  (2019) 118} [\href{https://arxiv.org/abs/1805.02671}{{\ttfamily
  1805.02671}}].

\bibitem{Choi19}
Y.~{Choi}, D.~{Nidever}, K.~{Olsen}, R.~{Blum}, G.~{Besla} and D.~{Zaritsky},
  \emph{{SMASHing the LMC: A Tidally-induced Warp in the Outer LMC and a Large
  Scale Reddening Map}},  in \emph{American Astronomical Society Meeting
  Abstracts \#233}, vol.~233 of \emph{American Astronomical Society Meeting
  Abstracts}, p.~416.06, Jan, 2019.

\bibitem{Khmelnitsky:2013lxt}
A.~Khmelnitsky and V.~Rubakov, \emph{{Pulsar timing signal from ultralight
  scalar dark matter}},
  \href{https://doi.org/10.1088/1475-7516/2014/02/019}{\emph{JCAP} {\bfseries
  1402} (2014) 019} [\href{https://arxiv.org/abs/1309.5888}{{\ttfamily
  1309.5888}}].

\bibitem{Porayko:2018sfa}
N.~K. Porayko et~al., \emph{{Parkes Pulsar Timing Array constraints on
  ultralight scalar-field dark matter}},
  \href{https://doi.org/10.1103/PhysRevD.98.102002}{\emph{Phys. Rev.}
  {\bfseries D98} (2018) 102002}
  [\href{https://arxiv.org/abs/1810.03227}{{\ttfamily 1810.03227}}].

\bibitem{watanabe2014integral}
K.~Watanabe, \emph{Integral transform techniques for {G}reen's function}.
  Springer, 2014.

\end{thebibliography}\endgroup
\end{document}